\documentclass[aps, prx, twocolumn, superscriptaddress, amsmath, amssymb, floatfix,longbibliography, notitlepage]{revtex4-2}
\usepackage{amsbsy}
\usepackage{amsfonts}
\usepackage{amsmath}
\usepackage{amstext}
\usepackage{amsthm}
\usepackage{amssymb}

\usepackage{graphicx}

\usepackage{epsfig}
\usepackage{epstopdf}
\usepackage{bm}
\usepackage{color}
\usepackage[dvipsnames]{xcolor}
\definecolor{prlblue}{rgb}{0.18,0.19,0.57}
\usepackage{multirow}
\usepackage[colorlinks]{hyperref}
\usepackage{cleveref}
\hypersetup{citecolor=prlblue, linkcolor=prlblue, filecolor=prlblue, urlcolor=prlblue}
\newcommand{\figref}[1]{Fig.\,\ref{#1}}
\newcommand{\secref}[1]{Sec.\,\ref{#1}}
\newcommand{\eqnref}[1]{Eq.\,\eqref{#1}}

\begin{document}
\graphicspath{{figures/}}

\title{Thermal Hall Effect and Neutral Spinons
in a Doped Mott Insulator }

\author{Zhi-Jian Song}
\affiliation{Institute for Advanced Study, Tsinghua University, Beijing 100084, China}

\author{Jia-Xin Zhang}
\email{zjx19@mails.tsinghua.edu.cn}
\affiliation{Institute for Advanced Study, Tsinghua University, Beijing 100084, China}

\author{Zheng-Yu Weng}
\affiliation{Institute for Advanced Study, Tsinghua University, Beijing 100084, China}

\date{\today}

\begin{abstract}
 In the pseudogap phase of the cuprate, a thermal Hall response of neutral objects has been recently detected experimentally, which continuously persists into the antiferromagnetic insulating phase. In this work, we study the transport properties of neutral spinons as the elementary excitation of a doped Mott insulator, which is governed by a mutual Chern-Simons topological gauge structure. We show that such a chiral spinon as a composite of an $S=1/2$ spin sitting at the core of a supercurrent vortex, can contribute to the thermal Hall effect, thermopower, and Hall effect due to its intrinsic transverse (cyclotron) motion under internal fictitious fluxes. In particular, the magnitudes of the transport coefficients are phenomenologically determined by two basic parameters: the doping concentration and $T_c$, quantitatively consistent with the experimental measurements including the signs and qualitative temperature and magnetic field dependence. Combined with the predictions of the spinon longitudinal transport properties, including the Nernst and spin Hall effects, a phenomenological description of the pseudogap phase is established as characterized by the neutral spinon excitations, which eventually become ``confined'' with an intrinsic superconducting transition at $T_c$. Finally, within this theoretical framework, the ``order to order'' phase transition between the superconducting and antiferromagnetic insulating phases are briefly discussed, with the thermal Hall monotonically increasing into the latter. 
\end{abstract}

\maketitle

\tableofcontents

\section{Introduction}
Transport measurements serve as a powerful tool to gain insight into the nature of elementary excitations in the cuprates \cite{Wen.Lee.2006z4, Zaanen.Keimer.2015}. The anomalous signals detected in these measurements are crucial for a systematic understanding at the microscopic level. For instance, the Hall number in the cuprates indicates a discontinuity at a doping $p^*$, which corresponds to the doping concentration at which the pseudogap (PG) phase terminates \cite{Taillefer.Proust.2019, Proust.Badoux.2016}. Within the PG phase, when $p<p^*$,  the Hall number aligns with the doping density $p$, which seemingly contrasts with free systems where the large Fermi surface encloses an area of $1+p$ as indicated experimentally at $p>p^*$. Previous studies have hypothesized that this discrepancy might stem from Fermi surface reconstruction due to antiferromagnetic (AFM) order with $Q=(\pi,\pi)$ \cite{Storey.Storey.2016} or in the absence of the explicit translation symmetry breaking due to strong correlations \cite{Sachdev.Zhang.2020, Zhang.Nikolaenko.2021}.

Furthermore, a linear magnetic-field dependent thermal Hall signal \cite{Taillefer.Grissonnanche.2020,Taillefer.Boulanger.2020,Taillefer.Grissonnanche.2019} in the family of the cuprate compounds has been recently observed at $p<p*$, extending to the AFM insulating phase. It is important to underscore that the experimental signal exhibits no effect for the magnetic field that is aligned parallel to the copper-oxide plane \cite{Taillefer.Boulanger.2020}
, which implies that the thermal Hall effect is originated from an orbit effect. Prior theoretical studies \cite{Lee.Katsura.2010, Lee.Han.2019} suggest that in the case of the cuprates, magnons on a square lattice will fail to yield a nonzero thermal Hall conductivity when subjected to either the Dzyaloshinskii-Moriya spin interaction or the localized formation of skyrmion defects. Additionally,  the phenomenological descriptions involve neutral excitations like spinons \cite{Sachdev.Guo.2020, Lee.Han.2019, Sachdev.Samajdar.2019, zhang_thermal_2023} and phonons \cite{Sachdev.Guo.2021, Sachdev.Guo.2022, Sun.Chen.2020} have been proposed. Phenomenologically a universal behavior with a scaling law was proposed \cite{Zhang.Yang.2020}. Besides the cuprates, the sizeable thermal Hall effect has been also found in spin-ice $\text{Tb}_2\text{Ti}_2\text{O}_7$ \cite{hirschberger_large_2015} and spin liquid $\text{RuCl}_3$ \cite{kasahara_unusual_2018} as well as the Kitaev materials \cite{watanabe_emergence_2016}. Moreover, both integer \cite{jezouin_quantum_2013,banerjee_observed_2017} and fractional \cite{banerjee_observation_2018} quantum Hall systems (QHE \& FQHE) offer a unique perspective, where the thermal Hall effect finds its explanation in the conformal field theory (CFT) of chiral edge modes \cite{kane_quantized_1997}. 

\begin{figure}[t]
	\centering
	\includegraphics[width=\linewidth]{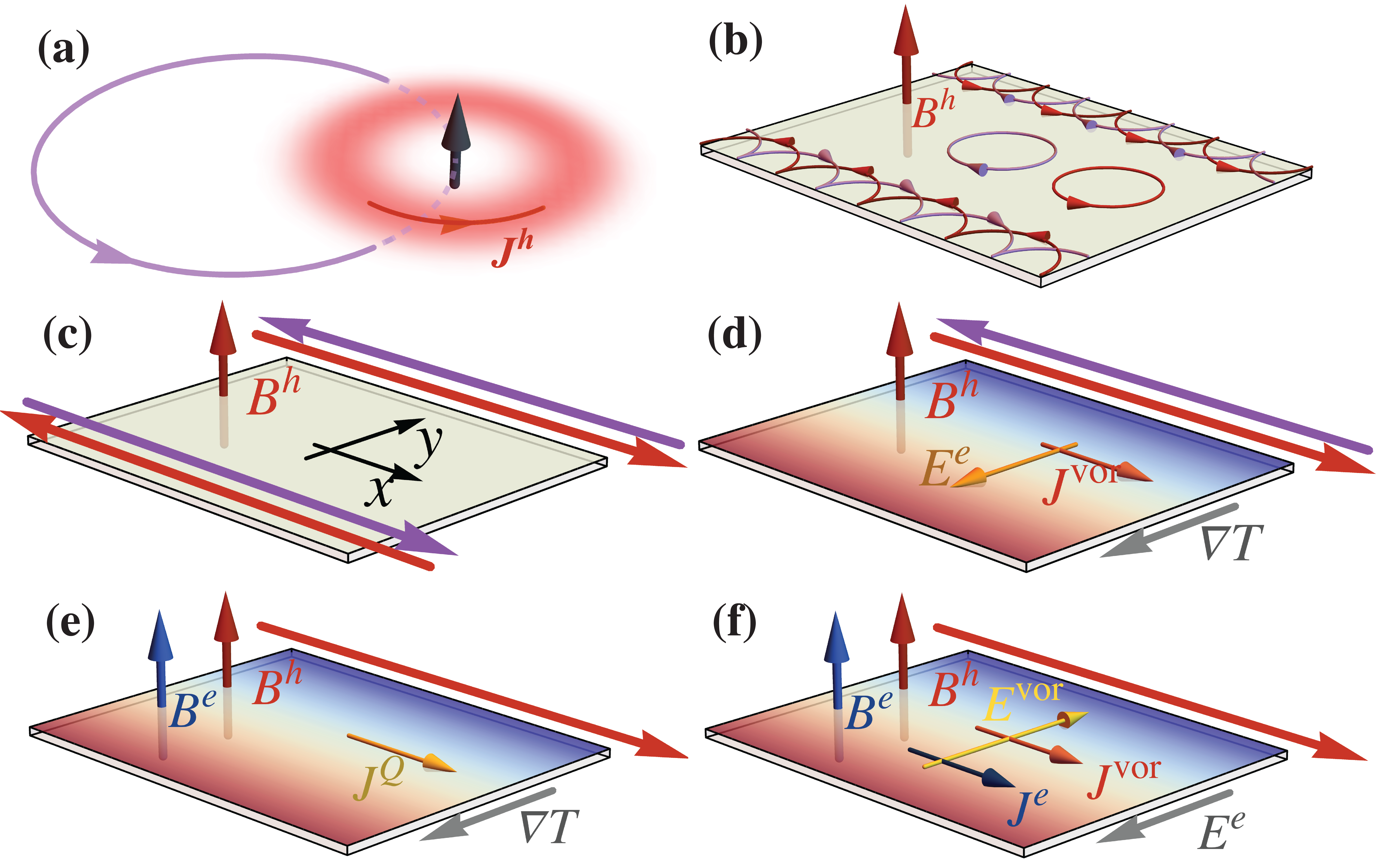}
	\caption{(a) A schematic illustration of a chiral spinon with a mutual Chern-Simons gauge structure: a neutral $S=1/2$ spin (black arrow) sitting at the core of an induced supercurrent vortex (red ring), which itself sees an intrinsic magnetic field $B^h$ to undergo a cyclotron motion (purple circle) self-consistently; (b) The semi-classical behavior of the chiral spinons under a uniform $B^h$ in the bulk and edges of the sample, and (c) the resultant vortex edge currents with opposite chiralities. Note that in the equilibrium state no time-reversal symmetry is broken as the opposite spins see the opposite sign of $B^h$ with the compensation of the opposite cyclotron motion and edge currents (red and purple with arrows);   (d) Temperature gradient $\nabla T$ breaks the equilibrium between the opposite edges, causing net vortex current $\boldsymbol{J}^{\text{vor}}$ and electric field in the sample, which contributes to a thermopower effect; (e) External perpendicular magnetic field $B^e$ (blue arrow) and the in-plane temperature gradient $\nabla T$ break the balance of the chirality of spinon-vortices to generatie a net thermal current $\boldsymbol{J}^Q$, which contributes to a thermal Hall effect; (f) An applied charge current $\boldsymbol{J}^{e}$ induces a force $\boldsymbol{E}^\text{vor}$ on the spinon-vortex, which generates a vortex current $\boldsymbol{J}^{\text{vor}}$ in the presence of an external magnetic field $B^e$ to produce a net electric field $E^e$. The resulting Hall effect gives rise to the Hall number precisely equal to the doping concentration $\delta$ at low temperature.}
	\label{fig_sketch}
\end{figure}

The transport measurements provide a direct probe into the contribution of the excitations that dominate the PG phase. The neutral spinon in the PG phase is an essential elementary excitation in the strongly correlated theories of doped Mott insulators.  Thus, if and how the neutral spinon participates in the thermal Hall and other transport phenomena become important issues that should be addressed very seriously. In this paper, we shall make a self-consistent study of spinon transport within the framework of the phase-string theory \cite{Weng.Weng.2011,Weng.Ma.2014}. The spinon predicted in this theory is different from either the slave-boson or slave-fermion approaches \cite{Wen.Lee.2006z4, Anderson.Baskaran.1987, Anderson.Anderson.1987, PhysRevB.38.316,Read.Sachdev.1991} due to the so-called phase-string effect \cite{Weng.Sheng.1996, Zaanen.Wu.2008, Lu.Weng_2023} hidden in the $t$-$J$ model upon doping, which is a topological Berry phase replacing the usual Fermi sign structure in the restricted Hilbert space. Specifically:

\begin{enumerate}
	\item Each spinon undergoes a cyclotron motion due to an intrinsic Berry curvature caused by the phase-string effect [cf. \figref{fig_sketch}(a)]. The time-reversal symmetry is retained in the absence of external magnetic fields as the opposite spins see the opposite fictitious fluxes with the opposite chiral edge currents [cf. \figref{fig_sketch}(b-c)]. 
    The system is distinct from the usual topological insulator \cite{Zhang.Bernevig.2006,Zhang.K.2007} in that all the spinons with opposite chiralities are RVB-paired in the ground state. 
    \item Each spinon is always locked with a charge-current vortex. Since an external perpendicular magnetic field must be balanced by the net (polarized) vortices, then unpaired (free) \emph{chiral} spinons must be generated from the RVB condensate, which contributes to the novel transport in the PG phase.    
    \item The cyclotron motion of the spinon and its locking with the charge vortex [cf. \figref{fig_sketch}(a)] are mathematically characterized by a mutual Chern-Simons gauge structure, which will contribute to unconventional transport phenomena, including the thermopower effect [cf. \figref{fig_sketch}(d)], thermal Hall [cf. \figref{fig_sketch}(e)], and Hall effect [cf. \figref{fig_sketch}(f)]. 
\end{enumerate}

We shall investigate the above spinon transport by using a semiclassical approach based on the mutual Chern-Simons gauge theory. The calculated results are essentially determined by the basic parameters of doping concentration as well as $T_c$, with the magnitudes comparable with the experimental measurements \cite{Taillefer.Proust.2019, Proust.Badoux.2016, Taillefer.Grissonnanche.2020,Taillefer.Boulanger.2020,Taillefer.Grissonnanche.2019, Taillefer.Daou.2009, Taillefer.Lizaire.2021}.
Also more physical implications arising from such neutral spinons as the elementary excitations will be briefly discussed in \secref{other}, which can explain the Nernst effect \cite{Ong.Wang.2001, Hardy.Wang.2002, Ong.Wang.2003} and the scaling relationship between $T_c$ and the spin resonance energy as observed in neutron measurements \cite{Dogan.Dai.1996, Aksay.Fong.1997, Keimer.He.20006wc, Colson.Gallais.2002}. Especially an ``order-to-order'' phase transition between the superconducting and AFM insulating phases can also naturally emerge within the same theoretical framework.

{ Finally, some general remarks on the microscopic theoretical description of the cuprate may be in order. The present fractionalization is based on a unified scenario \cite{Weng.Weng.2011,Weng.Ma.2014}, in which the close relationship between the AFM, superconducting, and PG phases, is realized via the phase-string effect induced by holons that effectively reduces the AFM long-range order into a short-range AFM ordered state at finite doping. The latter is superconducting in the ground state, while the phase coherence is destroyed by the proliferation of the free spinon-vortices as the elementary excitations above $T_c$ to give rise to a lower PG phase \cite{Weng.Weng.2011,Weng.Ma.2014}. Consequently, the experimental transport measurements in such a lower PG regime will serve as a crucial test of the unique behavior of the spinon-vortices as the building blocks of the fractionalization described by a mutual Chern-Simons gauge theory (see below). How those transport phenomena may be explained by distinct excitations in different theories with or without fractionalization can be highly revealing. As such, the essential role played by the spinon-vortex composite as to be explored in this paper will effectively distinguish the underlying fractionalization scheme from other approaches including the Fermi-liquid-like frameworks. For example, a recent Fermi-liquid Hartree-Fock calculation \cite{Laughlin.2014} also suggests a systematic phase diagram for the cuprate. Whether the transport properties contributed by the Landau quasiparticles may be comparable with the experiments, like the aforementioned thermal Hall effect, etc., would be intriguing. In general, given the mounting experimental data in the cuprate, a comparative and systematic theoretical investigation within any self-consistent scheme can be very useful to advance our understanding of the high-$T_c$ mechanism.  }

\section{Mutual Chern-Simons gauge theory of the doped Mott insulator}


\subsection{Topological gauge structure}
Phase-string theory of the doped Mott insulator is based on a nontrivial sign structure identified in both the $t$-$J$ model \cite{Weng.Sheng.1996, Zaanen.Wu.2008, Lu.Weng_2023} and the Hubbard model \cite{Weng.Zhang.2014, Xu.Weng_2023}, in which the conventional Fermi statistics of the electrons are replaced by the phase-string sign structure in the restricted Hilbert space of the lower (upper) Hubbard band. In the phase-string theory, such a sign structure is further precisely mapped to a topological gauge structure. The corresponding low-energy description involves the mutual Chern-Simons (MCS) gauge interaction between the spin and charge degrees of freedom \cite{Weng.Kou.2003yuc, kou_mutual-chern-simons_2005, Qi.Weng.2006, Weng.Qi.2007,Weng.Ye.2011, Weng.Ye.2012}, governed by the lattice Euclidean Lagrangian $L=L_h+L_s+L_{\mathrm{MCS}}$ as follows:
\begin{eqnarray}
	L_h&=&\sum_I h_I^{\dagger}\left[\partial_\tau-i A_0^s(I)-i  A_0^e (I)+\mu_h\right] h_I\notag\\
	&\;&-t_h \sum_{i\alpha}\left[h_I^{\dagger} h_{I+\alpha} e^{i \boldsymbol{A}_{\alpha}^s(I)+i  \boldsymbol{A}_{\alpha}^e(I) }+\text{h.c.}\right],\label{Lh}\\
	L_s&=&\sum_{i \sigma} b_{i \sigma}^{\dagger}\left[\partial_\tau-i \sigma A_0^h\left(i\right)+\lambda_b+\frac{1}{2} g \mu_B B^e \sigma\right] b_{i \sigma}\notag\\
	&\;&-J_s \sum_{i\alpha \sigma}\left[b_{i \sigma}^{\dagger} b_{i+\alpha, \bar \sigma}^{\dagger} e^{i \sigma \boldsymbol{A}_{\alpha}^h (i)}+\text {h.c.}\right],\label{Ls}\\
	L_{\mathrm{MCS}}&=&\frac{i}{\pi} \sum_i \epsilon^{\mu \nu \lambda} A_\mu^s(I) \partial_\nu A_\lambda^h (i)\label{LCS},
\end{eqnarray}
in which $L_h$ and $L_s$ describe the dynamics of the matter fields—bosonic spinless holon $h_I$, and bosonic neutral spinon $b_{i\sigma}$ (with $\bar \sigma\equiv-\sigma$), respectively. The indices $\alpha$ and $\beta$ denote only the spatial components: $x, y$, and the indices $\mu=(\tau, \boldsymbol{r})$ label the full time-space vector, with the indices $i$, $I$ representing the two-dimensional (2D) square lattice site and its dual lattice site, respectively.
{$\lambda_b$ and $\mu_h$ are the chemical potentials for the spinon $b$ and holon $h$, whose numbers are conserved, respectively. The hole doping concentration $\delta$ will be introduced via $\mu_h$. The renormalized hopping strength $t_h$ and effective spin antiferromagnetic (AFM) coupling $J_s$ are 
determined at a generalized mean-field level in Ref. \onlinecite{Weng.Ma.2014}. }
The external magnetic vector potential $\boldsymbol{A}^e$ with the field strength $B^e$ perpendicular to the 2D plane, interacts with the charge (holon) degree of freedom through the orbit effect (setting the charge equal to one) in Eq. (\ref{Lh}), and the spin degree of freedom via a Zeeman effect in Eq. (\ref{Ls}). 

Here the holon field $h$ and spinon field $b$ minimally couple to the gauge fields, $A^s_\mu$ and $A^h_\mu$, respectively, with the MCS topological structure given in \eqnref{LCS}. It implies that the holon (spinon) number $n_I^h$ ($n_i^b$) will determine the gauge-field strength of $A^h_\mu$ ($A^s_\mu$) as if each matter particle (holon or spinon) is attached to a fictitious $\pi$ flux tube visible only by a different species. This can be directly seen by considering the following equations of motion for $A_0^s$ and $A_0^h$, respectively:
\begin{eqnarray}
	\frac{\partial L}{\partial A_0^s(I)}=0 &\Rightarrow& \pi n_I^h=\epsilon^{\alpha \beta} \partial_\alpha \boldsymbol{A}_\beta^h(i)\equiv B^h,\label{conAs} \\
	\frac{\partial L}{\partial A_0^h(i)}=0 &\Rightarrow& \pi \sum_\sigma \sigma n_{i \sigma}^b=\epsilon^{\alpha \beta} \partial_\alpha \boldsymbol{A}_\beta^s(I)\equiv B^s.\label{conAh}
\end{eqnarray}

Similarly, by using the charge (holon) current $\boldsymbol{J}_\alpha^{h}(I)\equiv\partial {L}_{h}/\partial \boldsymbol{A}_\alpha^s(I)$ and spin current associated with the $b$-spinon: $\boldsymbol{J}_\alpha^\text{spin}(i)\equiv\partial {L}_{s}/\partial \boldsymbol{A}_\alpha^h(i)$, one has the following equations of motion for $\boldsymbol{A}_\alpha^s(I)$ and $\boldsymbol{A}_\alpha^h(i)$, respectively:
\begin{eqnarray}
	\frac{\partial L}{\partial \boldsymbol{A}_\alpha^h(i)}=0 &\Rightarrow& \pi \boldsymbol{J}_\alpha^{\operatorname{spin}}(i)=\epsilon_{\alpha \beta} \boldsymbol{E}_\beta^s(i),\label{Jspin}\\
	\frac{\partial L}{\partial \boldsymbol{A}_\alpha^s(I)}=0 &\Rightarrow& \pi \boldsymbol{J}_\alpha^{h}(I)= \epsilon_{\alpha \beta} \boldsymbol{E}_\beta^h(I),\label{Jh}
\end{eqnarray}
where $\boldsymbol{E}_\alpha^{s/h}=\partial_0 \boldsymbol{A}_\alpha^{s/h}-\partial_\alpha A_0^{s/h}$. Therefore, due to the $U(1)\times U(1)$ mutual Chern-Simons gauge structure, the conserved charge (holon) and spin density-currents are constrained to the internal gauge field strengths by the equations of motion in Eqs.(\ref{conAs})-(\ref{Jh}).  

\subsection{Low-temperature pseudogap phase.}
At half-filling with $n_I^h=0$, one has $\boldsymbol{A}_\beta^h(i)=0$, and $L\rightarrow L_s$ reduces to the Schwinger-boson mean-field state Lagrangian \cite{PhysRevB.38.316} that well describes the AFM phase. On the other hand, at finite doping, the Bose condensation of the bosonic holon field will define a low-temperature PG phase \cite{Qi.Weng.2006,Weng.Qi.2007}. 
As the holons are condensed, the total gauge fluctuations in $L_h$ of \eqnref{Lh} will be suppressed due to the Higgs mechanism, leading to 
\begin{equation}\label{As}
\boldsymbol{A}_{\alpha}^s(I)+ \boldsymbol{A}_{\alpha}^e(I)-2\pi m_\alpha(I)=0,
\end{equation} 
where $m_\alpha\in\mathbb{Z}$ comes from the compactness of the spatial components in \eqnref{Lh}. By using \eqnref{As}, the equations of motion \eqnref{conAh} and \eqnref{Jspin} can be reformulated as:
\begin{eqnarray}
	&\;&\pi \sum_\sigma \sigma n_{i \sigma}^b  - 2 \pi J_0^{2\pi}(i) +\Phi^e(i)  =0,\label{consHiggs} \\
    &\;&\pi \boldsymbol{J}_\alpha^{\text{spin}}(i)- 2 \pi \boldsymbol{J}_\alpha^{2\pi}(i)+\epsilon_{\alpha \beta} \boldsymbol{E}_\beta^e(i)=0,\label{JspinHiggs}
\end{eqnarray}
where $\Phi^e=\epsilon_{\alpha\beta}\Delta_\alpha  \boldsymbol{A}_\beta^e$ and $\boldsymbol{E}_\alpha^{e}=\partial_0 \boldsymbol{A}_\alpha^{e}-\partial_\alpha A_0^{e}$ represent the external magnetic flux and external electric field strength, respectively. Here, $J_0^{\mathrm{2\pi}} \equiv \epsilon^{\alpha \beta} \Delta_\alpha m_\beta \in \mathbb{Z}$ denotes the number of $2\pi$ vortices in the holon condensate, and  
{$\boldsymbol{J}_\alpha^{\mathrm{2\pi}} \equiv -\epsilon^{\alpha \beta} \partial_0 m_\beta$} 
represents the  current of the $2\pi$ vortices. In other words, \eqnref{consHiggs} corresponds to the fact that in the original holon language, each vortex with $J_0^{\mathrm{2\pi}}=\pm 1$ has a phase winding $\pm 2\pi$, while each spinon carries a half-vortex with a phase winding $\pm \pi$, known as the spinon-vortex \cite{Qi.Weng.2006, Weng.Qi.2007}.

In the ground state, when all vortices are in the confined phase \cite{Weng.Mei.20107w}, the superconducting phase coherence is realized with $\Phi^e=0$ in \eqnref{consHiggs} (i.e., the Meissner effect). Here the $b$-spinons are in the RVB pairing state according to \eqnref{Ls} and the $2\pi$ vortices of $J_0^{\mathrm{vor}}=\pm 1$ are also ``confined'' in vortex-antivortex pairs. In such an SC phase, a single spinon cannot be present in the bulk, but an $S=1$ excitation (totally with $\pm 2\pi$ vortex due to the double spinons) can be made since a $\mp 2\pi$ vortex of $J_0^{\mathrm{vor}}$ can be always bound to the $S=1$ excitation to make the total $\Phi^e=0$ in Eq. (\ref{consHiggs}). A minimal flux quantization condition of $\Phi^e=\pi$ ($= hc/2e\equiv \phi_0$ if the full units are restored) can be realized in \eqnref{consHiggs} with a single $b$-spinon trapped at the magnetic vortex core. The thermally excited free (unpaired) $b$-spinons can eventually destroy the Meissner effect with a uniform magnetic field penetrating the bulk according to \eqnref{consHiggs}, which disorders the SC phase coherence and leads to a Kosterlitz–Thouless (KT) like phase transition at [cf. more details in 
{Appendix~\ref{AppBTc}} 
and Ref. \onlinecite{Weng.Mei.20107w}]
\begin{equation}\label{Tc}
	T_c \approx E_s/3k_B,
\end{equation}
where $E_s$ is the lowest excited energy of the $b$-spinons, to be elaborated below. 

\begin{figure}[t]
	\centering    \includegraphics[width=\linewidth]{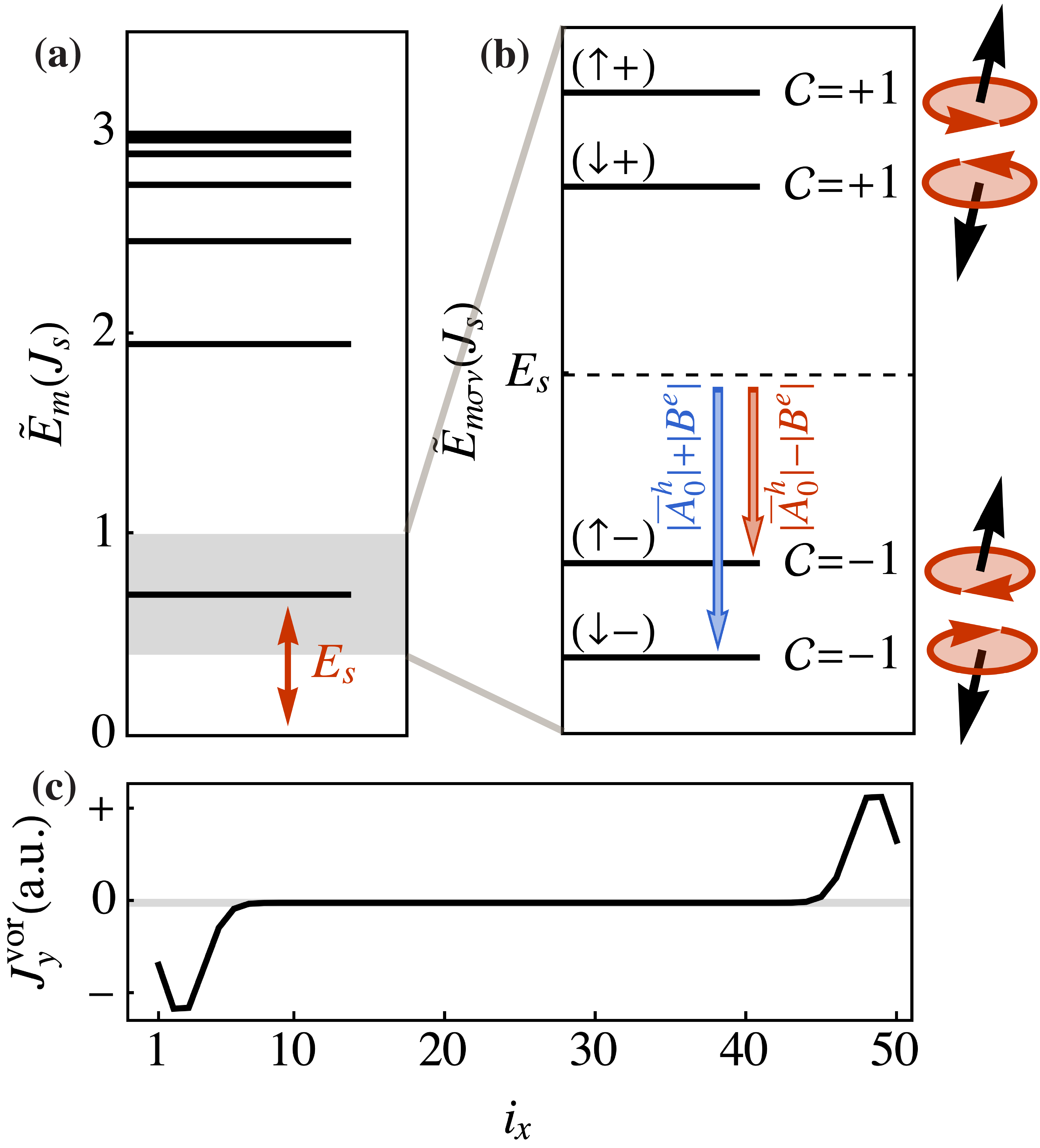}
	\caption{(a) The spinon energy levels $\tilde{E}_{m}$ (at $\delta=0.2$). The lowest excitations have an energy gap $E_s$ indicated by the red arrow; (b) The energy splitting of the lowest excitation level in the presence of a perpendicular magnetic field $B^e$. This corresponding window is marked by the gray region in (a). Each energy level in (b) is labeled with the quantum number $(\sigma\nu)$, with a corresponding diagram, i.e., a spinon trapped in the charge vortex core, illustrated on the right-hand side. Here the Chern number $\mathcal{C}$ and the splitting energy are also indicated; (c) The distribution of the vortex current $\boldsymbol{j}_\alpha^{\mathrm{vor}}(i) $ along the $y$ direction in the ground state, which is calculated on a sample with a periodic boundary condition along the $y$ direction and an open boundary condition along the $x$ direction. The length of the sample in the $x$ direction is $N_x=50$.}
	\label{fig_Eb4}
\end{figure}

At $T$ slightly above $T_c$, i.e., the lower PG regime, the conventional $2\pi$ vortices may remain well confined (vortex-antivortex paired) as their unpaired configuration would cost more free energy than that of the free $\pi$-spinon-vortices. To the leading order of approximation, one may then only focus on the spinon-vortex composites without considering the free $2\pi$ vortices in \eqnref{consHiggs} and \eqnref{JspinHiggs} unless the temperature is much higher than $T_c$ \cite{Weng.Qi.2007}. Note that a conventional $2\pi$ vortex can be still bound to a spinon to merely change the vorticity sign of the associated vortex as mentioned above. Namely, the low-energy elementary excitations consist of four types of excited (unpaired) $b$-spinons trapped in the vortex cores with quantum numbers of $\sigma=\pm 1$ and $\nu=\pm 1$, where $\sigma$ is the spin index and $\nu$ denotes the chirality of the vortex [illustrated on the right-hand-side of \figref{fig_Eb4}(b)]:
\begin{eqnarray}
	\sum_\sigma \sigma n_{i \sigma}^b  - 2 J_0^{2\pi}(i)&\Rightarrow& \sum_\nu \nu n_{i \sigma\nu}^b ,\\ 
 \boldsymbol{J}_\alpha^{\text{spin}}(i)- 2  \boldsymbol{J}_\alpha^{2\pi}(i)&\Rightarrow& \sum_\nu \nu \boldsymbol{J}_\alpha^\nu(i)\equiv \boldsymbol{J}_\alpha^{\text{vor}}(i),
\end{eqnarray}
where $n_{i \sigma\nu}^b$ denotes the number of excited free spinons with spin index $\sigma$ and vorticity $\nu$, and $\boldsymbol{J}_\alpha^{\mathrm{vor}}$ denotes the currents of the spinon-vortices, with $\boldsymbol{J}_\alpha^{\nu= \pm}$ representing the spinon current with $\pm$ chirality. 
Correspondingly, the equations of motion \eqnref{consHiggs} and \eqnref{JspinHiggs} of the mutual Chern-Simons gauge description reduce to the following forms
\begin{eqnarray}
	\sum_\nu \nu n_{i \sigma\nu}^b &=&-\pi^{-1}\Phi^e_i,\label{const}\\
 \boldsymbol{J}_\alpha^{\text{vor}}(i) &=&- \pi^{-1}\epsilon_{\alpha \beta} \boldsymbol{E}_\beta^e(i)\label{jE},
\end{eqnarray}



\begin{figure}[t]
	\centering
	\includegraphics[width=\linewidth]{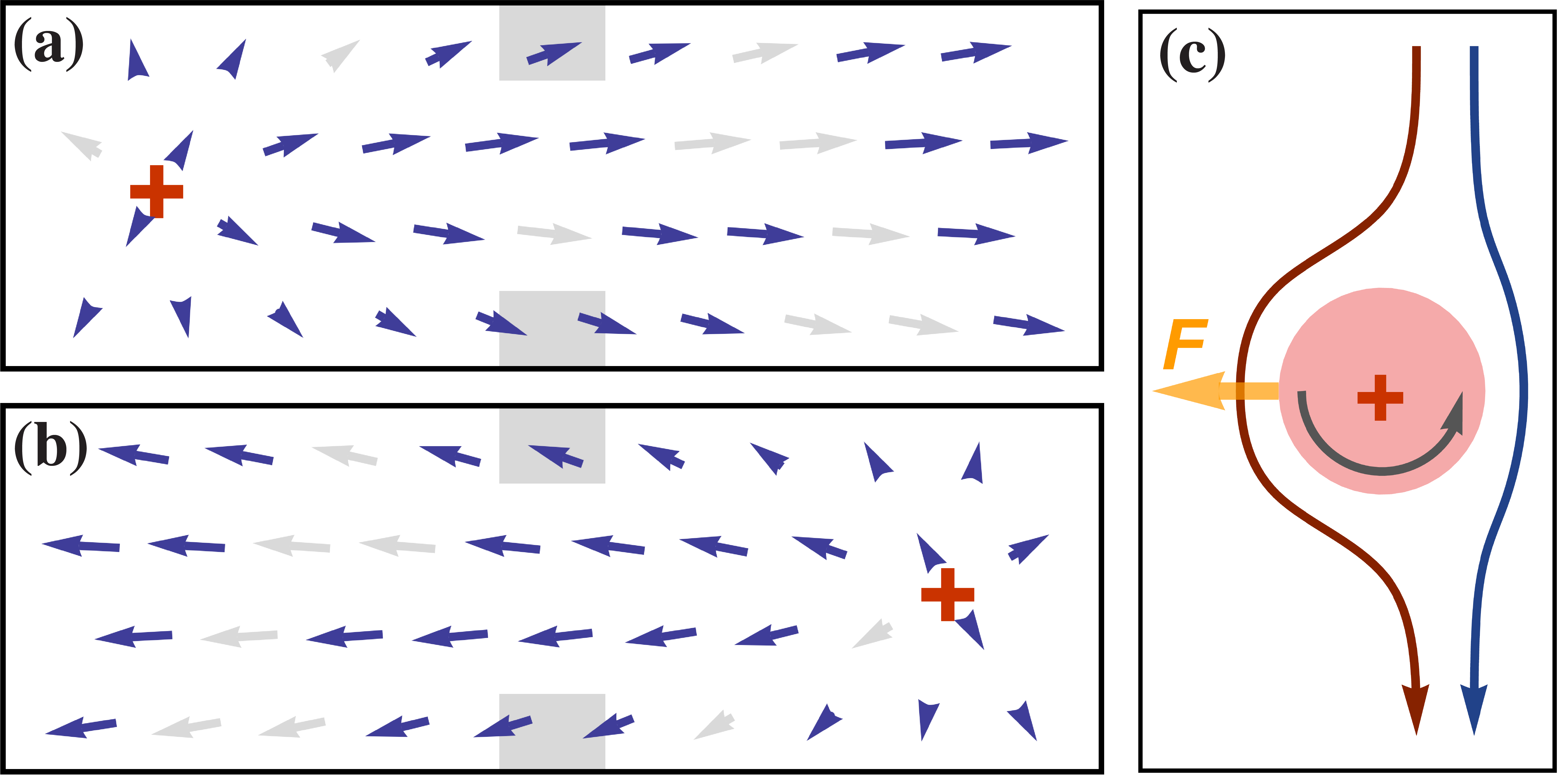}
	\caption{Schematic illustration of the spinon-vortex motion from one side of the sample [(a)] to the other [(b)]. This traverse along the horizontal direction results in a phase difference between the two opposite sides (indicated by grays) along the vertical direction changes by a $\pm 2\pi$ continuously, leading to an electric field given in \eqnref{jE}. Here the vortex core is denoted by a red ``$+$'' symbol, while the background local phases are represented by blue arrows; (c) The currents flowing along the two sides of a spinning entity, represented by a red disk (the arrow within the disk marks the direction of rotation). Here the red current exhibits a higher velocity compared to its blue counterpart, leading to a ``Magnus'' force $F$ exerting on the spinning entity (yellow arrow) as a fluid-dynamic interpretation of \eqnref{dualjE}.}
 
	\label{fig_vortex}
\end{figure}




Here, \eqnref{const} is actually the ``chirality-neutral'' condition, and \eqnref{jE} indicates that the current of the spinon-vortices $
\boldsymbol{J}_\alpha^{\mathrm{vor}}$ along one direction is induced by an external electric field along the perpendicular direction. Physically, the latter case can be interpreted as the steady vortex motion resulting in a $2\pi$ ``phase slip'' of the charge field between the opposite sides of perpendicular to the motion direction [cf. \figref{fig_vortex}(a) and (b)], and thereby generating an electric field, namely the Nernst effect (see in section \ref{Nernst}). 

Finally, the holon current $\boldsymbol{J}^h$ corresponds to the charge current, which can be denoted by $\boldsymbol{J}^e$ in the following. A spinon perceives the gauge field $\sigma A_\mu^h$ in \eqnref{Lh}, which results in \eqnref{Jh} where $\boldsymbol{E}^h$ is an effective ``electric'' field acting on the spinon of spin $\sigma=1$, which also denotes the vorticity of the original spinon-vortex composite. Note that a spinon-vortex of $\sigma=-1$ should experience an opposite force for the same direction of $\boldsymbol{J}^e$. Now such a spinon-vortex can be bound to a $\pm 2\pi$ vortex to change its vorticity to $\nu=\pm$, which becomes independent of $\sigma$ as given in \eqnref{const}. The force acting on the spinon-vortex of $\nu=+$ may then denoted by 
$\boldsymbol{E}^{\text {vor }} $ such that \eqnref{Jh} is rewritten as: 
\begin{equation}\label{dualjE}
	\boldsymbol{J}_\alpha^e = \pi^{-1} \epsilon_{\alpha \beta}\boldsymbol{E}_\beta^\text{vor}.
\end{equation}
Physically, such force acting on the vortex induced by the charge current along a direction perpendicular to it can be understood by drawing an analogy with the well-known ``Magnus effect'' \cite{POOLE2014355} in fluid dynamics, illustrated in \figref{fig_vortex}(c). In this semi-classical picture, a spinning object (analogous to the vortex) moving through a fluid (representative of the charge current) experiences a lateral force. This force arises from the differential fluid velocity on opposite sides of the spinning object, pushing it in a direction perpendicular to its motion.

Lastly, it is important to emphasize that the relationships given by \eqnref{jE} and \eqnref{dualjE} reflect the well-established concept of boson-vortex duality \cite{PhysRevB.39.2756,Wen1990}. Within this framework, the charge and vortex degrees of freedom can be interchanged, highlighting their mutual duality in the described context.


\section{Spinon Transport}

In the mutual Chern-Simions gauge description outlined above, the holon condensation will define the so-called lower PG phase, which is also known as the spontaneous vortex phase  (SVP) as the free $b$-spinons carry $\pm \pi$-vortices. It reaches an intrinsic superconducting phase coherence at a lower critical temperature $T_c$. As the basic elementary excitation, the $b$-spinon will dictate the lower PG or the SVP phase as well as the superconducting instability at $T_c$. The main task in this section is to explore the transport of the $b$-spinons, which can expose the physical consequences of the underlying mutual Chern-Simons gauge structure that the $b$-spinon is subjected to.

\subsection{Spinon excitation spectrum}

At the mean-field level, according to \eqnref{conAs}, the $b$-spinons in $L_s$ experience a uniform static gauge flux 
$\delta \pi$ flux per plaquette as the holons are condensed with $\langle n_I^h\rangle = \delta$, which gives rise to a Landau level like energy spectrum $E_m(\boldsymbol{k})$, with the lowest excited sector (LES) at $E_s$, as illustrated in \figref{fig_Eb4}(a) in the case of $\delta=0.2$ (with $\delta\equiv2p/q$ and $p,q \in \mathbb{Z}$ such that $p=1$ and $q=10$). 

In the presence of a perpendicular external magnetic field $B^e$, the spinon energy spectrum becomes [cf. more details in {Appendix~\ref{AppBTc}}]
\begin{equation}\label{A0h}
 \tilde E_{m\sigma\nu}(\boldsymbol{k}) \equiv E_m(\boldsymbol{k})+\sigma \frac{1}{2} g \mu_B B^e + \nu \bar{A}_0^h 
\end{equation} 
where the second term on the right is the usual Zeeman splitting for a spin-1/2 with the g-factor (usually taken as $2$). The third term originated from the temporal gauge $A_0^h$, which results in $i A_0^h\rightarrow \bar{A}_0^h$ in \eqnref{Ls} following a Wick rotation to enforce the constraint \eqnref{const} at $B^e\neq 0$.

\begin{figure}[t]
	\centering
	\includegraphics[width=\linewidth]{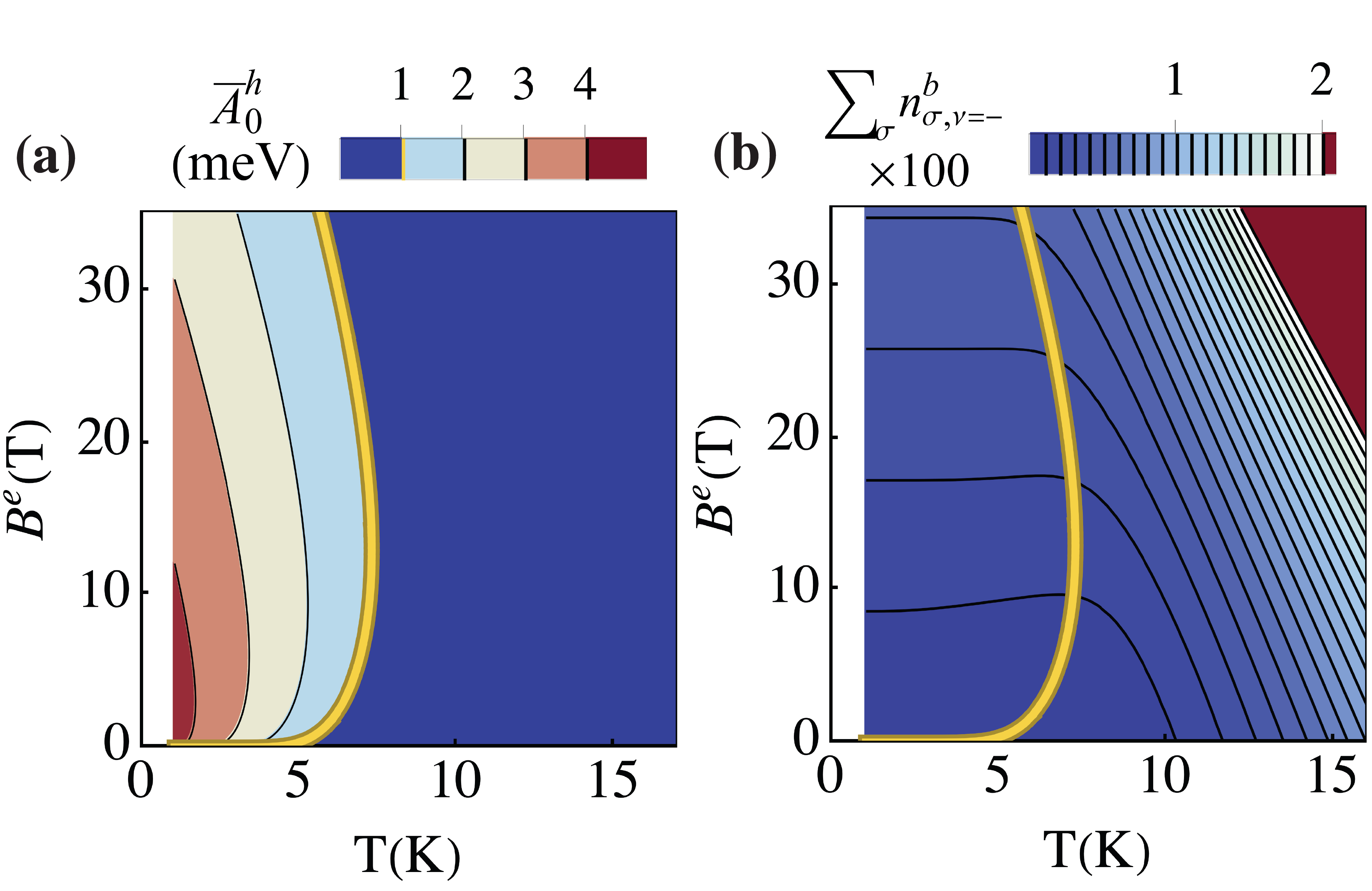}
	\caption{(a) The evolution of $\bar{A}_0^h$ with respect to temperature $T$ and magnetic field $B^e$; (b) The corresponding particle number of excited spinons, $\sum_\sigma n_{\sigma, v=-}^b$, versus $T$ and $B^e$ for a given chirality at lower energy. Here the two distinct temperature regions with different behavior are separated by the yellow line.}
	\label{fig_A0h}
\end{figure}

The mean-field effective Hamiltonian for the spinon-vortex composite may be written as $\tilde{H}_s=\sum_{m\sigma\nu\boldsymbol{k}}\tilde E_{m\sigma\nu}(\boldsymbol{k})\tilde{n}^b_{m\sigma\nu}(\boldsymbol{k})+E_0$ with $\tilde{n}^b_{m\sigma\nu}(\boldsymbol{k})$ denoting the number of the spinon-vortices as the elementary excitations and $E_0$ as the ground state energy.
As the external magnetic field $B^e$ is much weaker than the internal fictitious field $B^h=\delta \pi/a^2$ (with $a=3.8$\r{A} as the lattice constant), its effect mainly introduces a minor splitting via the last two terms in $\tilde E_{m\sigma\nu}$. When the temperature is not too high above $T_c$ [note that $E_s$ can be related to $T_c$ in \eqnref{Tc} explicitly], it is reasonable to project the Hilbert space into the LES around $E_m(\boldsymbol{k})=E_s$, which is split as shown in \figref{fig_Eb4}(b)  
\footnote{Note that the doping density can generally be expressed as $\delta/2=p/q$. In instances where $p=1$ the lowest Landau level for $\tilde{E}_m$ remains degenerate. However, further splitting can occur when $p\neq 1$. Despite this, the energy associated with such splitting is significantly smaller than the splitting effects induced by magnetic fields, specifically $\bar{A}_0^h$ and $|B^e|$. Consequently, these energy differences can be disregarded. As a result, all these modes are considered to reside within the lowest energy sector, denoted by $m\in \text{LES}$.}.
The mean-field parameter $\bar{A}_0^h$ can be explicitly determined by enforcing the constraint \eqnref{const} at $B^e\neq 0$, which 
is illustrated in \figref{fig_A0h}(a), and the corresponding low-energy excited spinon number $\sum_{\sigma}n_{\sigma\nu=-}^b$ in the LES is displayed in \figref{fig_A0h}(b). Both figures indicate the existence of two distinct temperature regions, separated by the yellow lines in \figref{fig_A0h}. In the high-temperature region,  the effect of $\bar{A}_0^h$ on the free spinon number is relatively small due to its small energy compared to $E_s$. On the other hand, at low temperatures, $\bar{A}_0^h$ dominates the lowest-energy excited level, causing the particle number of spinons to correlate linearly with the magnitude of the external field, but remain independent of temperature. The excited spinons will play a significant role in the transport behavior of the lower PG phase. 

\subsection{Transverse transport coefficients}

Importantly, due to the gauge field $B^h$, the $b$-spinon spectrum $\tilde{E}_{m\sigma \nu}$ carries nontrivial Berry curvatures $\boldsymbol{\Omega}_{m\sigma\nu}(\boldsymbol{k})=i\nabla_{\boldsymbol{k}} \times \left\langle u_{m\sigma\nu, \boldsymbol{k}}\left|\nabla_{\boldsymbol{k}}\right| u_{m\sigma\nu, \boldsymbol{k}}\right\rangle$, with $| u_{m\sigma\nu, \boldsymbol{k}}\rangle$ being the periodic part of the Bloch waves corresponding to the energy $\tilde{E}_{m\sigma\nu}(\boldsymbol{k})$. The nonzero Chern number $\mathcal{C}_{m\sigma\nu}= 2\pi\sum_{\boldsymbol{k}}  \Omega_{m\sigma\nu}^z(\boldsymbol{k})$ for each band within the LES is shown in \figref{fig_Eb4}(b), indicating that the sign of the Chern number depends solely on the chirality of the $b$-spinons, leading to $\sum_{m\in \text{LES}}\mathcal{C}_{m\sigma\nu}=2\nu$ \footnote{In the scenario where $p\neq 1$, the lowest excited level, denoted by energy $E_s$ as depicted in \figref{fig_Eb4}(a), undergoes a minor slitting, resulting in each split band carrying a more intricate Chern number. Nevertheless, the cumulative Chern number, represented by $\sum_{m \in \text{LES}} \mathcal{C}_{m \sigma \nu}=2\nu$, is derived from the summation across all these sub-bands.}. {The factor of 2 in this expression arises from the doubled density of states associated with the Bogoliubov quasiparticles of $b$-spinons.}. Physically, this is because the direction of the intrinsic magnetic field $B^h$ perceived by the spinons is solely determined by their vorticity sign. 

To study the transport properties for $b$-spinons, we base our approach on the semiclassical theory, analogous to the quantum
Hall effect in electron systems\cite{Niu.Xiao.2010}. We consider the $b$-spinon wave packet with a relatively determined center and momentum $(\boldsymbol{r}, \boldsymbol{k})$ with an intrinsic size, determined by the ``cyclotron length'' $a_c=a / \sqrt{\pi \delta}$ \cite{Weng.Mei.20107w}. The dynamics of such a wave packet is described by the semiclassical equation of motion, which includes the topological Berry phase term \cite{Niu.Xiao.2010}:
\begin{equation}\label{anov}
	\dot{\boldsymbol{r}}=\frac{1}{\hbar} \frac{\partial \tilde{E}_{m \sigma \nu}(\boldsymbol{k})}{\partial \boldsymbol{k}}-\dot{\boldsymbol{k}} \times \boldsymbol{\Omega}_{m \sigma \nu}(\boldsymbol{k})
\end{equation}
where $\hbar \dot{\boldsymbol{k}}=-\nabla U(\boldsymbol{r})$, and $U(\boldsymbol{r})$ is a confining potential that exists only near the boundary of the sample, which prevents the spinon wave packet from exiting the sample. On the edge along the $x$ direction, for example, the nontrivial Berry curvature produces an anomalous velocity $\dot{\boldsymbol{k}} \times \boldsymbol{\Omega}_{m \sigma \nu}(\boldsymbol{k})= -\hbar^{-1}\partial _y U(\boldsymbol{r})\Omega^z_{m \sigma \nu}(\boldsymbol{k})\hat x$ in \eqnref{anov}. Physically, this anomalous velocity arises from the fact that $b$-spinon perceives an intrinsic ``Lorentz force'' due to the uniform gauge field $B^h$ from the holons, the sign of which depends on the vorticity. Therefore, $b$-spinon undergoes a cyclotron motion in the bulk and a skipping orbit along the edge of the sample, as illustrated in \figref{fig_sketch}(b). It is crucial to note that both spinons carrying opposite chirality flow along the boundary. This scenario is reminiscent of the quantum spin Hall effect \cite{Zhang.K.2007, Zhang.Bernevig.2006}, where the electrons at the boundary form the currents with opposite chiralities for opposite spins. Also, in contrast to the chiral spin liquid with chiral edge modes \cite{Laughlin.Kalmeyer.1987, Wen.Wen.1990}, our effective description [as referenced in \eqnref{Lh}-(\ref{LCS})] maintains time-reversal symmetry. Here, to draw parallels and distinctions from previously observed phenomena, the behavior of the neutral spin within our framework may be termed the bosonic ``anomalous vortex Hall'' effect, which underscores the vortex edge current arising from the internal fictitious magnetic field. On the other hand, in the case of equilibrium, all the edge current in the sample cancels between one edge and the opposite edge shown in  \figref{fig_sketch}(c), resulting in no net current.

In the presence of either a spatially varying chemical potential $\mu$ or temperature $T$, a net edge current is contributed by the anomalous velocity of $b$-spinon, as shown in \figref{fig_sketch}(b)-(d). For instance, when there is a temperature gradient and a chemical potential gradient in the $y$ direction, the linear response of the chiral spinon current $\boldsymbol{J}^\nu$ and the heat current $\boldsymbol{J}_Q^\nu$, with their respective chirality $\nu$, can be expressed as \cite{Shi.Qin.2011, Shi.Qin.2012, Murakami.Matsumoto.2011, Murakami.Matsumoto.20111gl}
\begin{equation}\label{response}
	\left[\begin{array}{c}
		\boldsymbol{J}_x^{\nu} \\
	\left(\boldsymbol{J}_Q^{\nu}\right)_x
	\end{array}\right]=\boldsymbol{L}^{\mathrm{xy}\text{,}\nu}\left[\begin{array}{c}
	-\nabla_y \mu  \\
	T \nabla_y \frac{1}{T}
	\end{array}\right],
\end{equation}
Here, $\boldsymbol{L}^{\mathrm{xy}\text{,}\nu}$ signifies a $2\times 2$ matrix that represents the transverse transport coefficients. The parameter $\nu=\pm$ distinguishes between the different chiralities of spinon. The matrix elements are given by
\begin{eqnarray}
	L_{i j}^{x y,\nu}&\approx&-\frac{1}{\hbar V \beta^q } \sum_{m\in \text{LES}}\sum_{\sigma\boldsymbol{k}}  c_q(n_{m \sigma \nu}^b)\Omega_{m \sigma \nu}^z(\boldsymbol{k}) \notag \\
	&=&-\frac{\nu}{\hbar  \beta^q \pi} \sum_{\sigma}  c_q(n_{ \sigma \nu}^b)\label{Lij}
\end{eqnarray}
where $i,j=1,2$, $c_q(x)\equiv\int_0^{x}\left(\log \frac{1+t}{t}\right)^q d t $, $q=i+j-2$, and $n_{m \sigma \nu}^b=1 /\left(e^{\beta \tilde{E}_{m \sigma \nu}}-1\right)$ is the bosonic distribution function for $b$-spinons, which is independent of momentum, because $\tilde{E}_{m\sigma\nu}$ is the flat Landau-level band in our case. Note that in \eqnref{Lij}, as an approximation, we only sum over within the LES and use the relation $\sum_{m\in \text{LES}}\mathcal{C}_{m \sigma \nu}=2 \pi \sum_{\boldsymbol{k}} \Omega_{m \sigma \nu}^z(\boldsymbol{k})=2\nu$. In the following, we will investigate various transport measurements associated with the transverse transport coefficients $L_{i j}^{x y}$ in \eqnref{Lij} (for simplicity we shall drop the superscript $xy$ in the following such that $L_{i j}^{x y,\nu}\rightarrow L_{i j}^{\nu}$).

\section{Experimentally testable consequences}

\subsection{Thermopower}
\begin{figure}[t]
	\centering
	\includegraphics[width=0.7\linewidth]{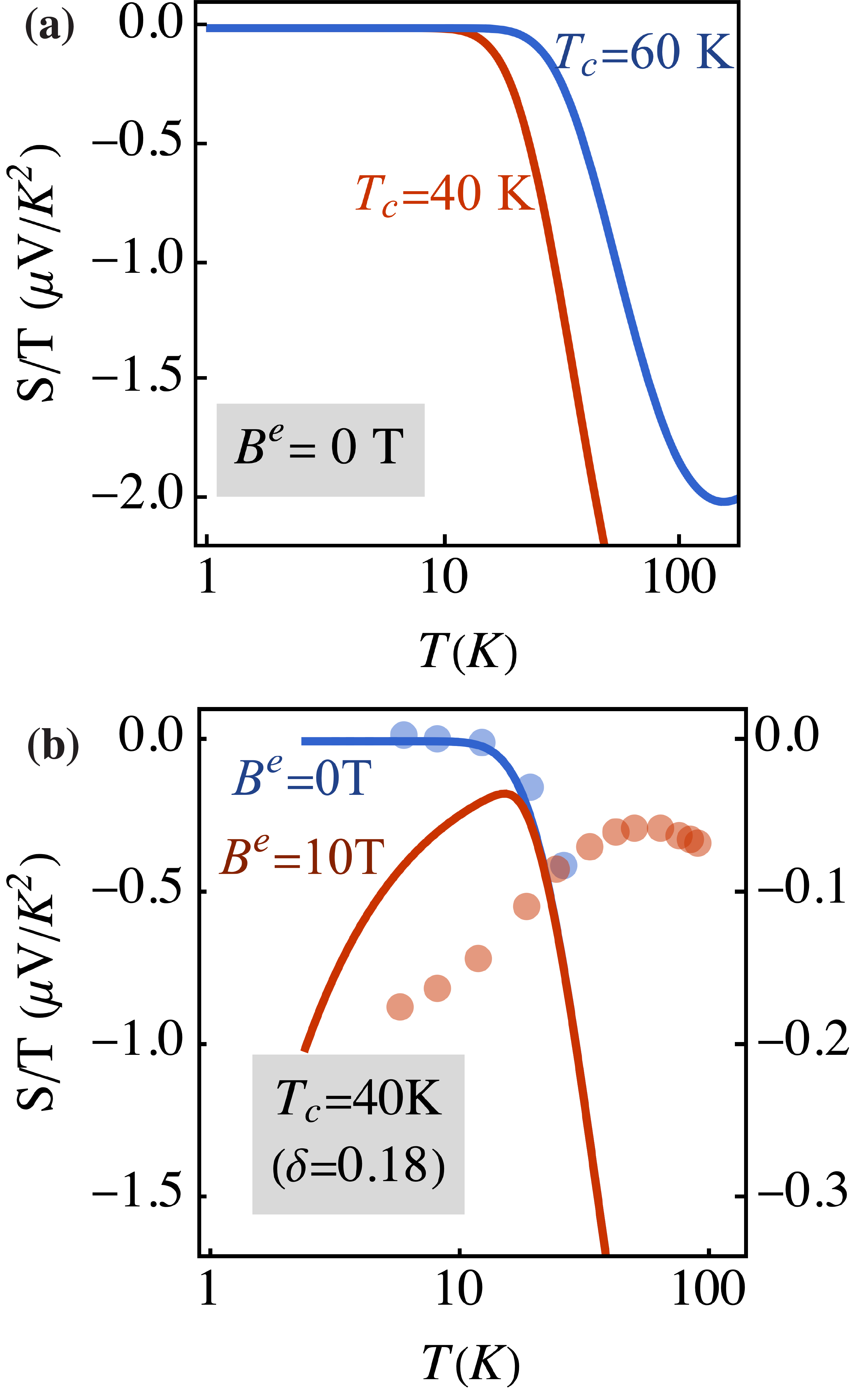}
	\caption{The evolution of the Seebeck coefficient with respect to temperature is depicted in (a), without the influence of a magnetic field, and in (b), at a doping density $\delta=0.18$ and a critical temperature $T_c=40K$. {Blue and red lines depict theoretical results under zero magnetic field and 10 T, respectively, while dots show experimental data \cite{Taillefer.Lizaire.2021}. Theoretical values are on the left axis, and experimental values on the right.}}
	\label{fig_see}
\end{figure}

As illustrated in \figref{fig_sketch}(c), due to the internal flux $B^h$, the $b$-spinons with opposite vorticities will propagate in opposite directions along the edges of the sample, such that there is a \emph{net} vortex current at each edge along the $x$ direction, which would be canceled out by the opposite edge current at $\nabla_y T=0$. Now
let us consider a temperature gradient $\nabla_y T$ applied along the $y$ direction. As depicted in \figref{fig_sketch}(d), the vortex current on one side of the boundary will be larger than on the higher-$T$ side, which will result in a finite total vortex current along the $x$ direction. Noting that $\boldsymbol{J}_x^{\text{vor}}=\boldsymbol{J}_x^{\nu=+}-\boldsymbol{J}_x^{\nu=-}$, where $\boldsymbol{J}_x^{\nu=\pm}$ represents the spinon current with $\pm$ chirality \cite{Weng.Qi.2007} as given by $\boldsymbol{J}_x^{\nu=\pm}=L_{12}^{\nu=\pm}\left(T \partial_y \frac{1}{T}\right)$ according to \eqnref{Lij}. Furthermore, according to \eqnref{jE}, the net vortex current $\boldsymbol{J}_x^{\text{vor}}$ along the $x$ direction can induce an electric field $\boldsymbol{E}_y^e$ along the $y$ direction (similar to the contribution to the Nernst effect as to be discussed later), which will contribute to a finite thermopower, with the Seebeck coefficient given by
\begin{equation}\label{S}
	S\equiv\frac{\boldsymbol{E}_y^e}{\nabla_y T}=-\frac{ k_B \phi_0 }{ \pi \hbar} \sum_{\sigma\nu}  c_1(n_{\sigma\nu}^b) 
\end{equation}
where $c_1(x) \equiv(1+x) \ln (1+x)-x \ln x$. Thus such a contribution of the $b$-spinon to the Seebeck coefficient is determined by the number of the excited $b$-spinons, $n_{\sigma\nu}^b$, which in turn is essentially governed by the lowest excited energy scale $E_s$ in \eqnref{A0h} at low temperatures. {Note that $E_s$ also determines $T_c$ according to Eq.(\ref{Tc}), which means that the sole free parameter in \eqnref{S} is effectively decided by $T_c$.}

A typical quantitative temperature-dependence of the Seebeck coefficient $S$ calculated using \eqnref{S} is shown in \figref{fig_see} at zero magnetic fields [(a)] and at finite $B$'s [(b)] in the overdoped regime. The overall $T$- and $B$-dependence and magnitude here, {including the negative sign,} are in agreement with the experimental measurements in the optimal and overdoped cuprates \cite{Taillefer.Daou.2009, Taillefer.Lizaire.2021}. 

{
It's worth noting that while the present manuscript focus on thermopower in the pseudogap regime, a previous research \cite{kou_self-localization_2005} also addressed thermopower in the lightly doped region, where the thermopower mainly arises from the entropy associated with energetically degenerate hole configurations in real-space distribution (the Heikes-like formula \cite{kou_self-localization_2005}), consistent with the experimental findings \cite{cooper_thermoelectric_1987, mandal_thermoelectric_1992, keshri_thermoelectric_1993}. In the optimal- and over-doped PG regime, however, this large positive Seebeck effect should be substantially reduced with the condensation of the holons and is neglected in \figref{fig_see}.
}

\begin{figure}[t]
	\centering
	\includegraphics[width=0.7\linewidth]{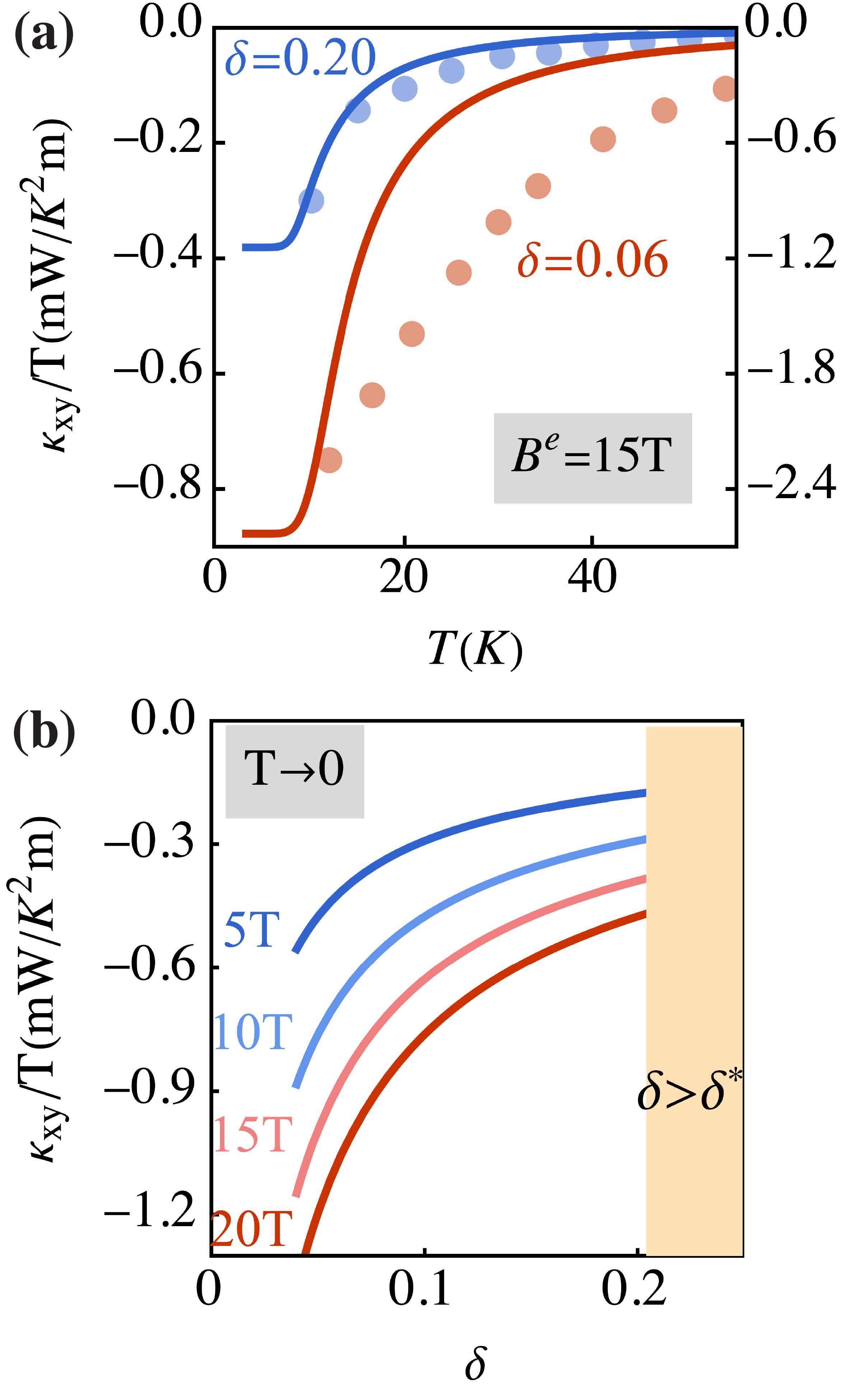}
	\caption{(a) The temperature evolution of the thermal-Hall coefficient when $B=15T$. 
    { The blue and red lines depict theoretical results for the overdoped and underdoped regimes, respectively. Corresponding experimental data points, colored accordingly, are sourced from \cite{Taillefer.Grissonnanche.2019}. Theoretical values are shown on the left axis, while experimental values are on the right.} (b) The doping evolution of the thermal-Hall coefficient as the temperature approaches zero. The thermal-Hall effect is predicted to revert to conventional Fermi liquid behaviors when the doping density $\delta$ is greater than the critical density $\delta^*$, as indicated by the yellow region.}
	\label{fig_thermal}
\end{figure}

\subsection{Thermal Hall}
Similarly, with a temperature gradient along the $y$ direction, we can also evaluate the net thermal current $(\boldsymbol{J}_Q)_x=(\boldsymbol{J}_Q^{\nu=+})_x+(\boldsymbol{J}_Q^{\nu=-})_x$ along the $x$ direction, as illustrated in \figref{fig_sketch}(e). Here, $\boldsymbol{J}_Q^{\nu=\pm}$ represents the thermal current contributed by spinons with $\pm$ chirality, expressed as $(\boldsymbol{J}_Q^{\nu= \pm})_x=L_{22}^{\nu= \pm}\left(T \partial_y \frac{1}{T}\right)$, according to \eqnref{Lij}. The thermal Hall conductivity is then given by\cite{Murakami.Matsumoto.2011, Murakami.Matsumoto.20111gl, Shi.Qin.2011, Shi.Qin.2012}:
\begin{eqnarray}\label{kappa}
	\kappa^{\mathrm{xy}}\equiv- \frac{J_Q}{\nabla_y T}=-\frac{ k_B^2 T}{ \pi \hbar } \sum_{\nu\sigma} \nu c_2(n_{\nu\sigma}^b)
\end{eqnarray}
where $c_2(x)=(x+1)(\ln \frac{1+x}{x})^2-(\ln x)^2-2 \operatorname{Li}_2(-x)-c$, with $\operatorname{Li}_2(z)$ being the polylogarithm function, and $c=\pi^2/3$ is a constant ensuring that $\kappa^{\mathrm{xy}}$ does not diverge as $T\rightarrow \infty$. 

It is crucial to note that, in the absence of an external magnetic field, the vanishing of both $B^e$ and $\bar A_0^h$ leads to the degeneracy of $\tilde E_{m\sigma\nu}$ with respect to chirality $\nu$, resulting in a zero value for $\kappa^{\mathrm{xy}}$ in \eqnref{kappa} due to the summation over $\nu$. Essentially, this outcome stems from the fact that the thermal current with opposite chirality flows in opposite directions along a boundary. Therefore, the preservation of total chirality to zero in a sample without an external magnetic field results in complete cancellations for the thermal current. Conversely, in the presence of an external magnetic field $B^e$, according to \eqnref{const}, the total chirality for $b$-spinons becomes finite, leading to a net thermal current along the boundary.

As a result, the evolution of thermal Hall conductivity $\kappa^{\mathrm{xy}}$ obtained by \eqnref{kappa} with respect to temperature is depicted in \figref{fig_thermal}(a). This evolution aligns with experimental results in terms of magnitude\cite{Taillefer.Grissonnanche.2020,Taillefer.Boulanger.2020,Taillefer.Grissonnanche.2019}, and it exhibits distinct behaviors across different temperature regions. In the high-temperature region, following the discussion about \eqnref{A0h}, $n_{\nu\sigma}^b$ is not sensitive to $\bar{A}^0_h$, leading to $\kappa^{\mathrm{xy}} / T=-\frac{B^e k_B^2}{\hbar \pi \phi_0 \delta}\left(\frac{3 T_c}{T}\right)^2$ [\eqnref{Tc} is used here]. 
On the other hand, in the low-temperature region where spontaneous (thermally excited) vortices are absent, \eqnref{const} reduces to $\sum_\sigma n_{\sigma\nu=-}^b\approx B^e a^2/\phi_0 \delta$. Here, all other energy levels remain unoccupied, leading to $\kappa_{xy}/T \rightarrow -2 k_B^2  c_2\left(B^e a^2/4\phi_0 \delta\right)/\hbar  \pi$  as $T$ approaches 0. 
The doping evolution of $\kappa_{xy}/T$ near zero temperature is presented in \figref{fig_thermal}(b).
 This evolution reveals enhanced signals in the underdoped regimes,
 corroborating the experimental measurements\cite{Taillefer.Grissonnanche.2020,Taillefer.Boulanger.2020,Taillefer.Grissonnanche.2019}. 
 Physically, this is because, under low doping conditions,  the degeneracy of the lowest Landau level of the spinons is reduced due to the small intrinsic magnetic field strength $\delta \pi$. 
  This reduction in degeneracy increases the average Berry curvature, 
 denoted as $\boldsymbol{\Omega}_{m\sigma \nu}$, 
 experienced by each spinon, 
 which enhances the anomalous velocity at the boundary, as indicated by \eqnref{anov}. 
 Therefore, the thermal Hall effect becomes more pronounced under low doping.
  However, $\kappa_{xy}/T$ will not truly diverge at $\delta\rightarrow 0$, 
due to the fact that the uniform holon condensation will either be broken down or form smaller domain structures when it is deeply in the AFM long-range ordered phase \cite{Taillefer.Grissonnanche.2019}. Our results for the thermal Hall conductivity differ from the bosonic scaling law in Ref. \onlinecite{Zhang.Yang.2020}, where Zhang et al. utilized the gapless bosons with a power-law Berry curvature. Here we emphasize that the intrinsic flux {$B^h$} leads to the Landau level structure, with the gap of the spinon-vortices constrained by \eqnref{const}, resulting in a decreasing gap as the temperature drops as illustrated in \figref{fig_A0h}. Notably, within the pseudogap regime, the role of the spinon-vortex is to neutralize the external magnetic field. Consequently, its Hall response is in the opposite direction compared to the charged quasiparticles manifesting in the Fermi liquid regime. This elucidates the observed sign change of the thermal Hall as the doping transitions into the pseudogap phase \cite{Taillefer.Grissonnanche.2019}.

It is important to note that our case does not involve the spontaneous breaking of time-reversal symmetry, a necessary condition to avoid the emergence of a hysteretic behavior not observed experimentally. Furthermore, according to \eqnref{const}, the total chirality carried by $b$-spinons is induced linearly with the applied magnetic field. This results in the linear-$B$ dependence of $\kappa^{\mathrm{xy}}$ in both distinct temperature regions, which aligns with the experimental measurement\cite{Taillefer.Grissonnanche.2020,Taillefer.Boulanger.2020,Taillefer.Grissonnanche.2019}.


\subsection{Hall effect}
According to \eqnref{dualjE}, driving a charge current  $\boldsymbol{J}_x^e$ along the $x$ direction induces an electric field $\boldsymbol{E}_y^{\mathrm{vor}}$ applied to the vortex along the $y$ direction. This electric field acts as the chemical potential gradient $\boldsymbol{E}_y^{\mathrm{vor}}=-\nabla_y \mu$ in \eqnref{response}. Since the $\pm$ vortices perceives the $\boldsymbol{E}_y^{\mathrm{vor}}$ in opposite directions, the response spinon current $\boldsymbol{J}_x^{\nu=\pm}=L_{11}^{\nu=\pm} \boldsymbol{E}_y^{\mathrm{vor}}$ leads to the vortex current $\boldsymbol{J}_x^{\text{vor}}=\sum_{\nu}\nu L_{11}^{\nu=\pm}\boldsymbol{E}_y^{\mathrm{vor}}$. From \eqnref{jE}, this induced vortex current further generates an electric field $\boldsymbol{E}_y^e$ along the $y$ direction, culminating in the Hall effects as illustrated in \figref{fig_sketch}(f). Therefore, the obtained Hall coefficient $R_H$ is given by
\begin{equation}\label{RH}
    R_H \equiv \frac{\boldsymbol{E}_y^e d}{\boldsymbol{J}^e_x B^e}= \frac{a^2 d}{e\delta},
\end{equation}
where $d$ denotes the lattice constants along the $z$-axis. We also employ the relation $c_0(x)=x$ and \eqnref{const}. 

Therefore, the Hall number calculated by \eqnref{RH} is given by $n_H=a^2 d/e R_H=\delta$, which is consistent with the experimental results\cite{Taillefer.Proust.2019, Proust.Badoux.2016}. Significantly, there exists a long-standing experimental puzzle wherein the charge carrier, as measured by the Hall number, correlates with the doped hole density $\delta$ in the PG. This contrasts with the $1+\delta$
measurement derived from the Fermi surface area observed through angle-resolved photoemission spectroscopy (ARPES)\cite{Damascelli.Plat.2005, Campuzano.Kaminski.2003, Campuzano.Chatterjee.2011}, seemingly deviating from the Luttinger sum rule. Our results offer a compelling explanation: chiral spinons primarily contribute to the Hall effects. In contrast, the entities forming the Fermi surface — Landau quasiparticles — display a negligible Hall effect signal due to their partially diminished weight (Fermi arcs) in the PG phase. 

Note that the Hall effect in this framework is primarily attributed by the edge states of chiral spinons. At elevated temperatures, the local phase coherence of holons can become further disrupted, rendering them incapable of sustaining condensation when the distance between spinon-vortices becomes comparable to that of the doped holes. In such a scenario, chiral spinons no longer experience the uniform static gauge flux emitted by holons, and thus cannot sustain their complete edge states. Consequently, the contribution of such channels to the Hall effect would diminish as the temperature rises, consistent with experimental measurements \cite{Taillefer.Proust.2019, Proust.Badoux.2016} {and numerical results \cite{AuerbachNPJ_hall_2023}}.

\subsection{Other properties of spinon-vortices}\label{other}

In the above subsections, the effects produced by the transverse motion of the chiral spinons with the MCS gauge structure have been explored. It is noted that the effects of the longitudinal motion of the same chiral spinons have been already studied previously \cite{Weng.Kou.2005,Qi.Weng.2006,Weng.Qi.2007}. Since the $b$-spinons are the elementary excitations that dictate the lower PG phase, for the sake of completeness, in the following we briefly discuss additional phenomena associated with the $b$-spinons.  


\subsubsection{Nernst effect} \label{Nernst}
When a temperature gradient is applied along the $y$ direction, our attention shifts from the transverse transport motion detailed in \eqnref{response}, to the longitudinal drift motion of spinons. To explore this, we introduce a viscosity constant, $\eta_s$, which allows us to determine the drift velocity $\boldsymbol{v}^b$ of chiral spinons using the equation $s_\phi \nabla T=-\eta_s \boldsymbol{v}^b$, where $s_\phi$ denotes the transport entropy carried by a spinon vortex. It's important to note that spinons of both chiralities are driven by a temperature gradient in the same direction along the $x$-axis, with the velocity being the same $\boldsymbol{v}^b$.

In the presence of an external magnetic field $B^e$, a vortex current $\boldsymbol{J}^\text{vor}_y$ is induced along the $y$-axis. As discussed in the thermopower section, the vortex current can be further expressed as $\boldsymbol{J}^\text{vor}= \rho_{\text{LES}}(n_{\sigma\nu=+}^b-n_{\sigma\nu=-}^b) \boldsymbol{v}^b$, {with $\rho_{\text{LES}}=\delta$ denoting the number of states for lowest energy sector besides spin and vorticity quantum number}, where the amplitude is proportional to the external magnetic field as per the chirality ``neutrality'' condition \eqnref{const}. This vortex current $\boldsymbol{J}^\text{vor}_y$ induces an electric field $\boldsymbol{E}^e_x$ along the $x$-axis, as dictated by \eqnref{jE}. This corresponds to the Nernst effects, with the signal defined by \cite{Qi.Weng.2006}:
\begin{equation}\label{eN}
    e_N=\frac{\boldsymbol{E}_y}{|\nabla_x T|} = B^e \frac{s_\phi}{\eta_s}.
\end{equation}

\begin{figure}[t]
	\centering
	\includegraphics[width=\linewidth]{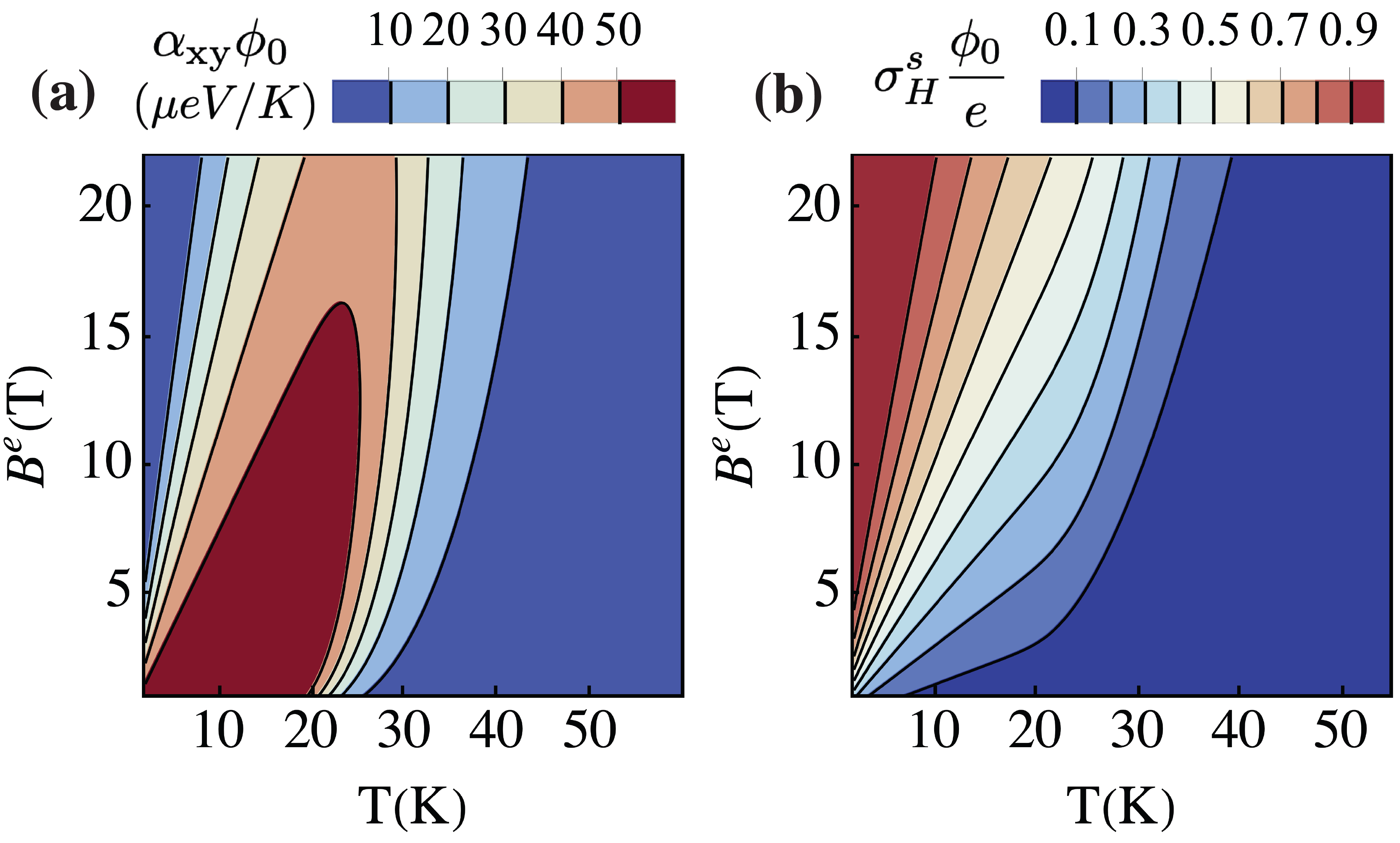}
	\caption{The temperature and magnetic field contour plot of $\alpha_{xy}\equiv e_N/\rho$ (a) and the spin Hall coefficient $\sigma_H^s$ (b).}
	\label{fig_Nen}
\end{figure}

To eliminate the viscosity $\eta_s$ in \eqnref{eN}, let us consider the longitudinal resistivity $\rho_e$ resulting from the drift motion of chiral spinons. According to \eqnref{dualjE}, driving a charge current $\boldsymbol{J}_x^e$ along the $x$ direction induces an ``electric field'' $\boldsymbol{E}_y^{\mathrm{vor}}$ on the vortex. In contrast to the temperature gradient, $\boldsymbol{E}_y^{\mathrm{vor}}$ prompts spinons of both chiralities to drift in opposite directions along the $y$-axis, in accordance with the relation $\boldsymbol{E}_y^\text{vor}=\pm \eta_s /\hbar c \boldsymbol{v}_y^b$. This results in a vortex current $\boldsymbol{J}^\text{vor}= \rho_{\text{LES}}n_{\sigma\nu=+}^b \boldsymbol{v}^b-\rho_{\text{LES}} n_{\sigma\nu=-}^b (-\boldsymbol{v}^b)= \rho_{\text{LES}}n^b \boldsymbol{v}^b$, where $n^b=\rho_{\text{LES}} \sum_{\sigma\nu} n_{\sigma \nu}^b$ is the total number of free $b$-spinons. Next, as derived from \eqnref{jE}, such vortex current $\boldsymbol{J}^\text{vor}_y$ induces the electric field $\boldsymbol{E}^e_x$ along the $x$ direction, leading to the longitudinal resistivity 
\begin{equation}\label{rhoe}
    \rho_e=\phi_0^2 n^b /\eta_s,
\end{equation}
where the contribution from quasiparticles is not included in this analysis. Lastly, the challenging-to-calculate viscosity $\eta_s$ is eliminated, yielding\cite{Qi.Weng.2006}:
\begin{equation}
    \alpha_{x y} \equiv \frac{e_N}{\rho}=\frac{B^e a^2 s_\phi}{\phi_0^2 n^b},
\end{equation}
where $\alpha_{x y}$ is the quantity introduced in Ref. \onlinecite{Ong.Wang.2001}. Within our framework, a unique aspect that sets it apart from a conventional BCS superconductor is the presence of a free $S = 1/2$ moment locked with the vortex core, which gives rise to the ``transport entropy''\cite{Hardy.Wang.2002} $s_\phi=k_B\{\ln \left[2 \cosh \left(\beta \mu_B B^e\right)\right]-\beta \mu_B B^e \tanh \left(\beta \mu_B B^e\right)\}$. The temperature and magnetic-field dependence of $\alpha_{x y}$ is illustrated in \figref{fig_Nen}(a). Its magnitude aligns quantitatively with experimental data \cite{Ong.Wang.2003, Ong.Wang.2001, Hardy.Wang.2002}, suggesting that the transport entropy due to the free moment in a spinon vortex can accurately replicate the Nernst signal observed experimentally.

\subsubsection{Spin Hall effect}
In the presence of a magnetic field, both the chirality and spin degrees of freedom become polarized through the orbit and Zeeman effects, respectively. In this scenario, a charge current $\boldsymbol{J}_x^e$ along the $x$ direction not only induces a vortex current $\boldsymbol{J}^\text{vor}_y$ along the $y$ direction—as previously discussed—but also generates a spin current. The latter can be expressed as $\boldsymbol{J}_\alpha^{\mathrm{s}}=\rho_{\text{LES}}(n_{\uparrow+}^b-n_{\downarrow+}^b) \boldsymbol{v}_\alpha+\rho_{\text{LES}}(n_{\uparrow-}^b-n_{\downarrow-}^b)(-\boldsymbol{v}_\alpha)=\rho_{\text{LES}}\sum_{\sigma \nu} \sigma \nu n_{\sigma v}^b \boldsymbol{v}_\alpha$. This results in the generation of a vortex current that accompanies a spin current, with the ratio defined as $\zeta\equiv J_\alpha^{s}/J_\alpha^\text{vor}=\rho_{\text{LES}}\sum_{\sigma \nu} \sigma \nu n_{\sigma \nu}^b/n^b$. As a consequence, the PG phase is predicted to exhibit a spin Hall effect, with the coefficient given by \cite{Weng.Kou.2005, Qi.Weng.2006}:
\begin{equation}
\sigma_H^s \equiv \frac{J_\alpha^s}{\boldsymbol{E}_\alpha^e}=\frac{e}{\phi_0} \zeta,
\end{equation}
of which the calculated magnitude is shown in \figref{fig_Nen}(b).

\subsubsection{Order-to-order phase transition}
At the units $\hbar=c=e=1$, \eqnref{rhoe} can be recast into a dual form:
\begin{equation}\label{mutualdual}
\sigma_e \sigma_s = \frac{1}{\pi^2},
\end{equation}
where $\sigma_e=1 / \rho_e$ represents the electrical conductance, while $\sigma_s\equiv \boldsymbol{J}^{\text {vor }}_\alpha/\boldsymbol{E}^{\text {vor }}_\alpha=n^b / \eta_s$ denotes the spinon conductance. Essentially, \eqnref{mutualdual} parallels the boson-vortex duality \cite{Wen1990, PhysRevB.39.2756}, in which both the Cooper pair and the superconductivity vortex perceive each other as vortices. As such, when one is in a superfluid state, the other resides in an insulating state.

Within the context of our work, the spinon (holon) carries the holon (spinon) vortex, thereby uniquely associating all vortices with quantum numbers. Based on \eqnref{mutualdual}, the superconductivity phase characterized by $\sigma_e\rightarrow \infty$, corresponds to an insulating phase for the spinon with $\sigma_s\rightarrow 0$. Moreover, when spinon condenses with $\sigma_s\rightarrow \infty$, indicating the establishment of antiferromagnetic long-range order, it triggers the proliferation of holon vortices, thereby resulting in an insulating phase in charge, i.e., $\sigma_e\rightarrow 0$. This sequence represents a novel type of ``order-to-order'' phase transition, widely investigated under the rubric of ``deconfined quantum critical point'' (DQCP) \cite{Senthil.Levin.2004, Fisher.Senthil.2004, Senthil.Wang.2017}.

\subsubsection{Relation between $T_c$ and resonance energy in INS}
The dynamic spin susceptibility, as observed via inelastic neutron scattering (INS), reveals the transition of the gapless spin-wave \cite{Fisk.Coldea.2001, Perring.Headings.2010} at the antiferromagnetic (AFM) wave vector $(\pi,\pi)$ to a gapped state upon disruption of the AFM long-range order. This spin excitation also manifests a resonance-like mode \cite{PhysRevLett.75.316, Keimer.He.20006wc, Bourges.Fauqu.2007, Keimer.Capogna.2007, Keimer.Fong.1999, Keimer.He.2002, Dai1999}  characterized by energy $E_g$, demonstrating a peak in the spin spectrum weight. When deviating slightly from the momentum $(\pi,\pi)$,  the resonance mode bifurcates and spans both higher and lower energies, resulting in the well-documented hourglass-shaped spectrum 
\cite{Greven.Chan.2016, Greven.Chan.2016noa, Keimer.P.2004, Dogan.Hayden.2004, Yamada.Tranquada.2004, Fujita.Sato.2020}. 

Within our proposed framework, the predominant low-lying spin spectrum weight originates from the LES of chiral spinons characterized by energy $E_s$. Furthermore, the $S=1$ spin excitation detected by INS is in fact a composite of two $S=1/2$ spinons, resulting in the resonance spin mode energy $E_g=2E_s$. A careful analysis \cite{Zhang.Weng.hg_2023} further validates that the spinon excitation discussed in our study is consistent with the observed hour-glass spin spectrum.
 
A key insight is the established relation between the resonance energy $E_g$
observed in INS and the superconducting critical temperature $T_c$. The relation [detailed derivation in {Appendix~\ref{AppBTc}} and Ref. \onlinecite{Weng.Mei.20107w}] is expressed as:
\begin{equation} \label{Tckappa}
    \kappa \equiv  \frac{E_g}{k_B T_c} \approx 6.45,
\end{equation}
 which aligns closely with the experimental measurement \cite{Dogan.Dai.1996, Aksay.Fong.1997, Keimer.He.20006wc, Colson.Gallais.2002} $\kappa^{\text{exp}} \approx 6$.

\section{Discussion}

One of the key hypotheses on the high-$T_c$ cuprate in a doped Mott insulator approach \cite{Wen.Lee.2006z4} is that the PG phase is a fractionalized novel state beyond the Landau Fermi-liquid description. In other words, it is the spinon and holon instead of the Landau quasiparticle that dictate the physics of the PG phase. The transport properties can provide a very powerful test of distinct hypotheses of the elementary excitations and thus the underlying states of matter.

In this work, we have specifically explored the transverse transport of the spinons in the phase-string formulation of the $t$-$J$ model. As the elementary excitations, here the spinon and holon are subjected to the mutual Chern-Simons gauge structure due to the phase-string effect in a doped Mott insulator, which preserves the time-reversal and parity symmetries in the absence of the external magnetic field. In the so-called lower PG phase, the holons remain Bose-condensed but the superconducting phase coherence is disordered by free spinon excitations until the ``confinement'' of the spinons below $T_c$. The time-reversal symmetry is retained because the opposite spins see the opposite directions of the fluxes and form the RVB pairing in the superconducting ground state. Here the transverse transport refers to the rotational motion of the spinons as the edge chiral currents under the internal statistical fictitious fluxes, which may be regarded as the bosonic ``anomalous vortex Hall'' effect. 

Both the neutral and electric Hall effects are exhibited in the presence of a perpendicular magnetic field.  
In contrast to the conventional Boltzmann transport of the Landau quasiparticles, the thermopower, thermal Hall, and the Hall effect studied here are all contributed by the chiral spinons, which are further locked with a vortex supercurrent via the mutual Chern-Simons gauge field in generating a transverse electric voltage. The magnitudes of the calculated transverse transport coefficients are intrinsically linked to the resonance-like energy scale of the chiral spinons, which can further determine \cite{Weng.Mei.20107w, Weng.Chen.2005, Zhang.Weng.hg_2023} the SC transition temperature, $T_c$, and be detected by the inelastic neutron scattering experiments \cite{{PhysRevLett.75.316, Keimer.He.20006wc, Keimer.Fong.1999, Keimer.He.2002, Dai1999}}. Previously the longitudinal transport of such chiral spinons has been shown to give rise to the Nernst effect \cite{Weng.Qi.2007}, the spin Hall effect \cite{Weng.Kou.2005} as briefly mentioned in \secref{other}.
Additionally,  in such a framework, an ``order-to-order'' phase transition between AFM insulating phase and SC phase is expected in the cuprates, which is worth further investigation in the future to establish a possible relationship with the DQCP \cite{Senthil.Levin.2004, Fisher.Senthil.2004, Senthil.Wang.2017}.



It is further noted that the origin of the thermal Hall effect in different studies \cite{Sachdev.Guo.2020, Lee.Han.2019, Sachdev.Samajdar.2019}, starting from the $\pi$-flux fermionic spinons, has been also attributed to the Berry curvatures of the spinon bands. However, without the external magnetic fields, the normal state of spinons \cite{Sachdev.Samajdar.2019} is usually conventional or topologically trivial. By contrast, here the external magnetic field merely shifts the balance number of the excited spinons with opposite chirality without changing the internal strong Berry curvatures introduced by the nontrivial topological (mutual Chern-Simons) gauge structure. The latter is intrinsically embedded in the pseudogap regime, describing the long-range entanglement between spin and charge degrees of freedom due to the phase-string effect in the doped Mott insulator. Additionally, some other studies attribute the enhancement of the thermal Hall signal to phonons through some extrinsic mechanisms\cite{Sachdev.Guo.2021, Sachdev.Guo.2022, Sun.Chen.2020}. It should be pointed out that bare phonons are not sensitive to the direction of external magnetic fields, but the experimentally observed thermal Hall coefficient in cuprates depends on the magnetic field component perpendicular to the copper oxide plane \cite{Taillefer.Boulanger.2020}. 

Importantly, all the transport results obtained in this work hinge on the robustness of the chiral spinon excitation, which is protected by the underlying bosonic RVB pairing. However, as the doping further increases beyond a critical point, i.e., $\delta>\delta^*$, the AFM correlation becomes too weak to preserve such an RVB pairing, leading to the breakdown of the pseudogap phase and the restoration of a Fermi liquid with a large Fermi surface \cite{Zhang.Weng_2022}, as has been suggested experimentally \cite{Balicas.Hussey.2003, Campuzano.Chatterjee.2011, Campuzano.Kaminski.2003, Damascelli.Plat.2005, Hussey.Vignolle.2008}. As a result, apparently, the present transport results should collapse with the contribution dominated by the quasiparticle excitations with the full Fermi surface restored in the overdoped regime at low temperatures. For instance, as indicated by experiments, the Hall number should change from $\delta$ to $1+\delta$ \cite{Taillefer.Proust.2019, Proust.Badoux.2016} and the thermal Hall coefficient should restore the behavior of the Wiedemann-Franz law \cite{Taillefer.Grissonnanche.2019}.  

Finally, we note that certain experiments have recently detected a signal of the thermal Hall effect along the z-axis in cuprates \cite{Taillefer.Grissonnanche.2020}. Our current study has been focused on the pure 2D and does not offer a quantitative explanation. We speculate that since the phase-string sign structure underlies the intrinsic Berry curvatures leading to the thermal Hall effect, its existence in any dimensions of a doped Mott insulator which has been rigorously proven \cite{Zaanen.Wu.2008} before, may be also responsible for the above-mentioned experimental observation beyond 2D. Technically, in realistic materials--- stacked copper oxide layers --- the interlayer coupling may cause the edge state of the spinons, as described in this paper, to undergo tunneling between different layers. A further study along this line will be worth proceeding elsewhere.


\begin{acknowledgments}
{\it Acknowledgments.---}
We acknowledge stimulating discussions with Long Zhang, Binghai Yan, Yuanming Lu, and Gang Li. Z. -J.S, J.-X.Z., and Z.-Y.W. are supported by MOST of China (Grant No. 2021YFA1402101) 
and NSF of China (Grant No. 12347107).
\end{acknowledgments}

\appendix


\renewcommand\thefigure{\thesection S\arabic{figure}}
\renewcommand\theequation{\thesection S\arabic{equation}}




\section{Effective Lagrangian of chiral spinons}
Upon the condensation of holons and following the decomposition $h_I=\sqrt{n_I^h} e^{i \theta_I}$ up to the second order of $\delta n_I^h = n_I^h-\bar{n}_I^h$ and $\theta_I$, one can integrate out the amplitude fluctuation $\delta n_I^h$. To account for the interaction between holons, we introduce an additional term, $u (n_I^h)^2/2$. With this, the Lagrangian \eqnref{Lh} can be recast as:

\begin{eqnarray}
	L_h &=& \sum_I \frac{1}{2 u}\left[\partial_\tau \theta_I-A_0^s(I)-e A_0^e\left(I\right)\right]^2 \notag \\
	& & -i\bar{n}^h \left[A_0^s\left(I\right)+A_0^e\left(I\right)\right] \notag \\
	& &+t_h \bar{n}^h\left[\Delta_\alpha \theta_I-A_\alpha^s(I)-\frac{e}{\hbar} A_\alpha^e(I)+2 \pi m_\alpha(I)\right]^2,\notag \\
\end{eqnarray}
in which the Villian expansion
\begin{equation}
    e^{\gamma \cos \Delta_\alpha \theta_I}\simeq \sum_{\{m_\alpha \in \mathbb{Z}\}} e^{-\frac{\gamma}{2}[\Delta_\alpha \theta_I+2 \pi m_\alpha(I)]^2}
\end{equation}
is applied. Proceeding further, integrating out the $A_0^s$ and $A_\alpha^s$ fields yields:
\begin{eqnarray}\label{LhCS}
	L_h +L_{\text{CS}} &=& \frac{u}{2 \pi^2}\left[B^h(i)-\pi \bar{n}^h\right]^2 \notag \\
	& & +\frac{1}{4 t_h \bar{n}^h \pi^2}\left(\boldsymbol{E}^h(i) \cdot z\right)_\alpha^2 \notag \\ 
	& & -\frac{i e}{\pi} \sum_i e^{\mu \nu \lambda} A_\mu^e \partial_v A_\lambda^h-\frac{i}{\pi} A_\lambda^h \epsilon^{\lambda \nu \mu} 2 \pi \partial_v m_\mu \notag \\ 
\end{eqnarray}
In conjunction with the $L_s$ term in \eqnref{Ls}, integrating out the $A_0^h$ produces a logarithmic attraction:
\begin{equation}
    -\frac{\pi n_h t_h}{2} \sum_{i \neq j} q_i \ln \left(\frac{\left|r_i-r_j\right|}{a}\right) q_j
\end{equation}
where $q_i \equiv \sum_\sigma n_{i, \sigma}^b+2 J_0^{2\pi}+\frac{\Phi^e}{\pi}$ represents the local total chirality. It's noteworthy that the logarithmic interaction can result in a divergent energy over long distances. Therefore, the condition for finite energy leads to the subsequent neutral condition:
\begin{equation}
    \sum_\sigma n_{i, \sigma}^b-2 J_0^{2\pi}+\frac{\Phi^e}{\pi}=0,
\end{equation}
which aligns with \eqnref{consHiggs} in the main text. Subsequently, let us introduce the vortex operator $\Phi_i^{\dagger} \equiv \boldsymbol{e}^{i \sum_{l \neq i} n_l^h \theta_i(l)}$. This operator satisfies the commutation relations:
$\left[\Phi_i^{\dagger}, J_0^{2\pi}(j)\right]=\Phi_i^{\dagger} \delta_{i, j}$ and $\left[\Phi_i, \Phi_j^{\dagger}\right]=0$ with $\Phi_i^{\dagger} \Phi_i=1$. Consequently, we can construct the four lowest energy vortex excitations, $b_{i \sigma \nu}^{\dagger}$, with winding numbers $\pm$, as follows: 
\begin{eqnarray}
    b_{i \uparrow +}^{\dagger}&\equiv& b_{i \uparrow +}^{\dagger},\;\;\;\;\;\;\;\;\; b_{i \uparrow -}^{\dagger}\equiv b_{i \uparrow +}^{\dagger} \Phi_i, \\
    b_{i \downarrow -}^{\dagger}&\equiv& b_{i \downarrow +}^{\dagger},\;\;\;\;\;\;\;\;\; 
    b_{i \downarrow +}^{\dagger}\equiv b_{i \downarrow +}^{\dagger} \Phi_i^\dagger.
\end{eqnarray}
These excitations adhere to the bosonic commutation relations. Hence, in conjunction with \eqnref{LhCS}, the Lagrangian characterizing the physical behavior of these low-lying excitations is denoted by $\tilde{L}=\tilde{L}_b+\tilde{L}_\text{MCS}$, which is expressed as:
\begin{eqnarray}
    \tilde{L}_s&=&\sum_{i \sigma \nu} b_{i \sigma \nu}^{\dagger}\left[\partial_\tau-i \nu A_0^h(i)+\lambda_b+\frac{1}{2} g \mu_B B^e \sigma\right] b_{i \sigma \nu} \notag \\ 
	& & -J_s \sum_{i \alpha \sigma \nu}\left[b_{i \sigma \nu}^{\dagger} b_{i+\alpha, \bar{\sigma} \bar{\nu}}^{\dagger} e^{i \nu \boldsymbol{A}_\alpha^h(i)}+\text { h.c. }\right]\label{Lsapp} \\
    \tilde{L}_{\mathrm{MCS}}&=&\sum_i \frac{u}{2 \pi^2}\left[B^h(i)-\pi \bar{n}^h\right]^2 
	+\frac{1}{4 t_h \bar{n}^h \pi^2}\left(\boldsymbol{E}^h(i) \cdot z\right)_\alpha^2 \notag \\
	& & -\frac{i e}{\pi} \sum_i \epsilon^{\mu \nu \lambda} A_\mu^h \partial_\nu A_\lambda^e\label{Ldualapp}.
\end{eqnarray}



\section{Determination of $T_c$} \label{AppBTc}
Utilizing the saddle point approximation for $\tilde{L}_h$, we have:
\begin{equation}
    B^h(i) \rightarrow \bar{B}^h=\pi \bar{n}^h, \;\;\;\;\;\;\;
	\boldsymbol{E}^h(i) \rightarrow \overline{\boldsymbol{E}}^h=0.
\end{equation}
Under this approximation, $b$-spinons encounter a uniform gauge field imparting a $\delta\pi$ flux per plaquette. Drawing parallels with the standard diagonalization approach found in the Hofstadter system, the Lagrangian \eqnref{Lsapp} can be reformulated as:
\begin{equation}\label{diagLs}
    \tilde{L}_s=\sum_{i \sigma \nu} b_{i \sigma \nu}^{\dagger}\left[\partial_\tau-i \nu A_0^h(i)\right] b_{i \sigma \nu}+\sum_{m\sigma \nu} E_m \gamma_{m \sigma \nu}^\dagger \gamma_{m \sigma \nu}
\end{equation}
with the $b$-spinons spectrum:
\begin{equation}\label{Emb}
	E_{m}^b=\sqrt{\lambda_b^{2}-(\xi_{m}^b)^{2}}
\end{equation}
via introducing the following Bogoliubov transformation:
\begin{equation}\label{bogo}
	b_{i \sigma\nu}=\sum_{m} \omega_{m \sigma\nu}(\boldsymbol{r}_i)\left(u_{m} \gamma_{m \sigma\nu}-v_{m} \gamma_{m\bar{\sigma}\bar{\nu}}^{\dagger}\right),
\end{equation}
where the coherent factors are given by
\begin{eqnarray}\label{Bogofactor}
u_{m}&=&\sqrt{\frac{1}{2}\left(1+\frac{\lambda_b}{E_{m}^b}\right)}
\notag\\
	v_{m}&=&\operatorname{sgn}\left(\xi_{m}^b\right) \sqrt{\frac{1}{2}\left(-1+\frac{\lambda_b}{E_{m}^b}\right)}.
\end{eqnarray}
Here, $\xi_{m}^b$ as well as $w_{m}(\boldsymbol{r}_i)\equiv w_{m \sigma\nu=+}(\boldsymbol{r}_i)=w_{m\sigma\bar{\nu}}^{*}(\boldsymbol{r}_i)$ in \eqnref{bogo} are the eigenfunctions and eigenvalues of the following equation:
\begin{equation}\label{Hbdiag}
	\xi_{m}^b \omega_{m }(\boldsymbol{r}_i)=-\frac{J \Delta^{s}}{2} \sum_{j=\text{NN}(i)} e^{i  A_{i j}^{h} }\omega_{m }(\boldsymbol{r}_j).
\end{equation}
\begin{figure}[t]
    \centering
    \includegraphics[width=\linewidth]{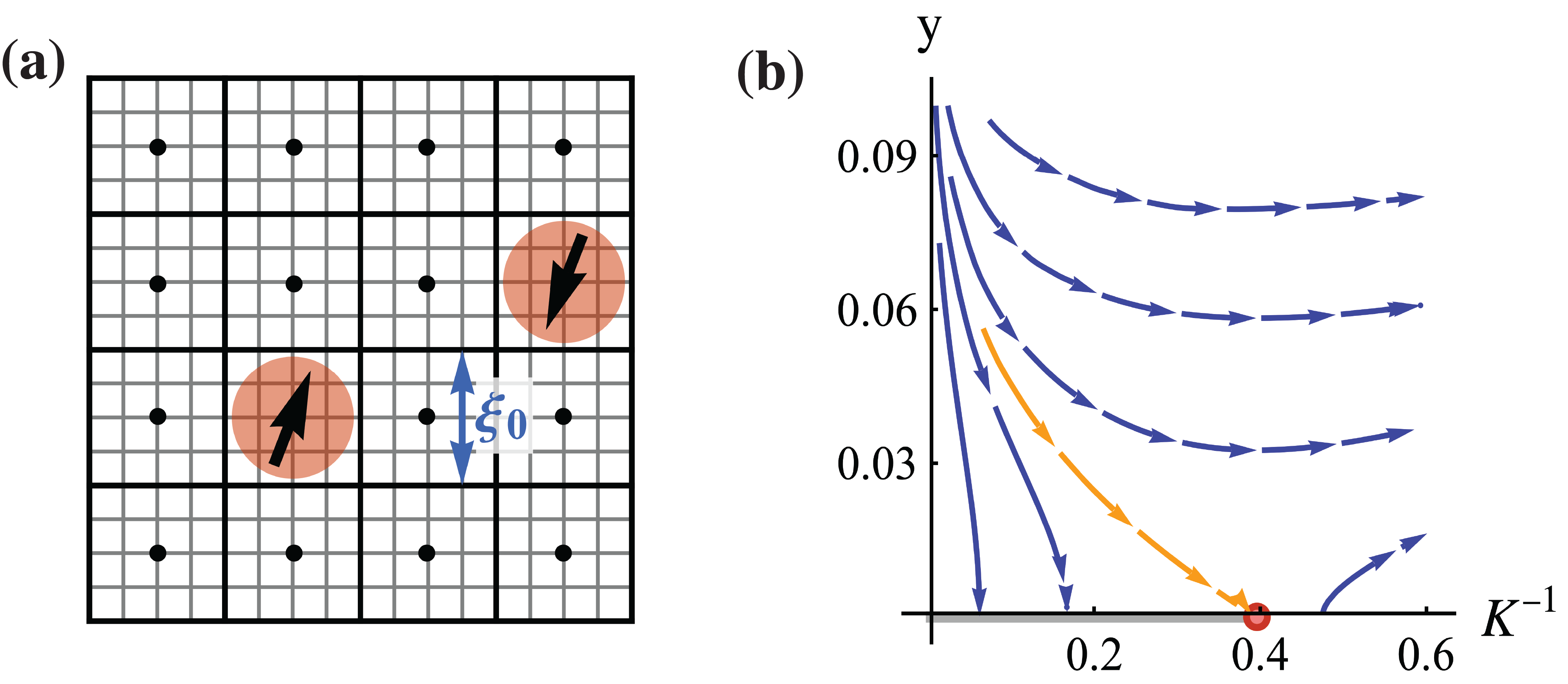}
    \caption{(a)The von Neumann lattice with the lattice constants $\xi_0$ is depicted with black lines. And the original lattice is illustrated with gray lines and has lattice constants $a$. The $b$-spinon wave packets, labeled by red disks, are positioned at the center $\boldsymbol{R}_m$ of the von Neumann lattice, marked by black points. (b) Illustration of the RG flow corresponding to \eqnref{RG1} and \eqnref{RG2}. Orange arrows indicate the separatrix, while the red point marks the fixed point $(K^*)^{-1}=\pi/8$ and $y^*=0$.}
    \label{fig_RG}
\end{figure}
The derived $b$-spinon dispersion $E_m^b$ in \eqnref{Emb} is depicted in \figref{fig_Eb4}(a), showcasing dispersionless, ``Landau-level-like'' discrete energy levels\cite{Zhang.Weng_2022, Weng.Chen.2005} with a gap $E_s$ (indicated by the red arrow). Our subsequent discussions focus primarily on the lowest Landau level (LLL) where $E_m = E_s$. Based on prior research\cite{Weng.Mei.20107w, Efros.Rashba.1997}, within the LLL, the $w_m$'s are termed as \emph{local} modes (LM). These have an intrinsic size on the scale of the ``cyclotron length'', given by $a_c = 1/\sqrt{\pi \delta}$. For simplicity, we consider the lattice constant to be $a=1$.

These spinon wave packets, represented by magnetic Wannier wave function $w_{m}(\mathbf{r})$, have an amplitude defined as:
\begin{equation}
    O_m(\boldsymbol{r}_i) \equiv\left|w_m(\boldsymbol{r}_i)\right|^2 \simeq \frac{a^2}{2 \pi a_c^2} \exp \left[-\frac{\left|\boldsymbol{r}_i-\boldsymbol{R}_m\right|^2}{2 a_c^2}\right].
\end{equation}
This peaks at the centers of a von Neumann lattice, with a lattice constant $\xi_0=\sqrt{2 \pi} a_c$ and the lattice site $\mathbf{R}_m$, illustrated in \figref{fig_RG}(a).

Combined with \eqnref{diagLs} and \eqnref{Ldualapp}, the saddle point effective Lagrangian becomes:
\begin{eqnarray}
    \tilde{L}_\text{eff} &=& \frac{1}{4 t_h \bar{n}^h \pi^2}\left(\nabla A_0^h\right)^2-i A_0^h \sum_i \sum_{\nu \sigma} \nu n_{i \sigma\nu}^b \notag \\
	& & +E_s \sum_{m \in \mathrm{LL}} \sum_{\sigma\nu} n_{m \sigma\nu}^b,
\end{eqnarray}
where the term $b_{i \sigma\nu}^{\dagger} \partial_\tau b_{i \sigma\nu}$ is disregarded under the assumption of a temperature high enough to preclude quantum fluctuations. Further, from \eqnref{bogo}, we obtain:
\begin{equation}
    \sum_{\sigma\nu} \nu n_{i \sigma\nu}^b=\sum_{m \in \mathrm{LL}} O_m(\boldsymbol{r}_i) \sum_{\sigma\nu} \nu n_{m \sigma\nu}.
\end{equation}
After integrating out $A_0^h$, the effective action is described as:
\begin{eqnarray}
    S_{\text {eff}}&\simeq& -\int_0^{\beta} d\tau\frac{\pi \bar{n}_h t_h}{2} \sum_{m \neq n} q\left(\boldsymbol{R}_m\right) \ln \left(\frac{\left|\boldsymbol{R}_m-\boldsymbol{R}_n\right|}{\xi_0}\right) q\left(\boldsymbol{R}_n\right) \notag \\
	& & +\int_0^{\beta} d\tau E_s \sum_m\left|q\left(\boldsymbol{R}_m\right)\right|^2\\
    &=& -\frac{\pi}{4} \frac{\rho_s}{k_B T} 
	\sum_{m \neq n} q\left(\boldsymbol{R}_m\right) 
	\ln \left(\frac{\left|\boldsymbol{R}_m-\boldsymbol{R}_n\right|}{\xi_0}\right)
	q\left(\boldsymbol{R}_n\right) \notag \\
	& & +\frac{E_s}{k_B T} \sum_m\left|q\left(\boldsymbol{R}_m\right)\right|^2,\label{2D}
\end{eqnarray}
where $t_h=\frac{\hbar^2}{2 m_h}$, the spin stiffness is denoted by $\rho_s \equiv \bar{n}_h / m_h$. In this context, \eqnref{2D} represents a two-dimensional neutral Coulomb plasma with unit charge $q=\pm 1$ situated on the von Neumann lattice. Following this, we can employ a standard approach to address the conventional KT transition. We introduce the reduced stiffness as $K=\rho_s / k_B T$ and define the effective fugacity of each vortex as $y \equiv e^{-E_s / k_B T}$.

The differential renormalization group (RG) equations are then expressed as:
\begin{eqnarray}
&\;&\frac{\mathrm{d} K^{-1}}{\mathrm{~d} l}=g^2 \pi^3 y^2+O\left(y^4\right),\label{RG1} \\
&\;&\frac{\mathrm{d} y}{\mathrm{~d} l}=\left(2-\frac{\pi}{4} K\right) y+O\left(y^3\right)\label{RG2},
\end{eqnarray}
where $g=4$ accounts for the degeneracy at each site $\boldsymbol{R}_m$ on the von Neumann lattice, arising from time reversal and bipartite lattice symmetries. An observation is that these RG equations remain valid even if we substitute $(K, y)$ with $(K 4, gy )$ in the conventional KT transition's RG equations. This substitution is attributed to the unit vorticity of each spinon vortex being $\pi$ instead of $2\pi$ of a conventional vortex, and $y$ is replaced by $gy$ because of the $g$ degeneracy for each site $\boldsymbol{R}_m$ on the von Neumann lattice.

The RG flow is illustrated in \figref{fig_RG}(b), which displays a separatrix, indicated by orange arrows, traversing the critical point $(K^*)^{-1}=\pi / 8$ and $y^*=0$. Points located above this separatrix tend toward larger values of both $K^{-1}$ and $y$, signifying a transition to the phase with unbound vortices corresponding to the SC disordered phase. Conversely, points below this separatrix are towards to the line $y = 0$. This indicates a prohibition on free vortex excitation at low temperatures and corresponds to the chiral spinon confinement in the SC phase. The separatrix flow intrinsically determines the SC critical temperature $T_c$, which can be found by solving:
\begin{equation}
    K_c^{-1}-\pi / 8=-\pi^2 y_c.
\end{equation}
Given the relationship $k_B T_c/\rho_s\ll\pi / 8$, we deduce:
\begin{equation}
    \frac{E_s}{k_B T_c} \simeq 3.22,
\end{equation}
which is consistent with \eqnref{Tc} in the main text. Moreover, by recognizing that $E_g=2E_s$, we can derive \eqnref{Tckappa}.

\section{Derivation of semi-classical transport coefficient}
\begin{figure}[t]
	\centering
\includegraphics[scale=0.4]{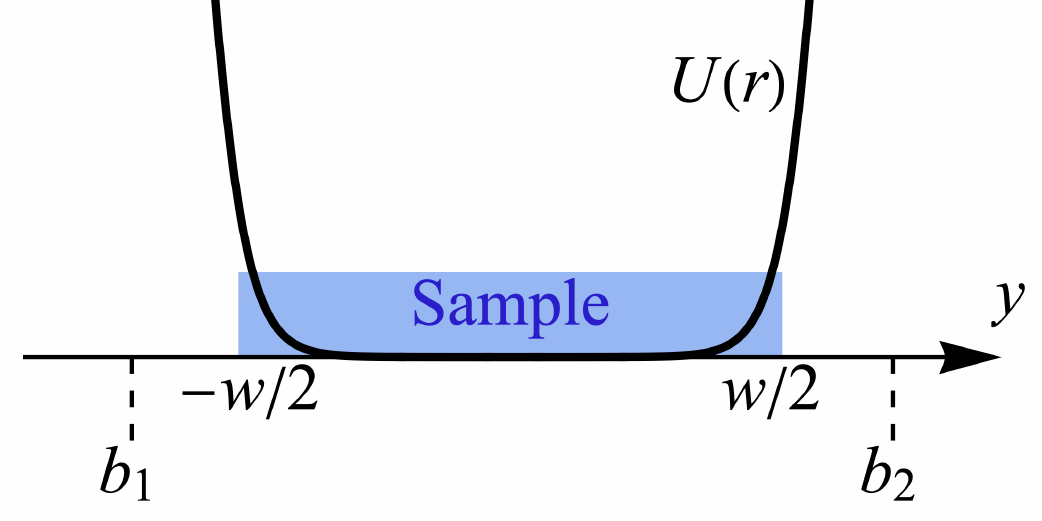}
	\caption{Depiction of a sample characterized by a width $w$, utilized in the determination of the edge current. The potential $U(\boldsymbol{r})$ acts as a confining mechanism within the sample.}
	\label{fig_sample}
\end{figure}
The velocity of a wave packet is characterized by the semiclassical equation of motion, incorporating the topological Berry phase term:
\begin{equation}\label{anorv}
\boldsymbol{v}\equiv\dot{\boldsymbol{r}}=\frac{1}{\hbar} \frac{\partial \epsilon_{n}(\boldsymbol{k})}{\partial \boldsymbol{k}}-\dot{\boldsymbol{k}} \times \boldsymbol{\Omega}_{n}(\boldsymbol{k})
\end{equation}
where $n$ denotes the band index, $\epsilon_n(\boldsymbol{k})$ represents the energy dispersion of the $n$-th band, and the Berry curvature in momentum space is given by $\boldsymbol{\Omega}_{n}(\boldsymbol{k})=i \nabla_{\boldsymbol{k}} \times\left\langle u_{n}(\boldsymbol{k})\left|\nabla_{\boldsymbol{k}}\right| u_{n}(\boldsymbol{k})\right\rangle$. It's worth noting that $\hbar \dot{\boldsymbol{k}}=-\nabla U(\boldsymbol{r})$, where $U(\boldsymbol{r})$ is a confining potential present only near the boundary of the sample. This potential ensures that the wave packet of particles remains within the sample, and its gradient applies a force on the particles.

Let us consider a sample delineated by a boundary, with a coordinate system as depicted in \figref{fig_sample}. Here, $w$ signifies the system's width in the $y$-direction, and $b_1, b_2$ represent the $y$-axis coordinates of the boundary where $U\left(b_1\right)=U\left(b_2\right)=\infty$. From the second term of \eqnref{anorv}, close to the sample's boundary, there emerges an edge current directed along the $x$-axis. The current density is obtained by summing up the local velocity $\boldsymbol{v}_x(y)$ and dividing it by the width:
\begin{eqnarray}
    J_x&=&\frac{1}{w} \int_{b_1}^{b_2} \frac{1}{ V}\boldsymbol{v}_x(y)dy\\
    &=& \frac{1}{w} \int_{b_1}^{b_2} \frac{1}{\hbar V} \sum_{k, m} n\left(\epsilon_{m}(\boldsymbol{k})+U(y) ; T\left(y\right)\right) \notag \\ 
	& & \left[\partial_y U(y)\right] \boldsymbol{\Omega}_{m}(\boldsymbol{k}) d y\\
    &\simeq&-\frac{1}{w} \sum_{\boldsymbol{k}, m}\int_{\epsilon_{m}(\boldsymbol{k})}^{\infty} \frac{1}{\hbar V}  n\left(\epsilon ; T\left(-\frac{w}{2}\right)\right) \boldsymbol{\Omega}_{m}(\boldsymbol{k}) d \epsilon \notag \\ 
	& & +\frac{1}{w} \sum_{\boldsymbol{k}, m}\int_{\epsilon_{m}(\boldsymbol{k})}^{\infty} \frac{1}{\hbar V}  n\left(\epsilon ; T\left(\frac{w}{2}\right)\right) \boldsymbol{\Omega}_{m}(\boldsymbol{k}) d \epsilon\\
    &=&\frac{1}{w} \int  \partial_y\left[\frac{1}{\hbar V} \sum_{\boldsymbol{k}, m} \int_{\epsilon_{m}(\boldsymbol{k})}^{\infty} n(\epsilon ; T(y)) \boldsymbol{\Omega}_{m}(\boldsymbol{k})d \epsilon\right] dy \notag \\
\end{eqnarray}
where the current here means the current of the particle number, $V$ is the area of the sample and $n(\epsilon;T)=\left[e^{(\epsilon-\mu)/k_B T}-1\right]^{-1}$ is the bosonic distribution. Analogously, the energy current derived from the edge current density is expressed as:
\begin{eqnarray}
    \left(J_E\right)_x^{\nabla T}&=& \frac{1}{w} \int \partial_y \left[ \frac{1}{\hbar V} \sum_{\boldsymbol{k}, m} \int_{\epsilon_n(\boldsymbol{k})}^{\infty} \epsilon \times n(\epsilon ; T(y)) \right. \notag \\
	& & \left. \boldsymbol{\Omega}_m(\boldsymbol{k}) d \epsilon \right] d y.
\end{eqnarray}

Given a temperature gradient along the $y$-direction, the gradient in particle distribution is formulated as:
\begin{equation}
    \partial_y n\left(\epsilon ; T(y)\right)=(\epsilon-\mu) \frac{\partial n}{\partial \epsilon} T \frac{\partial \beta}{\partial y},
\end{equation}
which leads to the edge current density as well as the energy current:
\begin{eqnarray}
    (J)_x^{\nabla T}&=&\frac{1}{w} \int \left[\frac{1}{\hbar V} \sum_{\boldsymbol{k}, m} \int_{\epsilon_{m}(\boldsymbol{k})}^{\infty}(\epsilon-\mu) \right. \notag \\
	& & \left. \frac{\partial n}{\partial \epsilon} T \frac{\partial \beta}{\partial y} \boldsymbol{\Omega}_{m}(\boldsymbol{k})d \epsilon\right]  dy 
	\label{J1}\\
    \left(J_E\right)_x^{\nabla T}&=&\frac{1}{w} \int \left[\frac{1}{\hbar V} \sum_{\boldsymbol{k}, m} \int_{\epsilon_{m}(\boldsymbol{k})}^{\infty} \epsilon(\epsilon-\mu) \right. \notag \\
	& & \left.  \frac{\partial n}{\partial \epsilon} T \frac{\partial \beta}{\partial y} \boldsymbol{\Omega}_{m}(\boldsymbol{k})d \epsilon\right]  dy\label{JE1}. 
\end{eqnarray}
Similarly, the chemical potential gradient along the $y$-direction can give the particle distribution gradient as follows:
\begin{eqnarray}
\partial_y n(\epsilon(y) ; T)=-\frac{\partial n(\epsilon ; T)}{\partial \epsilon} \frac{\partial \mu}{\partial y},
\end{eqnarray}
which results in the edge current density and the energy current:
\begin{eqnarray}
    (J)_x^{\nabla \mu}&=&-\frac{1}{w} \int \left[\frac{1}{\hbar V} \sum_{\boldsymbol{k}, m} \int_{\epsilon_{m}(\boldsymbol{k})}^{\infty} \right. \notag \\
	& & \left. \frac{\partial n(\epsilon ; T)}{\partial \epsilon} \frac{\partial \mu}{\partial y} \Omega_{n}(\boldsymbol{k})d \epsilon\right] dy
	\label{J2}\\
	\left(J_E\right)_x^{\nabla \mu}&=&-\frac{1}{w} \int d y\left[\frac{1}{\hbar V} \sum_{\boldsymbol{k}, m} \int_{\epsilon_{m}(\boldsymbol{k})}^{\infty} \right. \notag \\
	& & \left. \epsilon \frac{\partial n(\epsilon ; T)}{\partial \epsilon} \frac{\partial \mu}{\partial y} \boldsymbol{\Omega}_{n}(\boldsymbol{k})d \epsilon\right] dy\label{JE2} 
\end{eqnarray}
Now we define a heat current as $J_Q \equiv J_E-\mu J$ and write down the linear response of the current $J_x$ and heat current $(J_E)_x$ by combining \eqnref{J1}, \eqnref{JE1}, \eqnref{J2} and \eqnref{JE2}:
\begin{equation}
    \left[\begin{array}{c}
\boldsymbol{J}_x^\nu \\
\left(\boldsymbol{J}_Q^\nu\right)_x
\end{array}\right]=\boldsymbol{L}^{\mathrm{xy}}\left[\begin{array}{c}
-\nabla_y \mu \\
T \nabla_y \frac{1}{T}
\end{array}\right].
\end{equation}
Here, $\boldsymbol{L}^{\mathrm{xy}\text{,}\nu}$ signifies a $2\times 2$ matrix that represents the transverse transport coefficients. The parameter $\nu=\pm$ distinguishes between the different chiralities of spinon. The matrix elements are given by
\begin{eqnarray}
L_{i j}^{x y} = -\frac{1}{\hbar V \beta^q} \sum_{m } \sum_{\boldsymbol{k}} c_q\left[n \left( \epsilon_m(\boldsymbol{k});T \right) \right] \boldsymbol{\Omega}_{m}(\boldsymbol{k})
\end{eqnarray}
where $i,j=1,2$ and
\begin{equation}
    c_q[x] \equiv \int_0^x\left(\log \frac{1+t}{t}\right)^q d t,
\end{equation}
which is consistent with \eqnref{Lij} in the main text.

\bibliography{main}

\begin{thebibliography}{100}%
\makeatletter
\providecommand \@ifxundefined [1]{%
 \@ifx{#1\undefined}
}%
\providecommand \@ifnum [1]{%
 \ifnum #1\expandafter \@firstoftwo
 \else \expandafter \@secondoftwo
 \fi
}%
\providecommand \@ifx [1]{%
 \ifx #1\expandafter \@firstoftwo
 \else \expandafter \@secondoftwo
 \fi
}%
\providecommand \natexlab [1]{#1}%
\providecommand \enquote  [1]{``#1''}%
\providecommand \bibnamefont  [1]{#1}%
\providecommand \bibfnamefont [1]{#1}%
\providecommand \citenamefont [1]{#1}%
\providecommand \href@noop [0]{\@secondoftwo}%
\providecommand \href [0]{\begingroup \@sanitize@url \@href}%
\providecommand \@href[1]{\@@startlink{#1}\@@href}%
\providecommand \@@href[1]{\endgroup#1\@@endlink}%
\providecommand \@sanitize@url [0]{\catcode `\\12\catcode `\$12\catcode
  `\&12\catcode `\#12\catcode `\^12\catcode `\_12\catcode `\%12\relax}%
\providecommand \@@startlink[1]{}%
\providecommand \@@endlink[0]{}%
\providecommand \url  [0]{\begingroup\@sanitize@url \@url }%
\providecommand \@url [1]{\endgroup\@href {#1}{\urlprefix }}%
\providecommand \urlprefix  [0]{URL }%
\providecommand \Eprint [0]{\href }%
\providecommand \doibase [0]{https://doi.org/}%
\providecommand \selectlanguage [0]{\@gobble}%
\providecommand \bibinfo  [0]{\@secondoftwo}%
\providecommand \bibfield  [0]{\@secondoftwo}%
\providecommand \translation [1]{[#1]}%
\providecommand \BibitemOpen [0]{}%
\providecommand \bibitemStop [0]{}%
\providecommand \bibitemNoStop [0]{.\EOS\space}%
\providecommand \EOS [0]{\spacefactor3000\relax}%
\providecommand \BibitemShut  [1]{\csname bibitem#1\endcsname}%
\let\auto@bib@innerbib\@empty
\bibitem [{\citenamefont {Lee}\ \emph {et~al.}(2006)\citenamefont {Lee},
  \citenamefont {Nagaosa},\ and\ \citenamefont {Wen}}]{Wen.Lee.2006z4}%
  \BibitemOpen
  \bibfield  {author} {\bibinfo {author} {\bibfnamefont {P.~A.}\ \bibnamefont
  {Lee}}, \bibinfo {author} {\bibfnamefont {N.}~\bibnamefont {Nagaosa}},\ and\
  \bibinfo {author} {\bibfnamefont {X.-G.}\ \bibnamefont {Wen}},\ }\bibfield
  {title} {\bibinfo {title} {{Doping a Mott insulator: Physics of
  high-temperature superconductivity}},\ }\href
  {https://doi.org/10.1103/revmodphys.78.17} {\bibfield  {journal} {\bibinfo
  {journal} {Rev. Mod. Phys.}\ }\textbf {\bibinfo {volume} {78}},\ \bibinfo
  {pages} {17 } (\bibinfo {year} {2006})}\BibitemShut {NoStop}%
\bibitem [{\citenamefont {Keimer}\ \emph {et~al.}(2015)\citenamefont {Keimer},
  \citenamefont {Kivelson}, \citenamefont {Norman}, \citenamefont {Uchida},\
  and\ \citenamefont {Zaanen}}]{Zaanen.Keimer.2015}%
  \BibitemOpen
  \bibfield  {author} {\bibinfo {author} {\bibfnamefont {B.}~\bibnamefont
  {Keimer}}, \bibinfo {author} {\bibfnamefont {S.~A.}\ \bibnamefont
  {Kivelson}}, \bibinfo {author} {\bibfnamefont {M.~R.}\ \bibnamefont
  {Norman}}, \bibinfo {author} {\bibfnamefont {S.}~\bibnamefont {Uchida}},\
  and\ \bibinfo {author} {\bibfnamefont {J.}~\bibnamefont {Zaanen}},\
  }\bibfield  {title} {\bibinfo {title} {{From quantum matter to
  high-temperature superconductivity in copper oxides}},\ }\href
  {https://doi.org/10.1038/nature14165} {\bibfield  {journal} {\bibinfo
  {journal} {Nature}\ }\textbf {\bibinfo {volume} {518}},\ \bibinfo {pages}
  {179} (\bibinfo {year} {2015})}\BibitemShut {NoStop}%
\bibitem [{\citenamefont {Proust}\ and\ \citenamefont
  {Taillefer}(2019)}]{Taillefer.Proust.2019}%
  \BibitemOpen
  \bibfield  {author} {\bibinfo {author} {\bibfnamefont {C.}~\bibnamefont
  {Proust}}\ and\ \bibinfo {author} {\bibfnamefont {L.}~\bibnamefont
  {Taillefer}},\ }\bibfield  {title} {\bibinfo {title} {{The Remarkable
  Underlying Ground States of Cuprate Superconductors}},\ }\href
  {https://doi.org/10.1146/annurev-conmatphys-031218-013210} {\bibfield
  {journal} {\bibinfo  {journal} {Annu. Rev. Condens. Matter Phys.}\ }\textbf
  {\bibinfo {volume} {10}},\ \bibinfo {pages} {409} (\bibinfo {year}
  {2019})}\BibitemShut {NoStop}%
\bibitem [{\citenamefont {Badoux}\ \emph {et~al.}(2016)\citenamefont {Badoux},
  \citenamefont {Tabis}, \citenamefont {Laliberté}, \citenamefont
  {Grissonnanche}, \citenamefont {Vignolle}, \citenamefont {Vignolles},
  \citenamefont {Béard}, \citenamefont {Bonn}, \citenamefont {Hardy},
  \citenamefont {Liang}, \citenamefont {Doiron-Leyraud}, \citenamefont
  {Taillefer},\ and\ \citenamefont {Proust}}]{Proust.Badoux.2016}%
  \BibitemOpen
  \bibfield  {author} {\bibinfo {author} {\bibfnamefont {S.}~\bibnamefont
  {Badoux}}, \bibinfo {author} {\bibfnamefont {W.}~\bibnamefont {Tabis}},
  \bibinfo {author} {\bibfnamefont {F.}~\bibnamefont {Laliberté}}, \bibinfo
  {author} {\bibfnamefont {G.}~\bibnamefont {Grissonnanche}}, \bibinfo {author}
  {\bibfnamefont {B.}~\bibnamefont {Vignolle}}, \bibinfo {author}
  {\bibfnamefont {D.}~\bibnamefont {Vignolles}}, \bibinfo {author}
  {\bibfnamefont {J.}~\bibnamefont {Béard}}, \bibinfo {author} {\bibfnamefont
  {D.~A.}\ \bibnamefont {Bonn}}, \bibinfo {author} {\bibfnamefont {W.~N.}\
  \bibnamefont {Hardy}}, \bibinfo {author} {\bibfnamefont {R.}~\bibnamefont
  {Liang}}, \bibinfo {author} {\bibfnamefont {N.}~\bibnamefont
  {Doiron-Leyraud}}, \bibinfo {author} {\bibfnamefont {L.}~\bibnamefont
  {Taillefer}},\ and\ \bibinfo {author} {\bibfnamefont {C.}~\bibnamefont
  {Proust}},\ }\bibfield  {title} {\bibinfo {title} {{Change of carrier density
  at the pseudogap critical point of a cuprate superconductor}},\ }\href
  {https://doi.org/10.1038/nature16983} {\bibfield  {journal} {\bibinfo
  {journal} {Nature}\ }\textbf {\bibinfo {volume} {531}},\ \bibinfo {pages}
  {210} (\bibinfo {year} {2016})},\ \Eprint {https://arxiv.org/abs/1511.08162}
  {1511.08162} \BibitemShut {NoStop}%
\bibitem [{\citenamefont {Storey}(2016)}]{Storey.Storey.2016}%
  \BibitemOpen
  \bibfield  {author} {\bibinfo {author} {\bibfnamefont {J.~G.}\ \bibnamefont
  {Storey}},\ }\bibfield  {title} {\bibinfo {title} {{Hall effect and Fermi
  surface reconstruction via electron pockets in the high-Tc cuprates}},\
  }\href {https://doi.org/10.1209/0295-5075/113/27003} {\bibfield  {journal}
  {\bibinfo  {journal} {EPL}\ }\textbf {\bibinfo {volume} {113}},\ \bibinfo
  {pages} {27003} (\bibinfo {year} {2016})},\ \Eprint
  {https://arxiv.org/abs/1512.03112} {1512.03112} \BibitemShut {NoStop}%
\bibitem [{\citenamefont {Zhang}\ and\ \citenamefont
  {Sachdev}(2020)}]{Sachdev.Zhang.2020}%
  \BibitemOpen
  \bibfield  {author} {\bibinfo {author} {\bibfnamefont {Y.-H.}\ \bibnamefont
  {Zhang}}\ and\ \bibinfo {author} {\bibfnamefont {S.}~\bibnamefont
  {Sachdev}},\ }\bibfield  {title} {\bibinfo {title} {{From the pseudogap metal
  to the Fermi liquid using ancilla qubits}},\ }\href
  {https://doi.org/10.1103/physrevresearch.2.023172} {\bibfield  {journal}
  {\bibinfo  {journal} {Physical Review Research}\ }\textbf {\bibinfo {volume}
  {2}},\ \bibinfo {pages} {023172} (\bibinfo {year} {2020})},\ \Eprint
  {https://arxiv.org/abs/2001.09159} {2001.09159} \BibitemShut {NoStop}%
\bibitem [{\citenamefont {Nikolaenko}\ \emph {et~al.}(2021)\citenamefont
  {Nikolaenko}, \citenamefont {Tikhanovskaya}, \citenamefont {Sachdev},\ and\
  \citenamefont {Zhang}}]{Zhang.Nikolaenko.2021}%
  \BibitemOpen
  \bibfield  {author} {\bibinfo {author} {\bibfnamefont {A.}~\bibnamefont
  {Nikolaenko}}, \bibinfo {author} {\bibfnamefont {M.}~\bibnamefont
  {Tikhanovskaya}}, \bibinfo {author} {\bibfnamefont {S.}~\bibnamefont
  {Sachdev}},\ and\ \bibinfo {author} {\bibfnamefont {Y.-H.}\ \bibnamefont
  {Zhang}},\ }\bibfield  {title} {\bibinfo {title} {{Small to large Fermi
  surface transition in a single-band model using randomly coupled ancillas}},\
  }\href {https://doi.org/10.1103/physrevb.103.235138} {\bibfield  {journal}
  {\bibinfo  {journal} {Physical Review B}\ }\textbf {\bibinfo {volume}
  {103}},\ \bibinfo {pages} {235138} (\bibinfo {year} {2021})},\ \Eprint
  {https://arxiv.org/abs/2103.05009} {2103.05009} \BibitemShut {NoStop}%
\bibitem [{\citenamefont {Grissonnanche}\ \emph {et~al.}(2020)\citenamefont
  {Grissonnanche}, \citenamefont {Thériault}, \citenamefont {Gourgout},
  \citenamefont {Boulanger}, \citenamefont {Lefrançois}, \citenamefont
  {Ataei}, \citenamefont {Laliberté}, \citenamefont {Dion}, \citenamefont
  {Zhou}, \citenamefont {Pyon}, \citenamefont {Takayama}, \citenamefont
  {Takagi}, \citenamefont {Doiron-Leyraud},\ and\ \citenamefont
  {Taillefer}}]{Taillefer.Grissonnanche.2020}%
  \BibitemOpen
  \bibfield  {author} {\bibinfo {author} {\bibfnamefont {G.}~\bibnamefont
  {Grissonnanche}}, \bibinfo {author} {\bibfnamefont {S.}~\bibnamefont
  {Thériault}}, \bibinfo {author} {\bibfnamefont {A.}~\bibnamefont
  {Gourgout}}, \bibinfo {author} {\bibfnamefont {M.-E.}\ \bibnamefont
  {Boulanger}}, \bibinfo {author} {\bibfnamefont {E.}~\bibnamefont
  {Lefrançois}}, \bibinfo {author} {\bibfnamefont {A.}~\bibnamefont {Ataei}},
  \bibinfo {author} {\bibfnamefont {F.}~\bibnamefont {Laliberté}}, \bibinfo
  {author} {\bibfnamefont {M.}~\bibnamefont {Dion}}, \bibinfo {author}
  {\bibfnamefont {J.-S.}\ \bibnamefont {Zhou}}, \bibinfo {author}
  {\bibfnamefont {S.}~\bibnamefont {Pyon}}, \bibinfo {author} {\bibfnamefont
  {T.}~\bibnamefont {Takayama}}, \bibinfo {author} {\bibfnamefont
  {H.}~\bibnamefont {Takagi}}, \bibinfo {author} {\bibfnamefont
  {N.}~\bibnamefont {Doiron-Leyraud}},\ and\ \bibinfo {author} {\bibfnamefont
  {L.}~\bibnamefont {Taillefer}},\ }\bibfield  {title} {\bibinfo {title}
  {{Chiral phonons in the pseudogap phase of cuprates}},\ }\href
  {https://doi.org/10.1038/s41567-020-0965-y} {\bibfield  {journal} {\bibinfo
  {journal} {Nature Physics}\ }\textbf {\bibinfo {volume} {16}},\ \bibinfo
  {pages} {1108} (\bibinfo {year} {2020})},\ \Eprint
  {https://arxiv.org/abs/2003.00111} {2003.00111} \BibitemShut {NoStop}%
\bibitem [{\citenamefont {Boulanger}\ \emph {et~al.}(2020)\citenamefont
  {Boulanger}, \citenamefont {Grissonnanche}, \citenamefont {Badoux},
  \citenamefont {Allaire}, \citenamefont {Lefrançois}, \citenamefont {Legros},
  \citenamefont {Gourgout}, \citenamefont {Dion}, \citenamefont {Wang},
  \citenamefont {Chen}, \citenamefont {Liang}, \citenamefont {Hardy},
  \citenamefont {Bonn},\ and\ \citenamefont
  {Taillefer}}]{Taillefer.Boulanger.2020}%
  \BibitemOpen
  \bibfield  {author} {\bibinfo {author} {\bibfnamefont {M.-E.}\ \bibnamefont
  {Boulanger}}, \bibinfo {author} {\bibfnamefont {G.}~\bibnamefont
  {Grissonnanche}}, \bibinfo {author} {\bibfnamefont {S.}~\bibnamefont
  {Badoux}}, \bibinfo {author} {\bibfnamefont {A.}~\bibnamefont {Allaire}},
  \bibinfo {author} {\bibfnamefont {E.}~\bibnamefont {Lefrançois}}, \bibinfo
  {author} {\bibfnamefont {A.}~\bibnamefont {Legros}}, \bibinfo {author}
  {\bibfnamefont {A.}~\bibnamefont {Gourgout}}, \bibinfo {author}
  {\bibfnamefont {M.}~\bibnamefont {Dion}}, \bibinfo {author} {\bibfnamefont
  {C.~H.}\ \bibnamefont {Wang}}, \bibinfo {author} {\bibfnamefont {X.~H.}\
  \bibnamefont {Chen}}, \bibinfo {author} {\bibfnamefont {R.}~\bibnamefont
  {Liang}}, \bibinfo {author} {\bibfnamefont {W.~N.}\ \bibnamefont {Hardy}},
  \bibinfo {author} {\bibfnamefont {D.~A.}\ \bibnamefont {Bonn}},\ and\
  \bibinfo {author} {\bibfnamefont {L.}~\bibnamefont {Taillefer}},\ }\bibfield
  {title} {\bibinfo {title} {{Thermal Hall conductivity in the cuprate Mott
  insulators $\text{Nd}_2\text{CuO}_4$ and
  $\text{Sr}_2\text{CuO}_2\text{Cl}_2$}},\ }\href
  {https://doi.org/10.1038/s41467-020-18881-z} {\bibfield  {journal} {\bibinfo
  {journal} {Nature Communications}\ }\textbf {\bibinfo {volume} {11}},\
  \bibinfo {pages} {5325} (\bibinfo {year} {2020})},\ \Eprint
  {https://arxiv.org/abs/2007.05088} {2007.05088} \BibitemShut {NoStop}%
\bibitem [{\citenamefont {Grissonnanche}\ \emph {et~al.}(2019)\citenamefont
  {Grissonnanche}, \citenamefont {Legros}, \citenamefont {Badoux},
  \citenamefont {Lefrançois}, \citenamefont {Zatko}, \citenamefont {Lizaire},
  \citenamefont {Laliberté}, \citenamefont {Gourgout}, \citenamefont {Zhou},
  \citenamefont {Pyon}, \citenamefont {Takayama}, \citenamefont {Takagi},
  \citenamefont {Ono}, \citenamefont {Doiron-Leyraud},\ and\ \citenamefont
  {Taillefer}}]{Taillefer.Grissonnanche.2019}%
  \BibitemOpen
  \bibfield  {author} {\bibinfo {author} {\bibfnamefont {G.}~\bibnamefont
  {Grissonnanche}}, \bibinfo {author} {\bibfnamefont {A.}~\bibnamefont
  {Legros}}, \bibinfo {author} {\bibfnamefont {S.}~\bibnamefont {Badoux}},
  \bibinfo {author} {\bibfnamefont {E.}~\bibnamefont {Lefrançois}}, \bibinfo
  {author} {\bibfnamefont {V.}~\bibnamefont {Zatko}}, \bibinfo {author}
  {\bibfnamefont {M.}~\bibnamefont {Lizaire}}, \bibinfo {author} {\bibfnamefont
  {F.}~\bibnamefont {Laliberté}}, \bibinfo {author} {\bibfnamefont
  {A.}~\bibnamefont {Gourgout}}, \bibinfo {author} {\bibfnamefont {J.-S.}\
  \bibnamefont {Zhou}}, \bibinfo {author} {\bibfnamefont {S.}~\bibnamefont
  {Pyon}}, \bibinfo {author} {\bibfnamefont {T.}~\bibnamefont {Takayama}},
  \bibinfo {author} {\bibfnamefont {H.}~\bibnamefont {Takagi}}, \bibinfo
  {author} {\bibfnamefont {S.}~\bibnamefont {Ono}}, \bibinfo {author}
  {\bibfnamefont {N.}~\bibnamefont {Doiron-Leyraud}},\ and\ \bibinfo {author}
  {\bibfnamefont {L.}~\bibnamefont {Taillefer}},\ }\bibfield  {title} {\bibinfo
  {title} {{Giant thermal Hall conductivity in the pseudogap phase of cuprate
  superconductors}},\ }\href {https://doi.org/10.1038/s41586-019-1375-0}
  {\bibfield  {journal} {\bibinfo  {journal} {Nature}\ }\textbf {\bibinfo
  {volume} {571}},\ \bibinfo {pages} {376} (\bibinfo {year} {2019})},\ \Eprint
  {https://arxiv.org/abs/1901.03104} {1901.03104} \BibitemShut {NoStop}%
\bibitem [{\citenamefont {Katsura}\ \emph {et~al.}(2010)\citenamefont
  {Katsura}, \citenamefont {Nagaosa},\ and\ \citenamefont
  {Lee}}]{Lee.Katsura.2010}%
  \BibitemOpen
  \bibfield  {author} {\bibinfo {author} {\bibfnamefont {H.}~\bibnamefont
  {Katsura}}, \bibinfo {author} {\bibfnamefont {N.}~\bibnamefont {Nagaosa}},\
  and\ \bibinfo {author} {\bibfnamefont {P.~A.}\ \bibnamefont {Lee}},\
  }\bibfield  {title} {\bibinfo {title} {{Theory of the Thermal Hall Effect in
  Quantum Magnets}},\ }\href {https://doi.org/10.1103/physrevlett.104.066403}
  {\bibfield  {journal} {\bibinfo  {journal} {Physical Review Letters}\
  }\textbf {\bibinfo {volume} {104}},\ \bibinfo {pages} {066403} (\bibinfo
  {year} {2010})},\ \Eprint {https://arxiv.org/abs/0904.3427} {0904.3427}
  \BibitemShut {NoStop}%
\bibitem [{\citenamefont {Han}\ \emph {et~al.}(2019)\citenamefont {Han},
  \citenamefont {Park},\ and\ \citenamefont {Lee}}]{Lee.Han.2019}%
  \BibitemOpen
  \bibfield  {author} {\bibinfo {author} {\bibfnamefont {J.~H.}\ \bibnamefont
  {Han}}, \bibinfo {author} {\bibfnamefont {J.-H.}\ \bibnamefont {Park}},\ and\
  \bibinfo {author} {\bibfnamefont {P.~A.}\ \bibnamefont {Lee}},\ }\bibfield
  {title} {\bibinfo {title} {{Consideration of thermal Hall effect in undoped
  cuprates}},\ }\href {https://doi.org/10.1103/physrevb.99.205157} {\bibfield
  {journal} {\bibinfo  {journal} {Physical Review B}\ }\textbf {\bibinfo
  {volume} {99}},\ \bibinfo {pages} {205157} (\bibinfo {year} {2019})},\
  \Eprint {https://arxiv.org/abs/1903.01125} {1903.01125} \BibitemShut
  {NoStop}%
\bibitem [{\citenamefont {Guo}\ \emph {et~al.}(2020)\citenamefont {Guo},
  \citenamefont {Samajdar}, \citenamefont {Scheurer},\ and\ \citenamefont
  {Sachdev}}]{Sachdev.Guo.2020}%
  \BibitemOpen
  \bibfield  {author} {\bibinfo {author} {\bibfnamefont {H.}~\bibnamefont
  {Guo}}, \bibinfo {author} {\bibfnamefont {R.}~\bibnamefont {Samajdar}},
  \bibinfo {author} {\bibfnamefont {M.~S.}\ \bibnamefont {Scheurer}},\ and\
  \bibinfo {author} {\bibfnamefont {S.}~\bibnamefont {Sachdev}},\ }\bibfield
  {title} {\bibinfo {title} {{Gauge theories for the thermal Hall effect}},\
  }\href {https://doi.org/10.1103/physrevb.101.195126} {\bibfield  {journal}
  {\bibinfo  {journal} {Physical Review B}\ }\textbf {\bibinfo {volume}
  {101}},\ \bibinfo {pages} {195126} (\bibinfo {year} {2020})},\ \Eprint
  {https://arxiv.org/abs/2002.01947} {2002.01947} \BibitemShut {NoStop}%
\bibitem [{\citenamefont {Samajdar}\ \emph {et~al.}(2019)\citenamefont
  {Samajdar}, \citenamefont {Scheurer}, \citenamefont {Chatterjee},
  \citenamefont {Guo}, \citenamefont {Xu},\ and\ \citenamefont
  {Sachdev}}]{Sachdev.Samajdar.2019}%
  \BibitemOpen
  \bibfield  {author} {\bibinfo {author} {\bibfnamefont {R.}~\bibnamefont
  {Samajdar}}, \bibinfo {author} {\bibfnamefont {M.~S.}\ \bibnamefont
  {Scheurer}}, \bibinfo {author} {\bibfnamefont {S.}~\bibnamefont
  {Chatterjee}}, \bibinfo {author} {\bibfnamefont {H.}~\bibnamefont {Guo}},
  \bibinfo {author} {\bibfnamefont {C.}~\bibnamefont {Xu}},\ and\ \bibinfo
  {author} {\bibfnamefont {S.}~\bibnamefont {Sachdev}},\ }\bibfield  {title}
  {\bibinfo {title} {{Enhanced thermal Hall effect in the square-lattice Néel
  state}},\ }\href {https://doi.org/10.1038/s41567-019-0669-3} {\bibfield
  {journal} {\bibinfo  {journal} {Nature Physics}\ }\textbf {\bibinfo {volume}
  {15}},\ \bibinfo {pages} {1290} (\bibinfo {year} {2019})},\ \Eprint
  {https://arxiv.org/abs/1903.01992} {1903.01992} \BibitemShut {NoStop}%
\bibitem [{\citenamefont {Zhang}\ \emph
  {et~al.}(2023{\natexlab{a}})\citenamefont {Zhang}, \citenamefont {Gao},\ and\
  \citenamefont {Chen}}]{zhang_thermal_2023}%
  \BibitemOpen
  \bibfield  {author} {\bibinfo {author} {\bibfnamefont {X.-T.}\ \bibnamefont
  {Zhang}}, \bibinfo {author} {\bibfnamefont {Y.~H.}\ \bibnamefont {Gao}},\
  and\ \bibinfo {author} {\bibfnamefont {G.}~\bibnamefont {Chen}},\ }\href@noop
  {} {} (\bibinfo {year} {2023}{\natexlab{a}}),\ \Eprint
  {https://arxiv.org/abs/2305.04830} {arXiv:2305.04830} \BibitemShut {NoStop}%
\bibitem [{\citenamefont {Guo}\ and\ \citenamefont
  {Sachdev}(2021)}]{Sachdev.Guo.2021}%
  \BibitemOpen
  \bibfield  {author} {\bibinfo {author} {\bibfnamefont {H.}~\bibnamefont
  {Guo}}\ and\ \bibinfo {author} {\bibfnamefont {S.}~\bibnamefont {Sachdev}},\
  }\bibfield  {title} {\bibinfo {title} {{Extrinsic phonon thermal Hall
  transport from Hall viscosity}},\ }\href
  {https://doi.org/10.1103/physrevb.103.205115} {\bibfield  {journal} {\bibinfo
   {journal} {Physical Review B}\ }\textbf {\bibinfo {volume} {103}},\ \bibinfo
  {pages} {205115} (\bibinfo {year} {2021})},\ \Eprint
  {https://arxiv.org/abs/2103.02614} {2103.02614} \BibitemShut {NoStop}%
\bibitem [{\citenamefont {Guo}\ \emph {et~al.}(2022)\citenamefont {Guo},
  \citenamefont {Joshi},\ and\ \citenamefont {Sachdev}}]{Sachdev.Guo.2022}%
  \BibitemOpen
  \bibfield  {author} {\bibinfo {author} {\bibfnamefont {H.}~\bibnamefont
  {Guo}}, \bibinfo {author} {\bibfnamefont {D.~G.}\ \bibnamefont {Joshi}},\
  and\ \bibinfo {author} {\bibfnamefont {S.}~\bibnamefont {Sachdev}},\
  }\bibfield  {title} {\bibinfo {title} {{Resonant thermal Hall effect of
  phonons coupled to dynamical defects}},\ }\href
  {https://doi.org/10.1073/pnas.2215141119} {\bibfield  {journal} {\bibinfo
  {journal} {Proceedings of the National Academy of Sciences}\ }\textbf
  {\bibinfo {volume} {119}},\ \bibinfo {pages} {e2215141119} (\bibinfo {year}
  {2022})},\ \Eprint
  {https://arxiv.org/abs/https://www.pnas.org/doi/pdf/10.1073/pnas.2215141119}
  {https://www.pnas.org/doi/pdf/10.1073/pnas.2215141119} \BibitemShut {NoStop}%
\bibitem [{\citenamefont {Chen}\ \emph {et~al.}(2020)\citenamefont {Chen},
  \citenamefont {Kivelson},\ and\ \citenamefont {Sun}}]{Sun.Chen.2020}%
  \BibitemOpen
  \bibfield  {author} {\bibinfo {author} {\bibfnamefont {J.-Y.}\ \bibnamefont
  {Chen}}, \bibinfo {author} {\bibfnamefont {S.~A.}\ \bibnamefont {Kivelson}},\
  and\ \bibinfo {author} {\bibfnamefont {X.-Q.}\ \bibnamefont {Sun}},\
  }\bibfield  {title} {\bibinfo {title} {{Enhanced Thermal Hall Effect in
  Nearly Ferroelectric Insulators}},\ }\href
  {https://doi.org/10.1103/physrevlett.124.167601} {\bibfield  {journal}
  {\bibinfo  {journal} {Physical Review Letters}\ }\textbf {\bibinfo {volume}
  {124}},\ \bibinfo {pages} {167601} (\bibinfo {year} {2020})},\ \Eprint
  {https://arxiv.org/abs/1910.00018} {1910.00018} \BibitemShut {NoStop}%
\bibitem [{\citenamefont {Yang}\ \emph {et~al.}(2020)\citenamefont {Yang},
  \citenamefont {Zhang},\ and\ \citenamefont {Zhang}}]{Zhang.Yang.2020}%
  \BibitemOpen
  \bibfield  {author} {\bibinfo {author} {\bibfnamefont {Y.-f.}\ \bibnamefont
  {Yang}}, \bibinfo {author} {\bibfnamefont {G.-M.}\ \bibnamefont {Zhang}},\
  and\ \bibinfo {author} {\bibfnamefont {F.-C.}\ \bibnamefont {Zhang}},\
  }\bibfield  {title} {\bibinfo {title} {{Universal Behavior of the Thermal
  Hall Conductivity}},\ }\href {https://doi.org/10.1103/physrevlett.124.186602}
  {\bibfield  {journal} {\bibinfo  {journal} {Physical Review Letters}\
  }\textbf {\bibinfo {volume} {124}},\ \bibinfo {pages} {186602} (\bibinfo
  {year} {2020})},\ \Eprint {https://arxiv.org/abs/2001.08385} {2001.08385}
  \BibitemShut {NoStop}%
\bibitem [{\citenamefont {Hirschberger}\ \emph {et~al.}(2015)\citenamefont
  {Hirschberger}, \citenamefont {Krizan}, \citenamefont {Cava},\ and\
  \citenamefont {Ong}}]{hirschberger_large_2015}%
  \BibitemOpen
  \bibfield  {author} {\bibinfo {author} {\bibfnamefont {M.}~\bibnamefont
  {Hirschberger}}, \bibinfo {author} {\bibfnamefont {J.~W.}\ \bibnamefont
  {Krizan}}, \bibinfo {author} {\bibfnamefont {R.~J.}\ \bibnamefont {Cava}},\
  and\ \bibinfo {author} {\bibfnamefont {N.~P.}\ \bibnamefont {Ong}},\
  }\bibfield  {title} {\bibinfo {title} {{Large} thermal hall conductivity of
  neutral spin excitations in a frustrated quantum magnet},\ }\href
  {https://doi.org/10.1126/science.1257340} {\bibfield  {journal} {\bibinfo
  {journal} {Science}\ }\textbf {\bibinfo {volume} {348}},\ \bibinfo {pages}
  {106} (\bibinfo {year} {2015})}\BibitemShut {NoStop}%
\bibitem [{\citenamefont {Kasahara}\ \emph {et~al.}(2018)\citenamefont
  {Kasahara}, \citenamefont {Sugii}, \citenamefont {Ohnishi}, \citenamefont
  {Shimozawa}, \citenamefont {Yamashita}, \citenamefont {Kurita}, \citenamefont
  {Tanaka}, \citenamefont {Nasu}, \citenamefont {Motome}, \citenamefont
  {Shibauchi},\ and\ \citenamefont {Matsuda}}]{kasahara_unusual_2018}%
  \BibitemOpen
  \bibfield  {author} {\bibinfo {author} {\bibfnamefont {Y.}~\bibnamefont
  {Kasahara}}, \bibinfo {author} {\bibfnamefont {K.}~\bibnamefont {Sugii}},
  \bibinfo {author} {\bibfnamefont {T.}~\bibnamefont {Ohnishi}}, \bibinfo
  {author} {\bibfnamefont {M.}~\bibnamefont {Shimozawa}}, \bibinfo {author}
  {\bibfnamefont {M.}~\bibnamefont {Yamashita}}, \bibinfo {author}
  {\bibfnamefont {N.}~\bibnamefont {Kurita}}, \bibinfo {author} {\bibfnamefont
  {H.}~\bibnamefont {Tanaka}}, \bibinfo {author} {\bibfnamefont
  {J.}~\bibnamefont {Nasu}}, \bibinfo {author} {\bibfnamefont {Y.}~\bibnamefont
  {Motome}}, \bibinfo {author} {\bibfnamefont {T.}~\bibnamefont {Shibauchi}},\
  and\ \bibinfo {author} {\bibfnamefont {Y.}~\bibnamefont {Matsuda}},\
  }\bibfield  {title} {\bibinfo {title} {Unusual {Thermal} {Hall} {Effect} in a
  {Kitaev} {Spin} {Liquid} {Candidate} $\alpha-\text{RuCl}_3$},\ }\href
  {https://doi.org/10.1103/PhysRevLett.120.217205} {\bibfield  {journal}
  {\bibinfo  {journal} {Physical Review Letters}\ }\textbf {\bibinfo {volume}
  {120}},\ \bibinfo {pages} {217205} (\bibinfo {year} {2018})}\BibitemShut
  {NoStop}%
\bibitem [{\citenamefont {Watanabe}\ \emph {et~al.}(2016)\citenamefont
  {Watanabe}, \citenamefont {Sugii}, \citenamefont {Shimozawa}, \citenamefont
  {Suzuki}, \citenamefont {Yajima}, \citenamefont {Ishikawa}, \citenamefont
  {Hiroi}, \citenamefont {Shibauchi}, \citenamefont {Matsuda},\ and\
  \citenamefont {Yamashita}}]{watanabe_emergence_2016}%
  \BibitemOpen
  \bibfield  {author} {\bibinfo {author} {\bibfnamefont {D.}~\bibnamefont
  {Watanabe}}, \bibinfo {author} {\bibfnamefont {K.}~\bibnamefont {Sugii}},
  \bibinfo {author} {\bibfnamefont {M.}~\bibnamefont {Shimozawa}}, \bibinfo
  {author} {\bibfnamefont {Y.}~\bibnamefont {Suzuki}}, \bibinfo {author}
  {\bibfnamefont {T.}~\bibnamefont {Yajima}}, \bibinfo {author} {\bibfnamefont
  {H.}~\bibnamefont {Ishikawa}}, \bibinfo {author} {\bibfnamefont
  {Z.}~\bibnamefont {Hiroi}}, \bibinfo {author} {\bibfnamefont
  {T.}~\bibnamefont {Shibauchi}}, \bibinfo {author} {\bibfnamefont
  {Y.}~\bibnamefont {Matsuda}},\ and\ \bibinfo {author} {\bibfnamefont
  {M.}~\bibnamefont {Yamashita}},\ }\bibfield  {title} {\bibinfo {title}
  {Emergence of nontrivial magnetic excitations in a spin-liquid state of
  kagomé volborthite},\ }\href {https://doi.org/10.1073/pnas.1524076113}
  {\bibfield  {journal} {\bibinfo  {journal} {Proceedings of the National
  Academy of Sciences}\ }\textbf {\bibinfo {volume} {113}},\ \bibinfo {pages}
  {8653} (\bibinfo {year} {2016})}\BibitemShut {NoStop}%
\bibitem [{\citenamefont {Jezouin}\ \emph {et~al.}(2013)\citenamefont
  {Jezouin}, \citenamefont {Parmentier}, \citenamefont {Anthore}, \citenamefont
  {Gennser}, \citenamefont {Cavanna}, \citenamefont {Jin},\ and\ \citenamefont
  {Pierre}}]{jezouin_quantum_2013}%
  \BibitemOpen
  \bibfield  {author} {\bibinfo {author} {\bibfnamefont {S.}~\bibnamefont
  {Jezouin}}, \bibinfo {author} {\bibfnamefont {F.~D.}\ \bibnamefont
  {Parmentier}}, \bibinfo {author} {\bibfnamefont {A.}~\bibnamefont {Anthore}},
  \bibinfo {author} {\bibfnamefont {U.}~\bibnamefont {Gennser}}, \bibinfo
  {author} {\bibfnamefont {A.}~\bibnamefont {Cavanna}}, \bibinfo {author}
  {\bibfnamefont {Y.}~\bibnamefont {Jin}},\ and\ \bibinfo {author}
  {\bibfnamefont {F.}~\bibnamefont {Pierre}},\ }\bibfield  {title} {\bibinfo
  {title} {Quantum {Limit} of {Heat} {Flow} {Across} a {Single} {Electronic}
  {Channel}},\ }\href {https://doi.org/10.1126/science.1241912} {\bibfield
  {journal} {\bibinfo  {journal} {Science}\ }\textbf {\bibinfo {volume}
  {342}},\ \bibinfo {pages} {601} (\bibinfo {year} {2013})}\BibitemShut
  {NoStop}%
\bibitem [{\citenamefont {Banerjee}\ \emph {et~al.}(2017)\citenamefont
  {Banerjee}, \citenamefont {Heiblum}, \citenamefont {Rosenblatt},
  \citenamefont {Oreg}, \citenamefont {Feldman}, \citenamefont {Stern},\ and\
  \citenamefont {Umansky}}]{banerjee_observed_2017}%
  \BibitemOpen
  \bibfield  {author} {\bibinfo {author} {\bibfnamefont {M.}~\bibnamefont
  {Banerjee}}, \bibinfo {author} {\bibfnamefont {M.}~\bibnamefont {Heiblum}},
  \bibinfo {author} {\bibfnamefont {A.}~\bibnamefont {Rosenblatt}}, \bibinfo
  {author} {\bibfnamefont {Y.}~\bibnamefont {Oreg}}, \bibinfo {author}
  {\bibfnamefont {D.~E.}\ \bibnamefont {Feldman}}, \bibinfo {author}
  {\bibfnamefont {A.}~\bibnamefont {Stern}},\ and\ \bibinfo {author}
  {\bibfnamefont {V.}~\bibnamefont {Umansky}},\ }\bibfield  {title} {\bibinfo
  {title} {Observed quantization of anyonic heat flow},\ }\href
  {https://doi.org/10.1038/nature22052} {\bibfield  {journal} {\bibinfo
  {journal} {Nature}\ }\textbf {\bibinfo {volume} {545}},\ \bibinfo {pages}
  {75} (\bibinfo {year} {2017})}\BibitemShut {NoStop}%
\bibitem [{\citenamefont {Banerjee}\ \emph {et~al.}(2018)\citenamefont
  {Banerjee}, \citenamefont {Heiblum}, \citenamefont {Umansky}, \citenamefont
  {Feldman}, \citenamefont {Oreg},\ and\ \citenamefont
  {Stern}}]{banerjee_observation_2018}%
  \BibitemOpen
  \bibfield  {author} {\bibinfo {author} {\bibfnamefont {M.}~\bibnamefont
  {Banerjee}}, \bibinfo {author} {\bibfnamefont {M.}~\bibnamefont {Heiblum}},
  \bibinfo {author} {\bibfnamefont {V.}~\bibnamefont {Umansky}}, \bibinfo
  {author} {\bibfnamefont {D.~E.}\ \bibnamefont {Feldman}}, \bibinfo {author}
  {\bibfnamefont {Y.}~\bibnamefont {Oreg}},\ and\ \bibinfo {author}
  {\bibfnamefont {A.}~\bibnamefont {Stern}},\ }\bibfield  {title} {\bibinfo
  {title} {Observation of half-integer thermal {Hall} conductance},\ }\href
  {https://doi.org/10.1038/s41586-018-0184-1} {\bibfield  {journal} {\bibinfo
  {journal} {Nature}\ }\textbf {\bibinfo {volume} {559}},\ \bibinfo {pages}
  {205} (\bibinfo {year} {2018})}\BibitemShut {NoStop}%
\bibitem [{\citenamefont {Kane}\ and\ \citenamefont
  {Fisher}(1997)}]{kane_quantized_1997}%
  \BibitemOpen
  \bibfield  {author} {\bibinfo {author} {\bibfnamefont {C.~L.}\ \bibnamefont
  {Kane}}\ and\ \bibinfo {author} {\bibfnamefont {M.~P.~A.}\ \bibnamefont
  {Fisher}},\ }\bibfield  {title} {\bibinfo {title} {Quantized thermal
  transport in the fractional quantum {Hall} effect},\ }\href
  {https://doi.org/10.1103/PhysRevB.55.15832} {\bibfield  {journal} {\bibinfo
  {journal} {Physical Review B}\ }\textbf {\bibinfo {volume} {55}},\ \bibinfo
  {pages} {15832} (\bibinfo {year} {1997})}\BibitemShut {NoStop}%
\bibitem [{\citenamefont {Weng}(2011)}]{Weng.Weng.2011}%
  \BibitemOpen
  \bibfield  {author} {\bibinfo {author} {\bibfnamefont {Z.-Y.}\ \bibnamefont
  {Weng}},\ }\bibfield  {title} {\bibinfo {title} {{Superconducting ground
  state of a doped Mott insulator}},\ }\href
  {https://doi.org/10.1088/1367-2630/13/10/103039} {\bibfield  {journal}
  {\bibinfo  {journal} {New J. Phys.}\ }\textbf {\bibinfo {volume} {13}},\
  \bibinfo {pages} {103039 } (\bibinfo {year} {2011})}\BibitemShut {NoStop}%
\bibitem [{\citenamefont {Ma}\ \emph {et~al.}(2014)\citenamefont {Ma},
  \citenamefont {Ye},\ and\ \citenamefont {Weng}}]{Weng.Ma.2014}%
  \BibitemOpen
  \bibfield  {author} {\bibinfo {author} {\bibfnamefont {Y.}~\bibnamefont
  {Ma}}, \bibinfo {author} {\bibfnamefont {P.}~\bibnamefont {Ye}},\ and\
  \bibinfo {author} {\bibfnamefont {Z.-Y.}\ \bibnamefont {Weng}},\ }\bibfield
  {title} {\bibinfo {title} {{Low-temperature pseudogap phenomenon: precursor
  of high-$T_c$ superconductivity}},\ }\href
  {https://doi.org/10.1088/1367-2630/16/8/083039} {\bibfield  {journal}
  {\bibinfo  {journal} {New J. Phys.}\ }\textbf {\bibinfo {volume} {16}},\
  \bibinfo {pages} {083039} (\bibinfo {year} {2014})}\BibitemShut {NoStop}%
\bibitem [{\citenamefont {Baskaran}\ \emph {et~al.}(1987)\citenamefont
  {Baskaran}, \citenamefont {Zou},\ and\ \citenamefont
  {Anderson}}]{Anderson.Baskaran.1987}%
  \BibitemOpen
  \bibfield  {author} {\bibinfo {author} {\bibfnamefont {G.}~\bibnamefont
  {Baskaran}}, \bibinfo {author} {\bibfnamefont {Z.}~\bibnamefont {Zou}},\ and\
  \bibinfo {author} {\bibfnamefont {P.}~\bibnamefont {Anderson}},\ }\bibfield
  {title} {\bibinfo {title} {{The resonating valence bond state and high-$T_c$
  superconductivity — A mean field theory}},\ }\href
  {https://doi.org/10.1016/0038-1098(87)90642-9} {\bibfield  {journal}
  {\bibinfo  {journal} {Solid State Commun.}\ }\textbf {\bibinfo {volume}
  {63}},\ \bibinfo {pages} {973} (\bibinfo {year} {1987})}\BibitemShut
  {NoStop}%
\bibitem [{\citenamefont {Anderson}(1987)}]{Anderson.Anderson.1987}%
  \BibitemOpen
  \bibfield  {author} {\bibinfo {author} {\bibfnamefont {P.~W.}\ \bibnamefont
  {Anderson}},\ }\bibfield  {title} {\bibinfo {title} {{The Resonating Valence
  Bond State in $\mathrm{La_2CuO_4}$ and Superconductivity}},\ }\href
  {https://doi.org/10.1126/science.235.4793.1196} {\bibfield  {journal}
  {\bibinfo  {journal} {Science}\ }\textbf {\bibinfo {volume} {235}},\ \bibinfo
  {pages} {1196} (\bibinfo {year} {1987})}\BibitemShut {NoStop}%
\bibitem [{\citenamefont {Arovas}\ and\ \citenamefont
  {Auerbach}(1988)}]{PhysRevB.38.316}%
  \BibitemOpen
  \bibfield  {author} {\bibinfo {author} {\bibfnamefont {D.~P.}\ \bibnamefont
  {Arovas}}\ and\ \bibinfo {author} {\bibfnamefont {A.}~\bibnamefont
  {Auerbach}},\ }\bibfield  {title} {\bibinfo {title} {Functional integral
  theories of low-dimensional quantum heisenberg models},\ }\href
  {https://doi.org/10.1103/PhysRevB.38.316} {\bibfield  {journal} {\bibinfo
  {journal} {Phys. Rev. B}\ }\textbf {\bibinfo {volume} {38}},\ \bibinfo
  {pages} {316} (\bibinfo {year} {1988})}\BibitemShut {NoStop}%
\bibitem [{\citenamefont {Read}\ and\ \citenamefont
  {Sachdev}(1991)}]{Read.Sachdev.1991}%
  \BibitemOpen
  \bibfield  {author} {\bibinfo {author} {\bibfnamefont {N.}~\bibnamefont
  {Read}}\ and\ \bibinfo {author} {\bibfnamefont {S.}~\bibnamefont {Sachdev}},\
  }\bibfield  {title} {\bibinfo {title} {Large-n expansion for frustrated
  quantum antiferromagnets},\ }\href
  {https://doi.org/10.1103/PhysRevLett.66.1773} {\bibfield  {journal} {\bibinfo
   {journal} {Phys. Rev. Lett.}\ }\textbf {\bibinfo {volume} {66}},\ \bibinfo
  {pages} {1773} (\bibinfo {year} {1991})}\BibitemShut {NoStop}%
\bibitem [{\citenamefont {Sheng}\ \emph {et~al.}(1996)\citenamefont {Sheng},
  \citenamefont {Chen},\ and\ \citenamefont {Weng}}]{Weng.Sheng.1996}%
  \BibitemOpen
  \bibfield  {author} {\bibinfo {author} {\bibfnamefont {D.~N.}\ \bibnamefont
  {Sheng}}, \bibinfo {author} {\bibfnamefont {Y.~C.}\ \bibnamefont {Chen}},\
  and\ \bibinfo {author} {\bibfnamefont {Z.~Y.}\ \bibnamefont {Weng}},\
  }\bibfield  {title} {\bibinfo {title} {{Phase String Effect in a Doped
  Antiferromagnet}},\ }\href {https://doi.org/10.1103/physrevlett.77.5102}
  {\bibfield  {journal} {\bibinfo  {journal} {Phys. Rev. Lett.}\ }\textbf
  {\bibinfo {volume} {77}},\ \bibinfo {pages} {5102} (\bibinfo {year}
  {1996})}\BibitemShut {NoStop}%
\bibitem [{\citenamefont {Wu}\ \emph {et~al.}(2008)\citenamefont {Wu},
  \citenamefont {Weng},\ and\ \citenamefont {Zaanen}}]{Zaanen.Wu.2008}%
  \BibitemOpen
  \bibfield  {author} {\bibinfo {author} {\bibfnamefont {K.}~\bibnamefont
  {Wu}}, \bibinfo {author} {\bibfnamefont {Z.~Y.}\ \bibnamefont {Weng}},\ and\
  \bibinfo {author} {\bibfnamefont {J.}~\bibnamefont {Zaanen}},\ }\bibfield
  {title} {\bibinfo {title} {{Sign structure of the $t$-$J$ model}},\ }\href
  {https://doi.org/10.1103/physrevb.77.155102} {\bibfield  {journal} {\bibinfo
  {journal} {Phys. Rev. B}\ }\textbf {\bibinfo {volume} {77}},\ \bibinfo
  {pages} {155102} (\bibinfo {year} {2008})}\BibitemShut {NoStop}%
\bibitem [{\citenamefont {Lu}\ \emph {et~al.}(2023)\citenamefont {Lu},
  \citenamefont {Zhang}, \citenamefont {Gong}, \citenamefont {Sheng},\ and\
  \citenamefont {Weng}}]{Lu.Weng_2023}%
  \BibitemOpen
  \bibfield  {author} {\bibinfo {author} {\bibfnamefont {X.}~\bibnamefont
  {Lu}}, \bibinfo {author} {\bibfnamefont {J.-X.}\ \bibnamefont {Zhang}},
  \bibinfo {author} {\bibfnamefont {S.-S.}\ \bibnamefont {Gong}}, \bibinfo
  {author} {\bibfnamefont {D.~N.}\ \bibnamefont {Sheng}},\ and\ \bibinfo
  {author} {\bibfnamefont {Z.-Y.}\ \bibnamefont {Weng}},\ }\href@noop {} {}
  (\bibinfo {year} {2023}),\ \Eprint {https://arxiv.org/abs/2303.13498}
  {arXiv:2303.13498} \BibitemShut {NoStop}%
\bibitem [{\citenamefont {Bernevig}\ \emph {et~al.}(2006)\citenamefont
  {Bernevig}, \citenamefont {Hughes},\ and\ \citenamefont
  {Zhang}}]{Zhang.Bernevig.2006}%
  \BibitemOpen
  \bibfield  {author} {\bibinfo {author} {\bibfnamefont {B.~A.}\ \bibnamefont
  {Bernevig}}, \bibinfo {author} {\bibfnamefont {T.~L.}\ \bibnamefont
  {Hughes}},\ and\ \bibinfo {author} {\bibfnamefont {S.-C.}\ \bibnamefont
  {Zhang}},\ }\bibfield  {title} {\bibinfo {title} {{Quantum Spin Hall Effect
  and Topological Phase Transition in HgTe Quantum Wells}},\ }\href
  {https://doi.org/10.1126/science.1133734} {\bibfield  {journal} {\bibinfo
  {journal} {Science}\ }\textbf {\bibinfo {volume} {314}},\ \bibinfo {pages}
  {1757} (\bibinfo {year} {2006})}\BibitemShut {NoStop}%
\bibitem [{\citenamefont {König}\ \emph {et~al.}(2007)\citenamefont {König},
  \citenamefont {Wiedmann}, \citenamefont {Brüne}, \citenamefont {Roth},
  \citenamefont {Buhmann}, \citenamefont {Molenkamp}, \citenamefont {Qi},\ and\
  \citenamefont {Zhang}}]{Zhang.K.2007}%
  \BibitemOpen
  \bibfield  {author} {\bibinfo {author} {\bibfnamefont {M.}~\bibnamefont
  {König}}, \bibinfo {author} {\bibfnamefont {S.}~\bibnamefont {Wiedmann}},
  \bibinfo {author} {\bibfnamefont {C.}~\bibnamefont {Brüne}}, \bibinfo
  {author} {\bibfnamefont {A.}~\bibnamefont {Roth}}, \bibinfo {author}
  {\bibfnamefont {H.}~\bibnamefont {Buhmann}}, \bibinfo {author} {\bibfnamefont
  {L.~W.}\ \bibnamefont {Molenkamp}}, \bibinfo {author} {\bibfnamefont {X.-L.}\
  \bibnamefont {Qi}},\ and\ \bibinfo {author} {\bibfnamefont {S.-C.}\
  \bibnamefont {Zhang}},\ }\bibfield  {title} {\bibinfo {title} {{Quantum Spin
  Hall Insulator State in HgTe Quantum Wells}},\ }\href
  {https://doi.org/10.1126/science.1148047} {\bibfield  {journal} {\bibinfo
  {journal} {Science}\ }\textbf {\bibinfo {volume} {318}},\ \bibinfo {pages}
  {766} (\bibinfo {year} {2007})},\ \Eprint {https://arxiv.org/abs/0710.0582}
  {0710.0582} \BibitemShut {NoStop}%
\bibitem [{\citenamefont {Daou}\ \emph {et~al.}(2009)\citenamefont {Daou},
  \citenamefont {Cyr-Choinière}, \citenamefont {Laliberté}, \citenamefont
  {LeBoeuf}, \citenamefont {Doiron-Leyraud}, \citenamefont {Yan}, \citenamefont
  {Zhou}, \citenamefont {Goodenough},\ and\ \citenamefont
  {Taillefer}}]{Taillefer.Daou.2009}%
  \BibitemOpen
  \bibfield  {author} {\bibinfo {author} {\bibfnamefont {R.}~\bibnamefont
  {Daou}}, \bibinfo {author} {\bibfnamefont {O.}~\bibnamefont
  {Cyr-Choinière}}, \bibinfo {author} {\bibfnamefont {F.}~\bibnamefont
  {Laliberté}}, \bibinfo {author} {\bibfnamefont {D.}~\bibnamefont {LeBoeuf}},
  \bibinfo {author} {\bibfnamefont {N.}~\bibnamefont {Doiron-Leyraud}},
  \bibinfo {author} {\bibfnamefont {J.-Q.}\ \bibnamefont {Yan}}, \bibinfo
  {author} {\bibfnamefont {J.-S.}\ \bibnamefont {Zhou}}, \bibinfo {author}
  {\bibfnamefont {J.~B.}\ \bibnamefont {Goodenough}},\ and\ \bibinfo {author}
  {\bibfnamefont {L.}~\bibnamefont {Taillefer}},\ }\bibfield  {title} {\bibinfo
  {title} {{Thermopower across the stripe critical point of
  $\mathrm{La_{1.6-x}Nd_{0.4}Sr_xCuO_4}$: Evidence for a quantum critical point
  in a hole-doped high-$T_c$ superconductor}},\ }\href
  {https://doi.org/10.1103/physrevb.79.180505} {\bibfield  {journal} {\bibinfo
  {journal} {Phys. Rev. B}\ }\textbf {\bibinfo {volume} {79}},\ \bibinfo
  {pages} {180505} (\bibinfo {year} {2009})}\BibitemShut {NoStop}%
\bibitem [{\citenamefont {Lizaire}\ \emph {et~al.}(2021)\citenamefont
  {Lizaire}, \citenamefont {Legros}, \citenamefont {Gourgout}, \citenamefont
  {Benhabib}, \citenamefont {Badoux}, \citenamefont {Laliberté}, \citenamefont
  {Boulanger}, \citenamefont {Ataei}, \citenamefont {Grissonnanche},
  \citenamefont {LeBoeuf}, \citenamefont {Licciardello}, \citenamefont
  {Wiedmann}, \citenamefont {Ono}, \citenamefont {Raffy}, \citenamefont
  {Kawasaki}, \citenamefont {Zheng}, \citenamefont {Doiron-Leyraud},
  \citenamefont {Proust},\ and\ \citenamefont
  {Taillefer}}]{Taillefer.Lizaire.2021}%
  \BibitemOpen
  \bibfield  {author} {\bibinfo {author} {\bibfnamefont {M.}~\bibnamefont
  {Lizaire}}, \bibinfo {author} {\bibfnamefont {A.}~\bibnamefont {Legros}},
  \bibinfo {author} {\bibfnamefont {A.}~\bibnamefont {Gourgout}}, \bibinfo
  {author} {\bibfnamefont {S.}~\bibnamefont {Benhabib}}, \bibinfo {author}
  {\bibfnamefont {S.}~\bibnamefont {Badoux}}, \bibinfo {author} {\bibfnamefont
  {F.}~\bibnamefont {Laliberté}}, \bibinfo {author} {\bibfnamefont {M.-E.}\
  \bibnamefont {Boulanger}}, \bibinfo {author} {\bibfnamefont {A.}~\bibnamefont
  {Ataei}}, \bibinfo {author} {\bibfnamefont {G.}~\bibnamefont
  {Grissonnanche}}, \bibinfo {author} {\bibfnamefont {D.}~\bibnamefont
  {LeBoeuf}}, \bibinfo {author} {\bibfnamefont {S.}~\bibnamefont
  {Licciardello}}, \bibinfo {author} {\bibfnamefont {S.}~\bibnamefont
  {Wiedmann}}, \bibinfo {author} {\bibfnamefont {S.}~\bibnamefont {Ono}},
  \bibinfo {author} {\bibfnamefont {H.}~\bibnamefont {Raffy}}, \bibinfo
  {author} {\bibfnamefont {S.}~\bibnamefont {Kawasaki}}, \bibinfo {author}
  {\bibfnamefont {G.-Q.}\ \bibnamefont {Zheng}}, \bibinfo {author}
  {\bibfnamefont {N.}~\bibnamefont {Doiron-Leyraud}}, \bibinfo {author}
  {\bibfnamefont {C.}~\bibnamefont {Proust}},\ and\ \bibinfo {author}
  {\bibfnamefont {L.}~\bibnamefont {Taillefer}},\ }\bibfield  {title} {\bibinfo
  {title} {Transport signatures of the pseudogap critical point in the cuprate
  superconductor {$\text{Bi}_{2} \text{Sr}_{2-x} \text{La}_{x}
  \text{CuO}_{6+\delta}$}},\ }\href
  {https://doi.org/10.1103/physrevb.104.014515} {\bibfield  {journal} {\bibinfo
   {journal} {Phys. Rev. B}\ }\textbf {\bibinfo {volume} {104}},\ \bibinfo
  {pages} {014515} (\bibinfo {year} {2021})},\ \Eprint
  {https://arxiv.org/abs/2008.13692} {2008.13692} \BibitemShut {NoStop}%
\bibitem [{\citenamefont {Wang}\ \emph {et~al.}(2001)\citenamefont {Wang},
  \citenamefont {Xu}, \citenamefont {Kakeshita}, \citenamefont {Uchida},
  \citenamefont {Ono}, \citenamefont {Ando},\ and\ \citenamefont
  {Ong}}]{Ong.Wang.2001}%
  \BibitemOpen
  \bibfield  {author} {\bibinfo {author} {\bibfnamefont {Y.}~\bibnamefont
  {Wang}}, \bibinfo {author} {\bibfnamefont {Z.~A.}\ \bibnamefont {Xu}},
  \bibinfo {author} {\bibfnamefont {T.}~\bibnamefont {Kakeshita}}, \bibinfo
  {author} {\bibfnamefont {S.}~\bibnamefont {Uchida}}, \bibinfo {author}
  {\bibfnamefont {S.}~\bibnamefont {Ono}}, \bibinfo {author} {\bibfnamefont
  {Y.}~\bibnamefont {Ando}},\ and\ \bibinfo {author} {\bibfnamefont {N.~P.}\
  \bibnamefont {Ong}},\ }\bibfield  {title} {\bibinfo {title} {{Onset of the
  vortexlike Nernst signal above Tc in La2-xSrxCuO4 and Bi2Sr2-yLayCuO6}},\
  }\href {https://doi.org/10.1103/physrevb.64.224519} {\bibfield  {journal}
  {\bibinfo  {journal} {Physical Review B}\ }\textbf {\bibinfo {volume} {64}},\
  \bibinfo {pages} {224519} (\bibinfo {year} {2001})},\ \Eprint
  {https://arxiv.org/abs/cond-mat/0108242} {cond-mat/0108242} \BibitemShut
  {NoStop}%
\bibitem [{\citenamefont {Wang}\ \emph {et~al.}(2002)\citenamefont {Wang},
  \citenamefont {Ong}, \citenamefont {Xu}, \citenamefont {Kakeshita},
  \citenamefont {Uchida}, \citenamefont {Bonn}, \citenamefont {Liang},\ and\
  \citenamefont {Hardy}}]{Hardy.Wang.2002}%
  \BibitemOpen
  \bibfield  {author} {\bibinfo {author} {\bibfnamefont {Y.}~\bibnamefont
  {Wang}}, \bibinfo {author} {\bibfnamefont {N.~P.}\ \bibnamefont {Ong}},
  \bibinfo {author} {\bibfnamefont {Z.~A.}\ \bibnamefont {Xu}}, \bibinfo
  {author} {\bibfnamefont {T.}~\bibnamefont {Kakeshita}}, \bibinfo {author}
  {\bibfnamefont {S.}~\bibnamefont {Uchida}}, \bibinfo {author} {\bibfnamefont
  {D.~A.}\ \bibnamefont {Bonn}}, \bibinfo {author} {\bibfnamefont
  {R.}~\bibnamefont {Liang}},\ and\ \bibinfo {author} {\bibfnamefont {W.~N.}\
  \bibnamefont {Hardy}},\ }\bibfield  {title} {\bibinfo {title} {{High Field
  Phase Diagram of Cuprates Derived from the Nernst Effect}},\ }\href
  {https://doi.org/10.1103/physrevlett.88.257003} {\bibfield  {journal}
  {\bibinfo  {journal} {Physical Review Letters}\ }\textbf {\bibinfo {volume}
  {88}},\ \bibinfo {pages} {257003} (\bibinfo {year} {2002})},\ \Eprint
  {https://arxiv.org/abs/cond-mat/0205299} {cond-mat/0205299} \BibitemShut
  {NoStop}%
\bibitem [{\citenamefont {Wang}\ \emph {et~al.}(2003)\citenamefont {Wang},
  \citenamefont {Ono}, \citenamefont {Onose}, \citenamefont {Gu}, \citenamefont
  {Ando}, \citenamefont {Tokura}, \citenamefont {Uchida},\ and\ \citenamefont
  {Ong}}]{Ong.Wang.2003}%
  \BibitemOpen
  \bibfield  {author} {\bibinfo {author} {\bibfnamefont {Y.}~\bibnamefont
  {Wang}}, \bibinfo {author} {\bibfnamefont {S.}~\bibnamefont {Ono}}, \bibinfo
  {author} {\bibfnamefont {Y.}~\bibnamefont {Onose}}, \bibinfo {author}
  {\bibfnamefont {G.}~\bibnamefont {Gu}}, \bibinfo {author} {\bibfnamefont
  {Y.}~\bibnamefont {Ando}}, \bibinfo {author} {\bibfnamefont {Y.}~\bibnamefont
  {Tokura}}, \bibinfo {author} {\bibfnamefont {S.}~\bibnamefont {Uchida}},\
  and\ \bibinfo {author} {\bibfnamefont {N.~P.}\ \bibnamefont {Ong}},\
  }\bibfield  {title} {\bibinfo {title} {{Dependence of Upper Critical Field
  and Pairing Strength on Doping in Cuprates}},\ }\href
  {https://doi.org/10.1126/science.1078422} {\bibfield  {journal} {\bibinfo
  {journal} {Science}\ }\textbf {\bibinfo {volume} {299}},\ \bibinfo {pages}
  {86} (\bibinfo {year} {2003})}\BibitemShut {NoStop}%
\bibitem [{\citenamefont {Dai}\ \emph {et~al.}(1996)\citenamefont {Dai},
  \citenamefont {Yethiraj}, \citenamefont {Mook}, \citenamefont {Lindemer},\
  and\ \citenamefont {Dogan}}]{Dogan.Dai.1996}%
  \BibitemOpen
  \bibfield  {author} {\bibinfo {author} {\bibfnamefont {P.}~\bibnamefont
  {Dai}}, \bibinfo {author} {\bibfnamefont {M.}~\bibnamefont {Yethiraj}},
  \bibinfo {author} {\bibfnamefont {H.~A.}\ \bibnamefont {Mook}}, \bibinfo
  {author} {\bibfnamefont {T.~B.}\ \bibnamefont {Lindemer}},\ and\ \bibinfo
  {author} {\bibfnamefont {F.}~\bibnamefont {Dogan}},\ }\bibfield  {title}
  {\bibinfo {title} {{Magnetic Dynamics in Underdoped
  $\mathrm{YBa_2Cu_3O_{7-x}}$: Direct Observation of a Superconducting Gap}},\
  }\href {https://doi.org/10.1103/physrevlett.77.5425} {\bibfield  {journal}
  {\bibinfo  {journal} {Phys. Rev. Lett.}\ }\textbf {\bibinfo {volume} {77}},\
  \bibinfo {pages} {5425} (\bibinfo {year} {1996})}\BibitemShut {NoStop}%
\bibitem [{\citenamefont {Fong}\ \emph {et~al.}(1997)\citenamefont {Fong},
  \citenamefont {Keimer}, \citenamefont {Milius},\ and\ \citenamefont
  {Aksay}}]{Aksay.Fong.1997}%
  \BibitemOpen
  \bibfield  {author} {\bibinfo {author} {\bibfnamefont {H.~F.}\ \bibnamefont
  {Fong}}, \bibinfo {author} {\bibfnamefont {B.}~\bibnamefont {Keimer}},
  \bibinfo {author} {\bibfnamefont {D.~L.}\ \bibnamefont {Milius}},\ and\
  \bibinfo {author} {\bibfnamefont {I.~A.}\ \bibnamefont {Aksay}},\ }\bibfield
  {title} {\bibinfo {title} {{Superconductivity-Induced Anomalies in the Spin
  Excitation Spectra of Underdoped $\mathrm{YBa_2Cu_3O_{6+x}}$}},\ }\href
  {https://doi.org/10.1103/physrevlett.78.713} {\bibfield  {journal} {\bibinfo
  {journal} {Phys. Rev. Lett.}\ }\textbf {\bibinfo {volume} {78}},\ \bibinfo
  {pages} {713} (\bibinfo {year} {1997})}\BibitemShut {NoStop}%
\bibitem [{\citenamefont {He}\ \emph {et~al.}(2001)\citenamefont {He},
  \citenamefont {Sidis}, \citenamefont {Bourges}, \citenamefont {Gu},
  \citenamefont {Ivanov}, \citenamefont {Koshizuka}, \citenamefont {Liang},
  \citenamefont {Lin}, \citenamefont {Regnault}, \citenamefont {Schoenherr},\
  and\ \citenamefont {Keimer}}]{Keimer.He.20006wc}%
  \BibitemOpen
  \bibfield  {author} {\bibinfo {author} {\bibfnamefont {H.}~\bibnamefont
  {He}}, \bibinfo {author} {\bibfnamefont {Y.}~\bibnamefont {Sidis}}, \bibinfo
  {author} {\bibfnamefont {P.}~\bibnamefont {Bourges}}, \bibinfo {author}
  {\bibfnamefont {G.~D.}\ \bibnamefont {Gu}}, \bibinfo {author} {\bibfnamefont
  {A.}~\bibnamefont {Ivanov}}, \bibinfo {author} {\bibfnamefont
  {N.}~\bibnamefont {Koshizuka}}, \bibinfo {author} {\bibfnamefont
  {B.}~\bibnamefont {Liang}}, \bibinfo {author} {\bibfnamefont {C.~T.}\
  \bibnamefont {Lin}}, \bibinfo {author} {\bibfnamefont {L.~P.}\ \bibnamefont
  {Regnault}}, \bibinfo {author} {\bibfnamefont {E.}~\bibnamefont
  {Schoenherr}},\ and\ \bibinfo {author} {\bibfnamefont {B.}~\bibnamefont
  {Keimer}},\ }\bibfield  {title} {\bibinfo {title} {{Resonant Spin Excitation
  in an Overdoped High Temperature Superconductor}},\ }\href
  {https://doi.org/10.1103/physrevlett.86.1610} {\bibfield  {journal} {\bibinfo
   {journal} {Phys. Rev. Lett.}\ }\textbf {\bibinfo {volume} {86}},\ \bibinfo
  {pages} {1610} (\bibinfo {year} {2001})}\BibitemShut {NoStop}%
\bibitem [{\citenamefont {Gallais}\ \emph {et~al.}(2002)\citenamefont
  {Gallais}, \citenamefont {Sacuto}, \citenamefont {Bourges}, \citenamefont
  {Sidis}, \citenamefont {Forget},\ and\ \citenamefont
  {Colson}}]{Colson.Gallais.2002}%
  \BibitemOpen
  \bibfield  {author} {\bibinfo {author} {\bibfnamefont {Y.}~\bibnamefont
  {Gallais}}, \bibinfo {author} {\bibfnamefont {A.}~\bibnamefont {Sacuto}},
  \bibinfo {author} {\bibfnamefont {P.}~\bibnamefont {Bourges}}, \bibinfo
  {author} {\bibfnamefont {Y.}~\bibnamefont {Sidis}}, \bibinfo {author}
  {\bibfnamefont {A.}~\bibnamefont {Forget}},\ and\ \bibinfo {author}
  {\bibfnamefont {D.}~\bibnamefont {Colson}},\ }\bibfield  {title} {\bibinfo
  {title} {{Evidence for Two Distinct Energy Scales in the Raman Spectra of
  $\mathrm{YBa_2(Cu_{1-x}Ni_x)_3O_{6.95}}$}},\ }\href
  {https://doi.org/10.1103/physrevlett.88.177401} {\bibfield  {journal}
  {\bibinfo  {journal} {Phys. Rev. Lett.}\ }\textbf {\bibinfo {volume} {88}},\
  \bibinfo {pages} {177401} (\bibinfo {year} {2002})}\BibitemShut {NoStop}%
\bibitem [{\citenamefont {Laughlin}(2014)}]{Laughlin.2014}%
  \BibitemOpen
  \bibfield  {author} {\bibinfo {author} {\bibfnamefont {R.~B.}\ \bibnamefont
  {Laughlin}},\ }\bibfield  {title} {\bibinfo {title} {Hartree-{Fock}
  computation of the high- ${T}_c$ cuprate phase diagram},\ }\href
  {https://doi.org/10.1103/PhysRevB.89.035134} {\bibfield  {journal} {\bibinfo
  {journal} {Physical Review B}\ }\textbf {\bibinfo {volume} {89}},\ \bibinfo
  {pages} {035134} (\bibinfo {year} {2014})}\BibitemShut {NoStop}%
\bibitem [{\citenamefont {Zhang}\ and\ \citenamefont
  {Weng}(2014)}]{Weng.Zhang.2014}%
  \BibitemOpen
  \bibfield  {author} {\bibinfo {author} {\bibfnamefont {L.}~\bibnamefont
  {Zhang}}\ and\ \bibinfo {author} {\bibfnamefont {Z.-Y.}\ \bibnamefont
  {Weng}},\ }\bibfield  {title} {\bibinfo {title} {{Sign structure, electron
  fractionalization, and emergent gauge description of the Hubbard model}},\
  }\href {https://doi.org/10.1103/physrevb.90.165120} {\bibfield  {journal}
  {\bibinfo  {journal} {Phys. Rev. B}\ }\textbf {\bibinfo {volume} {90}},\
  \bibinfo {pages} {165120 } (\bibinfo {year} {2014})}\BibitemShut {NoStop}%
\bibitem [{\citenamefont {Xu}\ \emph {et~al.}(2023)\citenamefont {Xu},
  \citenamefont {Zhu}, \citenamefont {Wu},\ and\ \citenamefont
  {Weng}}]{Xu.Weng_2023}%
  \BibitemOpen
  \bibfield  {author} {\bibinfo {author} {\bibfnamefont {J.-S.}\ \bibnamefont
  {Xu}}, \bibinfo {author} {\bibfnamefont {Z.}~\bibnamefont {Zhu}}, \bibinfo
  {author} {\bibfnamefont {K.}~\bibnamefont {Wu}},\ and\ \bibinfo {author}
  {\bibfnamefont {Z.-Y.}\ \bibnamefont {Weng}},\ }\href@noop {} {} (\bibinfo
  {year} {2023}),\ \Eprint {https://arxiv.org/abs/2306.11096}
  {arXiv:2306.11096} \BibitemShut {NoStop}%
\bibitem [{\citenamefont {Kou}\ and\ \citenamefont
  {Weng}(2003)}]{Weng.Kou.2003yuc}%
  \BibitemOpen
  \bibfield  {author} {\bibinfo {author} {\bibfnamefont {S.-P.}\ \bibnamefont
  {Kou}}\ and\ \bibinfo {author} {\bibfnamefont {Z.-Y.}\ \bibnamefont {Weng}},\
  }\bibfield  {title} {\bibinfo {title} {Topological gauge structure and phase
  diagram for weakly doped antiferromagnets},\ }\href
  {https://doi.org/10.1103/PhysRevLett.90.157003} {\bibfield  {journal}
  {\bibinfo  {journal} {Phys. Rev. Lett.}\ }\textbf {\bibinfo {volume} {90}},\
  \bibinfo {pages} {157003} (\bibinfo {year} {2003})}\BibitemShut {NoStop}%
\bibitem [{\citenamefont {Kou}\ \emph {et~al.}(2005{\natexlab{a}})\citenamefont
  {Kou}, \citenamefont {Qi},\ and\ \citenamefont
  {Weng}}]{kou_mutual-chern-simons_2005}%
  \BibitemOpen
  \bibfield  {author} {\bibinfo {author} {\bibfnamefont {S.-P.}\ \bibnamefont
  {Kou}}, \bibinfo {author} {\bibfnamefont {X.-L.}\ \bibnamefont {Qi}},\ and\
  \bibinfo {author} {\bibfnamefont {Z.-Y.}\ \bibnamefont {Weng}},\ }\bibfield
  {title} {\bibinfo {title} {Mutual-{Chern}-{Simons} effective theory of doped
  antiferromagnets},\ }\href {https://doi.org/10.1103/PhysRevB.71.235102}
  {\bibfield  {journal} {\bibinfo  {journal} {Physical Review B}\ }\textbf
  {\bibinfo {volume} {71}},\ \bibinfo {pages} {235102} (\bibinfo {year}
  {2005}{\natexlab{a}})}\BibitemShut {NoStop}%
\bibitem [{\citenamefont {Weng}\ and\ \citenamefont {Qi}(2006)}]{Qi.Weng.2006}%
  \BibitemOpen
  \bibfield  {author} {\bibinfo {author} {\bibfnamefont {Z.-Y.}\ \bibnamefont
  {Weng}}\ and\ \bibinfo {author} {\bibfnamefont {X.-L.}\ \bibnamefont {Qi}},\
  }\bibfield  {title} {\bibinfo {title} {{Lower pseudogap phase of Mott
  insulators: A spin/vortex liquid state}},\ }\href
  {https://doi.org/10.1103/physrevb.74.144518} {\bibfield  {journal} {\bibinfo
  {journal} {Physical Review B}\ }\textbf {\bibinfo {volume} {74}},\ \bibinfo
  {pages} {144518} (\bibinfo {year} {2006})},\ \Eprint
  {https://arxiv.org/abs/cond-mat/0603097} {cond-mat/0603097} \BibitemShut
  {NoStop}%
\bibitem [{\citenamefont {Qi}\ and\ \citenamefont {Weng}(2007)}]{Weng.Qi.2007}%
  \BibitemOpen
  \bibfield  {author} {\bibinfo {author} {\bibfnamefont {X.-L.}\ \bibnamefont
  {Qi}}\ and\ \bibinfo {author} {\bibfnamefont {Z.-Y.}\ \bibnamefont {Weng}},\
  }\bibfield  {title} {\bibinfo {title} {{Mutual Chern-Simons gauge theory of
  spontaneous vortex phase}},\ }\href
  {https://doi.org/10.1103/physrevb.76.104502} {\bibfield  {journal} {\bibinfo
  {journal} {Phys. Rev. B}\ }\textbf {\bibinfo {volume} {76}},\ \bibinfo
  {pages} {104502} (\bibinfo {year} {2007})}\BibitemShut {NoStop}%
\bibitem [{\citenamefont {Ye}\ \emph {et~al.}(2011)\citenamefont {Ye},
  \citenamefont {Tian}, \citenamefont {Qi},\ and\ \citenamefont
  {Weng}}]{Weng.Ye.2011}%
  \BibitemOpen
  \bibfield  {author} {\bibinfo {author} {\bibfnamefont {P.}~\bibnamefont
  {Ye}}, \bibinfo {author} {\bibfnamefont {C.-S.}\ \bibnamefont {Tian}},
  \bibinfo {author} {\bibfnamefont {X.-L.}\ \bibnamefont {Qi}},\ and\ \bibinfo
  {author} {\bibfnamefont {Z.-Y.}\ \bibnamefont {Weng}},\ }\bibfield  {title}
  {\bibinfo {title} {{Confinement-Deconfinement Interplay in Quantum Phases of
  Doped Mott Insulators}},\ }\href
  {https://doi.org/10.1103/physrevlett.106.147002} {\bibfield  {journal}
  {\bibinfo  {journal} {Phys. Rev. Lett.}\ }\textbf {\bibinfo {volume} {106}},\
  \bibinfo {pages} {147002} (\bibinfo {year} {2011})},\ \Eprint
  {https://arxiv.org/abs/1007.2507} {1007.2507} \BibitemShut {NoStop}%
\bibitem [{\citenamefont {Ye}\ \emph {et~al.}(2012)\citenamefont {Ye},
  \citenamefont {Tian}, \citenamefont {Qi},\ and\ \citenamefont
  {Weng}}]{Weng.Ye.2012}%
  \BibitemOpen
  \bibfield  {author} {\bibinfo {author} {\bibfnamefont {P.}~\bibnamefont
  {Ye}}, \bibinfo {author} {\bibfnamefont {C.-S.}\ \bibnamefont {Tian}},
  \bibinfo {author} {\bibfnamefont {X.-L.}\ \bibnamefont {Qi}},\ and\ \bibinfo
  {author} {\bibfnamefont {Z.-Y.}\ \bibnamefont {Weng}},\ }\bibfield  {title}
  {\bibinfo {title} {{Electron fractionalization and unconventional order
  parameters of the t-J model}},\ }\href
  {https://doi.org/10.1016/j.nuclphysb.2011.09.019} {\bibfield  {journal}
  {\bibinfo  {journal} {Nuclear Physics B}\ }\textbf {\bibinfo {volume}
  {854}},\ \bibinfo {pages} {815} (\bibinfo {year} {2012})},\ \Eprint
  {https://arxiv.org/abs/1106.1223} {1106.1223} \BibitemShut {NoStop}%
\bibitem [{\citenamefont {Mei}\ and\ \citenamefont
  {Weng}(2010)}]{Weng.Mei.20107w}%
  \BibitemOpen
  \bibfield  {author} {\bibinfo {author} {\bibfnamefont {J.~W.}\ \bibnamefont
  {Mei}}\ and\ \bibinfo {author} {\bibfnamefont {Z.~Y.}\ \bibnamefont {Weng}},\
  }\bibfield  {title} {\bibinfo {title} {{Spin-roton excitations in the cuprate
  superconductors}},\ }\href {https://doi.org/10.1103/physrevb.81.014507}
  {\bibfield  {journal} {\bibinfo  {journal} {Phys. Rev. B}\ }\textbf {\bibinfo
  {volume} {81}},\ \bibinfo {pages} {014507 } (\bibinfo {year}
  {2010})}\BibitemShut {NoStop}%
\bibitem [{\citenamefont {Poole}\ \emph {et~al.}(2014)\citenamefont {Poole},
  \citenamefont {Prozorov}, \citenamefont {Farach},\ and\ \citenamefont
  {Creswick}}]{POOLE2014355}%
  \BibitemOpen
  \bibfield  {author} {\bibinfo {author} {\bibfnamefont {C.~P.}\ \bibnamefont
  {Poole}}, \bibinfo {author} {\bibfnamefont {R.}~\bibnamefont {Prozorov}},
  \bibinfo {author} {\bibfnamefont {H.~A.}\ \bibnamefont {Farach}},\ and\
  \bibinfo {author} {\bibfnamefont {R.~J.}\ \bibnamefont {Creswick}},\
  }\bibfield  {title} {\bibinfo {title} {9 - type ii superconductivity},\ }in\
  \href {https://doi.org/https://doi.org/10.1016/B978-0-12-409509-0.00009-3}
  {\emph {\bibinfo {booktitle} {Superconductivity (Third Edition)}}},\ \bibinfo
  {editor} {edited by\ \bibinfo {editor} {\bibfnamefont {C.~P.}\ \bibnamefont
  {Poole}}, \bibinfo {editor} {\bibfnamefont {R.}~\bibnamefont {Prozorov}},
  \bibinfo {editor} {\bibfnamefont {H.~A.}\ \bibnamefont {Farach}},\ and\
  \bibinfo {editor} {\bibfnamefont {R.~J.}\ \bibnamefont {Creswick}}}\
  (\bibinfo  {publisher} {Elsevier},\ \bibinfo {address} {London},\ \bibinfo
  {year} {2014})\ \bibinfo {edition} {third edition}\ ed.,\ pp.\ \bibinfo
  {pages} {355--424}\BibitemShut {NoStop}%
\bibitem [{\citenamefont {Fisher}\ and\ \citenamefont
  {Lee}(1989)}]{PhysRevB.39.2756}%
  \BibitemOpen
  \bibfield  {author} {\bibinfo {author} {\bibfnamefont {M.~P.~A.}\
  \bibnamefont {Fisher}}\ and\ \bibinfo {author} {\bibfnamefont {D.~H.}\
  \bibnamefont {Lee}},\ }\bibfield  {title} {\bibinfo {title} {Correspondence
  between two-dimensional bosons and a bulk superconductor in a magnetic
  field},\ }\href {https://doi.org/10.1103/PhysRevB.39.2756} {\bibfield
  {journal} {\bibinfo  {journal} {Phys. Rev. B}\ }\textbf {\bibinfo {volume}
  {39}},\ \bibinfo {pages} {2756} (\bibinfo {year} {1989})}\BibitemShut
  {NoStop}%
\bibitem [{\citenamefont {Wen}\ and\ \citenamefont {Zee}(1990)}]{Wen1990}%
  \BibitemOpen
  \bibfield  {author} {\bibinfo {author} {\bibfnamefont {X.~G.}\ \bibnamefont
  {Wen}}\ and\ \bibinfo {author} {\bibfnamefont {A.}~\bibnamefont {Zee}},\
  }\bibfield  {title} {\bibinfo {title} {Universal conductance at the
  superconductor-insulator},\ }\href
  {https://doi.org/10.1142/S0217979290000206} {\bibfield  {journal} {\bibinfo
  {journal} {Int. J. Mod. Phys. B}\ }\textbf {\bibinfo {volume} {04}},\
  \bibinfo {pages} {437} (\bibinfo {year} {1990})}\BibitemShut {NoStop}%
\bibitem [{Note1()}]{Note1}%
  \BibitemOpen
  \bibinfo {note} {Note that the doping density can generally be expressed as
  $\delta /2=p/q$. In instances where $p=1$ the lowest Landau level for
  $\protect \tilde {E}_m$ remains degenerate. However, further splitting can
  occur when $p\protect \neq 1$. Despite this, the energy associated with such
  splitting is significantly smaller than the splitting effects induced by
  magnetic fields, specifically $\protect \bar {A}_0^h$ and $|B^e|$.
  Consequently, these energy differences can be disregarded. As a result, all
  these modes are considered to reside within the lowest energy sector, denoted
  by $m\in \protect \text {LES}$.}\BibitemShut {Stop}%
\bibitem [{Note2()}]{Note2}%
  \BibitemOpen
  \bibinfo {note} {In the scenario where $p\protect \neq 1$, the lowest excited
  level, denoted by energy $E_s$ as depicted in Fig.\protect \,\ref
  {fig_Eb4}(a), undergoes a minor slitting, resulting in each split band
  carrying a more intricate Chern number. Nevertheless, the cumulative Chern
  number, represented by $\DOTSB \sum@ \slimits@ _{m \in \protect \text {LES}}
  \protect \mathcal {C}_{m \sigma \nu }=2\nu $, is derived from the summation
  across all these sub-bands.}\BibitemShut {Stop}%
\bibitem [{\citenamefont {Xiao}\ \emph {et~al.}(2010)\citenamefont {Xiao},
  \citenamefont {Chang},\ and\ \citenamefont {Niu}}]{Niu.Xiao.2010}%
  \BibitemOpen
  \bibfield  {author} {\bibinfo {author} {\bibfnamefont {D.}~\bibnamefont
  {Xiao}}, \bibinfo {author} {\bibfnamefont {M.-C.}\ \bibnamefont {Chang}},\
  and\ \bibinfo {author} {\bibfnamefont {Q.}~\bibnamefont {Niu}},\ }\bibfield
  {title} {\bibinfo {title} {{Berry phase effects on electronic properties}},\
  }\href {https://doi.org/10.1103/revmodphys.82.1959} {\bibfield  {journal}
  {\bibinfo  {journal} {Reviews of Modern Physics}\ }\textbf {\bibinfo {volume}
  {82}},\ \bibinfo {pages} {1959} (\bibinfo {year} {2010})},\ \Eprint
  {https://arxiv.org/abs/0907.2021} {0907.2021} \BibitemShut {NoStop}%
\bibitem [{\citenamefont {Kalmeyer}\ and\ \citenamefont
  {Laughlin}(1987)}]{Laughlin.Kalmeyer.1987}%
  \BibitemOpen
  \bibfield  {author} {\bibinfo {author} {\bibfnamefont {V.}~\bibnamefont
  {Kalmeyer}}\ and\ \bibinfo {author} {\bibfnamefont {R.~B.}\ \bibnamefont
  {Laughlin}},\ }\bibfield  {title} {\bibinfo {title} {{Equivalence of the
  resonating-valence-bond and fractional quantum Hall states}},\ }\href
  {https://doi.org/10.1103/physrevlett.59.2095} {\bibfield  {journal} {\bibinfo
   {journal} {Physical Review Letters}\ }\textbf {\bibinfo {volume} {59}},\
  \bibinfo {pages} {2095} (\bibinfo {year} {1987})}\BibitemShut {NoStop}%
\bibitem [{\citenamefont {Wen}(1990)}]{Wen.Wen.1990}%
  \BibitemOpen
  \bibfield  {author} {\bibinfo {author} {\bibfnamefont {X.~G.}\ \bibnamefont
  {Wen}},\ }\bibfield  {title} {\bibinfo {title} {{Chiral Luttinger liquid and
  the edge excitations in the fractional quantum Hall states}},\ }\href
  {https://doi.org/10.1103/physrevb.41.12838} {\bibfield  {journal} {\bibinfo
  {journal} {Physical Review B}\ }\textbf {\bibinfo {volume} {41}},\ \bibinfo
  {pages} {12838} (\bibinfo {year} {1990})}\BibitemShut {NoStop}%
\bibitem [{\citenamefont {Qin}\ \emph {et~al.}(2011)\citenamefont {Qin},
  \citenamefont {Niu},\ and\ \citenamefont {Shi}}]{Shi.Qin.2011}%
  \BibitemOpen
  \bibfield  {author} {\bibinfo {author} {\bibfnamefont {T.}~\bibnamefont
  {Qin}}, \bibinfo {author} {\bibfnamefont {Q.}~\bibnamefont {Niu}},\ and\
  \bibinfo {author} {\bibfnamefont {J.}~\bibnamefont {Shi}},\ }\bibfield
  {title} {\bibinfo {title} {{Energy Magnetization and the Thermal Hall
  Effect}},\ }\href {https://doi.org/10.1103/physrevlett.107.236601} {\bibfield
   {journal} {\bibinfo  {journal} {Physical Review Letters}\ }\textbf {\bibinfo
  {volume} {107}},\ \bibinfo {pages} {236601} (\bibinfo {year} {2011})},\
  \Eprint {https://arxiv.org/abs/1108.3879} {1108.3879} \BibitemShut {NoStop}%
\bibitem [{\citenamefont {Qin}\ \emph {et~al.}(2012)\citenamefont {Qin},
  \citenamefont {Zhou},\ and\ \citenamefont {Shi}}]{Shi.Qin.2012}%
  \BibitemOpen
  \bibfield  {author} {\bibinfo {author} {\bibfnamefont {T.}~\bibnamefont
  {Qin}}, \bibinfo {author} {\bibfnamefont {J.}~\bibnamefont {Zhou}},\ and\
  \bibinfo {author} {\bibfnamefont {J.}~\bibnamefont {Shi}},\ }\bibfield
  {title} {\bibinfo {title} {{Berry curvature and the phonon Hall effect}},\
  }\href {https://doi.org/10.1103/physrevb.86.104305} {\bibfield  {journal}
  {\bibinfo  {journal} {Physical Review B}\ }\textbf {\bibinfo {volume} {86}},\
  \bibinfo {pages} {104305} (\bibinfo {year} {2012})},\ \Eprint
  {https://arxiv.org/abs/1111.1322} {1111.1322} \BibitemShut {NoStop}%
\bibitem [{\citenamefont {Matsumoto}\ and\ \citenamefont
  {Murakami}(2011{\natexlab{a}})}]{Murakami.Matsumoto.2011}%
  \BibitemOpen
  \bibfield  {author} {\bibinfo {author} {\bibfnamefont {R.}~\bibnamefont
  {Matsumoto}}\ and\ \bibinfo {author} {\bibfnamefont {S.}~\bibnamefont
  {Murakami}},\ }\bibfield  {title} {\bibinfo {title} {{Rotational motion of
  magnons and the thermal Hall effect}},\ }\href
  {https://doi.org/10.1103/physrevb.84.184406} {\bibfield  {journal} {\bibinfo
  {journal} {Physical Review B}\ }\textbf {\bibinfo {volume} {84}},\ \bibinfo
  {pages} {184406} (\bibinfo {year} {2011}{\natexlab{a}})},\ \Eprint
  {https://arxiv.org/abs/1106.1987} {1106.1987} \BibitemShut {NoStop}%
\bibitem [{\citenamefont {Matsumoto}\ and\ \citenamefont
  {Murakami}(2011{\natexlab{b}})}]{Murakami.Matsumoto.20111gl}%
  \BibitemOpen
  \bibfield  {author} {\bibinfo {author} {\bibfnamefont {R.}~\bibnamefont
  {Matsumoto}}\ and\ \bibinfo {author} {\bibfnamefont {S.}~\bibnamefont
  {Murakami}},\ }\bibfield  {title} {\bibinfo {title} {{Theoretical Prediction
  of a Rotating Magnon Wave Packet in Ferromagnets}},\ }\href
  {https://doi.org/10.1103/physrevlett.106.197202} {\bibfield  {journal}
  {\bibinfo  {journal} {Physical Review Letters}\ }\textbf {\bibinfo {volume}
  {106}},\ \bibinfo {pages} {197202} (\bibinfo {year} {2011}{\natexlab{b}})},\
  \Eprint {https://arxiv.org/abs/1103.1221} {1103.1221} \BibitemShut {NoStop}%
\bibitem [{\citenamefont {Kou}\ and\ \citenamefont
  {Weng}(2005)}]{kou_self-localization_2005}%
  \BibitemOpen
  \bibfield  {author} {\bibinfo {author} {\bibfnamefont {S.-P.}\ \bibnamefont
  {Kou}}\ and\ \bibinfo {author} {\bibfnamefont {Z.-Y.}\ \bibnamefont {Weng}},\
  }\bibfield  {title} {\bibinfo {title} {Self-localization of holes in a
  lightly doped {Mott} insulator},\ }\href
  {https://doi.org/10.1140/epjb/e2005-00300-7} {\bibfield  {journal} {\bibinfo
  {journal} {The European Physical Journal B}\ }\textbf {\bibinfo {volume}
  {47}},\ \bibinfo {pages} {37} (\bibinfo {year} {2005})}\BibitemShut {NoStop}%
\bibitem [{\citenamefont {Cooper}\ \emph {et~al.}(1987)\citenamefont {Cooper},
  \citenamefont {Alavi}, \citenamefont {Zhou}, \citenamefont {Beyermann},\ and\
  \citenamefont {Grüner}}]{cooper_thermoelectric_1987}%
  \BibitemOpen
  \bibfield  {author} {\bibinfo {author} {\bibfnamefont {J.~R.}\ \bibnamefont
  {Cooper}}, \bibinfo {author} {\bibfnamefont {B.}~\bibnamefont {Alavi}},
  \bibinfo {author} {\bibfnamefont {L.-W.}\ \bibnamefont {Zhou}}, \bibinfo
  {author} {\bibfnamefont {W.~P.}\ \bibnamefont {Beyermann}},\ and\ \bibinfo
  {author} {\bibfnamefont {G.}~\bibnamefont {Grüner}},\ }\bibfield  {title}
  {\bibinfo {title} {Thermoelectric power of some high-$t_c$ oxides},\ }\href
  {https://doi.org/10.1103/PhysRevB.35.8794} {\bibfield  {journal} {\bibinfo
  {journal} {Physical Review B}\ }\textbf {\bibinfo {volume} {35}},\ \bibinfo
  {pages} {8794} (\bibinfo {year} {1987})}\BibitemShut {NoStop}%
\bibitem [{\citenamefont {Mandal}\ \emph {et~al.}(1992)\citenamefont {Mandal},
  \citenamefont {Keshri}, \citenamefont {Mandal}, \citenamefont {Poddar},
  \citenamefont {Das},\ and\ \citenamefont
  {Ghosh}}]{mandal_thermoelectric_1992}%
  \BibitemOpen
  \bibfield  {author} {\bibinfo {author} {\bibfnamefont {J.~B.}\ \bibnamefont
  {Mandal}}, \bibinfo {author} {\bibfnamefont {S.}~\bibnamefont {Keshri}},
  \bibinfo {author} {\bibfnamefont {P.}~\bibnamefont {Mandal}}, \bibinfo
  {author} {\bibfnamefont {A.}~\bibnamefont {Poddar}}, \bibinfo {author}
  {\bibfnamefont {A.~N.}\ \bibnamefont {Das}},\ and\ \bibinfo {author}
  {\bibfnamefont {B.}~\bibnamefont {Ghosh}},\ }\bibfield  {title} {\bibinfo
  {title} {Thermoelectric power of the $\text{Bi}_2 \text{Sr}_2 \text{Ca}_{1-x}
  \text{Y}_x \text{CuO}_{8+y} $ ($x=0-1.0$) system},\ }\href
  {https://doi.org/10.1103/PhysRevB.46.11840} {\bibfield  {journal} {\bibinfo
  {journal} {Physical Review B}\ }\textbf {\bibinfo {volume} {46}},\ \bibinfo
  {pages} {11840} (\bibinfo {year} {1992})}\BibitemShut {NoStop}%
\bibitem [{\citenamefont {Keshri}\ \emph {et~al.}(1993)\citenamefont {Keshri},
  \citenamefont {Mandal}, \citenamefont {Mandal}, \citenamefont {Poddar},
  \citenamefont {Das},\ and\ \citenamefont
  {Ghosh}}]{keshri_thermoelectric_1993}%
  \BibitemOpen
  \bibfield  {author} {\bibinfo {author} {\bibfnamefont {S.}~\bibnamefont
  {Keshri}}, \bibinfo {author} {\bibfnamefont {J.~B.}\ \bibnamefont {Mandal}},
  \bibinfo {author} {\bibfnamefont {P.}~\bibnamefont {Mandal}}, \bibinfo
  {author} {\bibfnamefont {A.}~\bibnamefont {Poddar}}, \bibinfo {author}
  {\bibfnamefont {A.~N.}\ \bibnamefont {Das}},\ and\ \bibinfo {author}
  {\bibfnamefont {B.}~\bibnamefont {Ghosh}},\ }\bibfield  {title} {\bibinfo
  {title} {Thermoelectric power of $\text{Tl}_2 \text{Ba}_2 \text{Ca}_{1-x}
  \text{Y}_x \text{CuO}_{6+y} $ ( $0 \le x \le 0.6$ ) samples},\ }\href
  {https://doi.org/10.1103/PhysRevB.47.9048} {\bibfield  {journal} {\bibinfo
  {journal} {Physical Review B}\ }\textbf {\bibinfo {volume} {47}},\ \bibinfo
  {pages} {9048} (\bibinfo {year} {1993})}\BibitemShut {NoStop}%
\bibitem [{\citenamefont {Plate}\ \emph {et~al.}(2005)\citenamefont {Plate},
  \citenamefont {Mottershead}, \citenamefont {Elfimov}, \citenamefont {Peets},
  \citenamefont {Liang}, \citenamefont {Bonn}, \citenamefont {Hardy},
  \citenamefont {Chiuzbaian}, \citenamefont {Falub}, \citenamefont {Shi},
  \citenamefont {Patthey},\ and\ \citenamefont
  {Damascelli}}]{Damascelli.Plat.2005}%
  \BibitemOpen
  \bibfield  {author} {\bibinfo {author} {\bibfnamefont {M.}~\bibnamefont
  {Plate}}, \bibinfo {author} {\bibfnamefont {J.~D.~F.}\ \bibnamefont
  {Mottershead}}, \bibinfo {author} {\bibfnamefont {I.~S.}\ \bibnamefont
  {Elfimov}}, \bibinfo {author} {\bibfnamefont {D.~C.}\ \bibnamefont {Peets}},
  \bibinfo {author} {\bibfnamefont {R.}~\bibnamefont {Liang}}, \bibinfo
  {author} {\bibfnamefont {D.~A.}\ \bibnamefont {Bonn}}, \bibinfo {author}
  {\bibfnamefont {W.~N.}\ \bibnamefont {Hardy}}, \bibinfo {author}
  {\bibfnamefont {S.}~\bibnamefont {Chiuzbaian}}, \bibinfo {author}
  {\bibfnamefont {M.}~\bibnamefont {Falub}}, \bibinfo {author} {\bibfnamefont
  {M.}~\bibnamefont {Shi}}, \bibinfo {author} {\bibfnamefont {L.}~\bibnamefont
  {Patthey}},\ and\ \bibinfo {author} {\bibfnamefont {A.}~\bibnamefont
  {Damascelli}},\ }\bibfield  {title} {\bibinfo {title} {{Fermi Surface and
  Quasiparticle Excitations of Overdoped $\mathrm{Tl_2Ba_2CuO_{6+\delta}}$}},\
  }\href {https://doi.org/10.1103/physrevlett.95.077001} {\bibfield  {journal}
  {\bibinfo  {journal} {Phys. Rev. Lett.}\ }\textbf {\bibinfo {volume} {95}},\
  \bibinfo {pages} {077001} (\bibinfo {year} {2005})}\BibitemShut {NoStop}%
\bibitem [{\citenamefont {Kaminski}\ \emph {et~al.}(2003)\citenamefont
  {Kaminski}, \citenamefont {Rosenkranz}, \citenamefont {Fretwell},
  \citenamefont {Li}, \citenamefont {Raffy}, \citenamefont {Randeria},
  \citenamefont {Norman},\ and\ \citenamefont
  {Campuzano}}]{Campuzano.Kaminski.2003}%
  \BibitemOpen
  \bibfield  {author} {\bibinfo {author} {\bibfnamefont {A.}~\bibnamefont
  {Kaminski}}, \bibinfo {author} {\bibfnamefont {S.}~\bibnamefont
  {Rosenkranz}}, \bibinfo {author} {\bibfnamefont {H.~M.}\ \bibnamefont
  {Fretwell}}, \bibinfo {author} {\bibfnamefont {Z.~Z.}\ \bibnamefont {Li}},
  \bibinfo {author} {\bibfnamefont {H.}~\bibnamefont {Raffy}}, \bibinfo
  {author} {\bibfnamefont {M.}~\bibnamefont {Randeria}}, \bibinfo {author}
  {\bibfnamefont {M.~R.}\ \bibnamefont {Norman}},\ and\ \bibinfo {author}
  {\bibfnamefont {J.~C.}\ \bibnamefont {Campuzano}},\ }\bibfield  {title}
  {\bibinfo {title} {{Crossover from Coherent to Incoherent Electronic
  Excitations in the Normal State of $\mathrm{Bi_2Sr_2CaCu_2O_{8+\delta}}$}},\
  }\href {https://doi.org/10.1103/physrevlett.90.207003} {\bibfield  {journal}
  {\bibinfo  {journal} {Phys. Rev. Lett.}\ }\textbf {\bibinfo {volume} {90}},\
  \bibinfo {pages} {207003} (\bibinfo {year} {2003})}\BibitemShut {NoStop}%
\bibitem [{\citenamefont {Chatterjee}\ \emph {et~al.}(2011)\citenamefont
  {Chatterjee}, \citenamefont {Ai}, \citenamefont {Zhao}, \citenamefont
  {Rosenkranz}, \citenamefont {Kaminski}, \citenamefont {Raffy}, \citenamefont
  {Li}, \citenamefont {Kadowaki}, \citenamefont {Randeria}, \citenamefont
  {Norman},\ and\ \citenamefont {Campuzano}}]{Campuzano.Chatterjee.2011}%
  \BibitemOpen
  \bibfield  {author} {\bibinfo {author} {\bibfnamefont {U.}~\bibnamefont
  {Chatterjee}}, \bibinfo {author} {\bibfnamefont {D.}~\bibnamefont {Ai}},
  \bibinfo {author} {\bibfnamefont {J.}~\bibnamefont {Zhao}}, \bibinfo {author}
  {\bibfnamefont {S.}~\bibnamefont {Rosenkranz}}, \bibinfo {author}
  {\bibfnamefont {A.}~\bibnamefont {Kaminski}}, \bibinfo {author}
  {\bibfnamefont {H.}~\bibnamefont {Raffy}}, \bibinfo {author} {\bibfnamefont
  {Z.}~\bibnamefont {Li}}, \bibinfo {author} {\bibfnamefont {K.}~\bibnamefont
  {Kadowaki}}, \bibinfo {author} {\bibfnamefont {M.}~\bibnamefont {Randeria}},
  \bibinfo {author} {\bibfnamefont {M.~R.}\ \bibnamefont {Norman}},\ and\
  \bibinfo {author} {\bibfnamefont {J.~C.}\ \bibnamefont {Campuzano}},\
  }\bibfield  {title} {\bibinfo {title} {{Electronic phase diagram of
  high-temperature copper oxide superconductors}},\ }\href
  {https://doi.org/10.1073/pnas.1101008108} {\bibfield  {journal} {\bibinfo
  {journal} {Proc. Natl. Acad. Sci.}\ }\textbf {\bibinfo {volume} {108}},\
  \bibinfo {pages} {9346} (\bibinfo {year} {2011})}\BibitemShut {NoStop}%
\bibitem [{\citenamefont {Khait}\ \emph {et~al.}(2023)\citenamefont {Khait},
  \citenamefont {Bhattacharyya}, \citenamefont {Samanta},\ and\ \citenamefont
  {Auerbach}}]{AuerbachNPJ_hall_2023}%
  \BibitemOpen
  \bibfield  {author} {\bibinfo {author} {\bibfnamefont {I.}~\bibnamefont
  {Khait}}, \bibinfo {author} {\bibfnamefont {S.}~\bibnamefont
  {Bhattacharyya}}, \bibinfo {author} {\bibfnamefont {A.}~\bibnamefont
  {Samanta}},\ and\ \bibinfo {author} {\bibfnamefont {A.}~\bibnamefont
  {Auerbach}},\ }\bibfield  {title} {\bibinfo {title} {Hall anomalies of the
  doped {Mott} insulator},\ }\href {https://doi.org/10.1038/s41535-023-00611-5}
  {\bibfield  {journal} {\bibinfo  {journal} {npj Quantum Materials}\ }\textbf
  {\bibinfo {volume} {8}},\ \bibinfo {pages} {75} (\bibinfo {year}
  {2023})}\BibitemShut {NoStop}%
\bibitem [{\citenamefont {Kou}\ \emph {et~al.}(2005{\natexlab{b}})\citenamefont
  {Kou}, \citenamefont {Qi},\ and\ \citenamefont {Weng}}]{Weng.Kou.2005}%
  \BibitemOpen
  \bibfield  {author} {\bibinfo {author} {\bibfnamefont {S.-P.}\ \bibnamefont
  {Kou}}, \bibinfo {author} {\bibfnamefont {X.-L.}\ \bibnamefont {Qi}},\ and\
  \bibinfo {author} {\bibfnamefont {Z.-Y.}\ \bibnamefont {Weng}},\ }\bibfield
  {title} {\bibinfo {title} {{Spin Hall effect in a doped Mott insulator}},\
  }\href {https://doi.org/10.1103/physrevb.72.165114} {\bibfield  {journal}
  {\bibinfo  {journal} {Physical Review B}\ }\textbf {\bibinfo {volume} {72}},\
  \bibinfo {pages} {165114} (\bibinfo {year} {2005}{\natexlab{b}})},\ \Eprint
  {https://arxiv.org/abs/cond-mat/0412146} {cond-mat/0412146} \BibitemShut
  {NoStop}%
\bibitem [{\citenamefont {Levin}\ and\ \citenamefont
  {Senthil}(2004)}]{Senthil.Levin.2004}%
  \BibitemOpen
  \bibfield  {author} {\bibinfo {author} {\bibfnamefont {M.}~\bibnamefont
  {Levin}}\ and\ \bibinfo {author} {\bibfnamefont {T.}~\bibnamefont
  {Senthil}},\ }\bibfield  {title} {\bibinfo {title} {{Deconfined quantum
  criticality and Néel order via dimer disorder}},\ }\href
  {https://doi.org/10.1103/physrevb.70.220403} {\bibfield  {journal} {\bibinfo
  {journal} {Physical Review B}\ }\textbf {\bibinfo {volume} {70}},\ \bibinfo
  {pages} {220403} (\bibinfo {year} {2004})},\ \Eprint
  {https://arxiv.org/abs/cond-mat/0405702} {cond-mat/0405702} \BibitemShut
  {NoStop}%
\bibitem [{\citenamefont {Senthil}\ \emph {et~al.}(2004)\citenamefont
  {Senthil}, \citenamefont {Vishwanath}, \citenamefont {Balents}, \citenamefont
  {Sachdev},\ and\ \citenamefont {Fisher}}]{Fisher.Senthil.2004}%
  \BibitemOpen
  \bibfield  {author} {\bibinfo {author} {\bibfnamefont {T.}~\bibnamefont
  {Senthil}}, \bibinfo {author} {\bibfnamefont {A.}~\bibnamefont {Vishwanath}},
  \bibinfo {author} {\bibfnamefont {L.}~\bibnamefont {Balents}}, \bibinfo
  {author} {\bibfnamefont {S.}~\bibnamefont {Sachdev}},\ and\ \bibinfo {author}
  {\bibfnamefont {M.~P.~A.}\ \bibnamefont {Fisher}},\ }\bibfield  {title}
  {\bibinfo {title} {{Deconfined Quantum Critical Points}},\ }\href
  {https://doi.org/10.1126/science.1091806} {\bibfield  {journal} {\bibinfo
  {journal} {Science}\ }\textbf {\bibinfo {volume} {303}},\ \bibinfo {pages}
  {1490} (\bibinfo {year} {2004})},\ \Eprint
  {https://arxiv.org/abs/cond-mat/0311326} {cond-mat/0311326} \BibitemShut
  {NoStop}%
\bibitem [{\citenamefont {Wang}\ \emph {et~al.}(2017)\citenamefont {Wang},
  \citenamefont {Nahum}, \citenamefont {Metlitski}, \citenamefont {Xu},\ and\
  \citenamefont {Senthil}}]{Senthil.Wang.2017}%
  \BibitemOpen
  \bibfield  {author} {\bibinfo {author} {\bibfnamefont {C.}~\bibnamefont
  {Wang}}, \bibinfo {author} {\bibfnamefont {A.}~\bibnamefont {Nahum}},
  \bibinfo {author} {\bibfnamefont {M.~A.}\ \bibnamefont {Metlitski}}, \bibinfo
  {author} {\bibfnamefont {C.}~\bibnamefont {Xu}},\ and\ \bibinfo {author}
  {\bibfnamefont {T.}~\bibnamefont {Senthil}},\ }\bibfield  {title} {\bibinfo
  {title} {{Deconfined Quantum Critical Points: Symmetries and Dualities}},\
  }\href {https://doi.org/10.1103/physrevx.7.031051} {\bibfield  {journal}
  {\bibinfo  {journal} {Physical Review X}\ }\textbf {\bibinfo {volume} {7}},\
  \bibinfo {pages} {031051} (\bibinfo {year} {2017})},\ \Eprint
  {https://arxiv.org/abs/1703.02426} {1703.02426} \BibitemShut {NoStop}%
\bibitem [{\citenamefont {Coldea}\ \emph {et~al.}(2001)\citenamefont {Coldea},
  \citenamefont {Hayden}, \citenamefont {Aeppli}, \citenamefont {Perring},
  \citenamefont {Frost}, \citenamefont {Mason}, \citenamefont {Cheong},\ and\
  \citenamefont {Fisk}}]{Fisk.Coldea.2001}%
  \BibitemOpen
  \bibfield  {author} {\bibinfo {author} {\bibfnamefont {R.}~\bibnamefont
  {Coldea}}, \bibinfo {author} {\bibfnamefont {S.~M.}\ \bibnamefont {Hayden}},
  \bibinfo {author} {\bibfnamefont {G.}~\bibnamefont {Aeppli}}, \bibinfo
  {author} {\bibfnamefont {T.~G.}\ \bibnamefont {Perring}}, \bibinfo {author}
  {\bibfnamefont {C.~D.}\ \bibnamefont {Frost}}, \bibinfo {author}
  {\bibfnamefont {T.~E.}\ \bibnamefont {Mason}}, \bibinfo {author}
  {\bibfnamefont {S.-W.}\ \bibnamefont {Cheong}},\ and\ \bibinfo {author}
  {\bibfnamefont {Z.}~\bibnamefont {Fisk}},\ }\bibfield  {title} {\bibinfo
  {title} {{Spin Waves and Electronic Interactions in $\mathrm{La_2 CuO_4}$}},\
  }\href {https://doi.org/10.1103/physrevlett.86.5377} {\bibfield  {journal}
  {\bibinfo  {journal} {Physical Review Letters}\ }\textbf {\bibinfo {volume}
  {86}},\ \bibinfo {pages} {5377} (\bibinfo {year} {2001})},\ \Eprint
  {https://arxiv.org/abs/cond-mat/0006384} {cond-mat/0006384} \BibitemShut
  {NoStop}%
\bibitem [{\citenamefont {Headings}\ \emph {et~al.}(2010)\citenamefont
  {Headings}, \citenamefont {Hayden}, \citenamefont {Coldea},\ and\
  \citenamefont {Perring}}]{Perring.Headings.2010}%
  \BibitemOpen
  \bibfield  {author} {\bibinfo {author} {\bibfnamefont {N.~S.}\ \bibnamefont
  {Headings}}, \bibinfo {author} {\bibfnamefont {S.~M.}\ \bibnamefont
  {Hayden}}, \bibinfo {author} {\bibfnamefont {R.}~\bibnamefont {Coldea}},\
  and\ \bibinfo {author} {\bibfnamefont {T.~G.}\ \bibnamefont {Perring}},\
  }\bibfield  {title} {\bibinfo {title} {{Anomalous High-Energy Spin
  Excitations in the High-Tc Superconductor-Parent Antiferromagnet
  $\mathrm{La_2 CuO_4}$}},\ }\href
  {https://doi.org/10.1103/physrevlett.105.247001} {\bibfield  {journal}
  {\bibinfo  {journal} {Physical Review Letters}\ }\textbf {\bibinfo {volume}
  {105}},\ \bibinfo {pages} {247001} (\bibinfo {year} {2010})},\ \Eprint
  {https://arxiv.org/abs/1009.2915} {1009.2915} \BibitemShut {NoStop}%
\bibitem [{\citenamefont {Fong}\ \emph {et~al.}(1995)\citenamefont {Fong},
  \citenamefont {Keimer}, \citenamefont {Anderson}, \citenamefont {Reznik},
  \citenamefont {Do\ifmmode~\breve{g}\else \u{g}\fi{}an},\ and\ \citenamefont
  {Aksay}}]{PhysRevLett.75.316}%
  \BibitemOpen
  \bibfield  {author} {\bibinfo {author} {\bibfnamefont {H.~F.}\ \bibnamefont
  {Fong}}, \bibinfo {author} {\bibfnamefont {B.}~\bibnamefont {Keimer}},
  \bibinfo {author} {\bibfnamefont {P.~W.}\ \bibnamefont {Anderson}}, \bibinfo
  {author} {\bibfnamefont {D.}~\bibnamefont {Reznik}}, \bibinfo {author}
  {\bibfnamefont {F.}~\bibnamefont {Do\ifmmode~\breve{g}\else \u{g}\fi{}an}},\
  and\ \bibinfo {author} {\bibfnamefont {I.~A.}\ \bibnamefont {Aksay}},\
  }\bibfield  {title} {\bibinfo {title} {{Phonon and Magnetic Neutron
  Scattering at 41 $\mathrm{meV}$ in $\mathrm{YBa_2 Cu_3 O_7}$}},\ }\href
  {https://doi.org/10.1103/PhysRevLett.75.316} {\bibfield  {journal} {\bibinfo
  {journal} {Phys. Rev. Lett.}\ }\textbf {\bibinfo {volume} {75}},\ \bibinfo
  {pages} {316} (\bibinfo {year} {1995})}\BibitemShut {NoStop}%
\bibitem [{\citenamefont {Fauque}\ \emph {et~al.}(2007)\citenamefont {Fauque},
  \citenamefont {Sidis}, \citenamefont {Capogna}, \citenamefont {Ivanov},
  \citenamefont {Hradil}, \citenamefont {Ulrich}, \citenamefont {Rykov},
  \citenamefont {Keimer},\ and\ \citenamefont {Bourges}}]{Bourges.Fauqu.2007}%
  \BibitemOpen
  \bibfield  {author} {\bibinfo {author} {\bibfnamefont {B.}~\bibnamefont
  {Fauque}}, \bibinfo {author} {\bibfnamefont {Y.}~\bibnamefont {Sidis}},
  \bibinfo {author} {\bibfnamefont {L.}~\bibnamefont {Capogna}}, \bibinfo
  {author} {\bibfnamefont {A.}~\bibnamefont {Ivanov}}, \bibinfo {author}
  {\bibfnamefont {K.}~\bibnamefont {Hradil}}, \bibinfo {author} {\bibfnamefont
  {C.}~\bibnamefont {Ulrich}}, \bibinfo {author} {\bibfnamefont {A.~I.}\
  \bibnamefont {Rykov}}, \bibinfo {author} {\bibfnamefont {B.}~\bibnamefont
  {Keimer}},\ and\ \bibinfo {author} {\bibfnamefont {P.}~\bibnamefont
  {Bourges}},\ }\bibfield  {title} {\bibinfo {title} {{Dispersion of the odd
  magnetic resonant mode in near-optimally doped $\mathrm{Bi_2Sr_2CaCu_2O_{8+
  \delta } }$}},\ }\href {https://doi.org/10.1103/physrevb.76.214512}
  {\bibfield  {journal} {\bibinfo  {journal} {Phys. Rev. B}\ }\textbf {\bibinfo
  {volume} {76}},\ \bibinfo {pages} {214512} (\bibinfo {year}
  {2007})}\BibitemShut {NoStop}%
\bibitem [{\citenamefont {Capogna}\ \emph {et~al.}(2007)\citenamefont
  {Capogna}, \citenamefont {Fauque}, \citenamefont {Sidis}, \citenamefont
  {Ulrich}, \citenamefont {Bourges}, \citenamefont {Pailhes}, \citenamefont
  {Ivanov}, \citenamefont {Tallon}, \citenamefont {Liang}, \citenamefont {Lin},
  \citenamefont {Rykov},\ and\ \citenamefont {Keimer}}]{Keimer.Capogna.2007}%
  \BibitemOpen
  \bibfield  {author} {\bibinfo {author} {\bibfnamefont {L.}~\bibnamefont
  {Capogna}}, \bibinfo {author} {\bibfnamefont {B.}~\bibnamefont {Fauque}},
  \bibinfo {author} {\bibfnamefont {Y.}~\bibnamefont {Sidis}}, \bibinfo
  {author} {\bibfnamefont {C.}~\bibnamefont {Ulrich}}, \bibinfo {author}
  {\bibfnamefont {P.}~\bibnamefont {Bourges}}, \bibinfo {author} {\bibfnamefont
  {S.}~\bibnamefont {Pailhes}}, \bibinfo {author} {\bibfnamefont
  {A.}~\bibnamefont {Ivanov}}, \bibinfo {author} {\bibfnamefont {J.~L.}\
  \bibnamefont {Tallon}}, \bibinfo {author} {\bibfnamefont {B.}~\bibnamefont
  {Liang}}, \bibinfo {author} {\bibfnamefont {C.~T.}\ \bibnamefont {Lin}},
  \bibinfo {author} {\bibfnamefont {A.~I.}\ \bibnamefont {Rykov}},\ and\
  \bibinfo {author} {\bibfnamefont {B.}~\bibnamefont {Keimer}},\ }\bibfield
  {title} {\bibinfo {title} {{Odd and even magnetic resonant modes in highly
  overdoped $\mathrm{Bi_2Sr_2CaCu_2O_{8+\delta}}$}},\ }\href
  {https://doi.org/10.1103/physrevb.75.060502} {\bibfield  {journal} {\bibinfo
  {journal} {Phys. Rev. B}\ }\textbf {\bibinfo {volume} {75}},\ \bibinfo
  {pages} {060502(R)} (\bibinfo {year} {2007})}\BibitemShut {NoStop}%
\bibitem [{\citenamefont {Fong}\ \emph {et~al.}(1999)\citenamefont {Fong},
  \citenamefont {Bourges}, \citenamefont {Sidis}, \citenamefont {Regnault},
  \citenamefont {Ivanov}, \citenamefont {Gu}, \citenamefont {Koshizuka},\ and\
  \citenamefont {Keimer}}]{Keimer.Fong.1999}%
  \BibitemOpen
  \bibfield  {author} {\bibinfo {author} {\bibfnamefont {H.~F.}\ \bibnamefont
  {Fong}}, \bibinfo {author} {\bibfnamefont {P.}~\bibnamefont {Bourges}},
  \bibinfo {author} {\bibfnamefont {Y.}~\bibnamefont {Sidis}}, \bibinfo
  {author} {\bibfnamefont {L.~P.}\ \bibnamefont {Regnault}}, \bibinfo {author}
  {\bibfnamefont {A.}~\bibnamefont {Ivanov}}, \bibinfo {author} {\bibfnamefont
  {G.~D.}\ \bibnamefont {Gu}}, \bibinfo {author} {\bibfnamefont
  {N.}~\bibnamefont {Koshizuka}},\ and\ \bibinfo {author} {\bibfnamefont
  {B.}~\bibnamefont {Keimer}},\ }\bibfield  {title} {\bibinfo {title} {{Neutron
  scattering from magnetic excitations in
  $\mathrm{Bi_2Sr_2CaCu_2O_{8+\delta}}$}},\ }\href
  {https://doi.org/10.1038/19255} {\bibfield  {journal} {\bibinfo  {journal}
  {Nature}\ }\textbf {\bibinfo {volume} {398}},\ \bibinfo {pages} {588}
  (\bibinfo {year} {1999})}\BibitemShut {NoStop}%
\bibitem [{\citenamefont {He}\ \emph {et~al.}(2002)\citenamefont {He},
  \citenamefont {Bourges}, \citenamefont {Sidis}, \citenamefont {Ulrich},
  \citenamefont {Regnault}, \citenamefont {Pailhes}, \citenamefont
  {Berzigiarova}, \citenamefont {Kolesnikov},\ and\ \citenamefont
  {Keimer}}]{Keimer.He.2002}%
  \BibitemOpen
  \bibfield  {author} {\bibinfo {author} {\bibfnamefont {H.}~\bibnamefont
  {He}}, \bibinfo {author} {\bibfnamefont {P.}~\bibnamefont {Bourges}},
  \bibinfo {author} {\bibfnamefont {Y.}~\bibnamefont {Sidis}}, \bibinfo
  {author} {\bibfnamefont {C.}~\bibnamefont {Ulrich}}, \bibinfo {author}
  {\bibfnamefont {L.~P.}\ \bibnamefont {Regnault}}, \bibinfo {author}
  {\bibfnamefont {S.}~\bibnamefont {Pailhes}}, \bibinfo {author} {\bibfnamefont
  {N.~S.}\ \bibnamefont {Berzigiarova}}, \bibinfo {author} {\bibfnamefont
  {N.~N.}\ \bibnamefont {Kolesnikov}},\ and\ \bibinfo {author} {\bibfnamefont
  {B.}~\bibnamefont {Keimer}},\ }\bibfield  {title} {\bibinfo {title}
  {{Magnetic Resonant Mode in the Single-Layer High-Temperature Superconductor
  $\mathrm{Tl_2Ba_2CuO_{6+\delta}}$}},\ }\href
  {https://doi.org/10.1126/science.1067877} {\bibfield  {journal} {\bibinfo
  {journal} {Science}\ }\textbf {\bibinfo {volume} {295}},\ \bibinfo {pages}
  {1045} (\bibinfo {year} {2002})}\BibitemShut {NoStop}%
\bibitem [{\citenamefont {Dai}\ \emph {et~al.}(1999)\citenamefont {Dai},
  \citenamefont {Mook}, \citenamefont {Hayden}, \citenamefont {Aeppli},
  \citenamefont {Perring}, \citenamefont {Hunt},\ and\ \citenamefont
  {Dogan}}]{Dai1999}%
  \BibitemOpen
  \bibfield  {author} {\bibinfo {author} {\bibfnamefont {P.}~\bibnamefont
  {Dai}}, \bibinfo {author} {\bibfnamefont {H.~A.}\ \bibnamefont {Mook}},
  \bibinfo {author} {\bibfnamefont {S.~M.}\ \bibnamefont {Hayden}}, \bibinfo
  {author} {\bibfnamefont {G.}~\bibnamefont {Aeppli}}, \bibinfo {author}
  {\bibfnamefont {T.~G.}\ \bibnamefont {Perring}}, \bibinfo {author}
  {\bibfnamefont {R.~D.}\ \bibnamefont {Hunt}},\ and\ \bibinfo {author}
  {\bibfnamefont {F.}~\bibnamefont {Dogan}},\ }\bibfield  {title} {\bibinfo
  {title} {The magnetic excitation spectrum and thermodynamics of high-$t_c$
  superconductors},\ }\href {https://doi.org/10.1126/science.284.5418.1344}
  {\bibfield  {journal} {\bibinfo  {journal} {Science}\ }\textbf {\bibinfo
  {volume} {284}},\ \bibinfo {pages} {1344} (\bibinfo {year}
  {1999})}\BibitemShut {NoStop}%
\bibitem [{\citenamefont {Chan}\ \emph
  {et~al.}(2016{\natexlab{a}})\citenamefont {Chan}, \citenamefont {Dorow},
  \citenamefont {Mangin-Thro}, \citenamefont {Tang}, \citenamefont {Ge},
  \citenamefont {Veit}, \citenamefont {Yu}, \citenamefont {Zhao}, \citenamefont
  {Christianson}, \citenamefont {Park}, \citenamefont {Sidis}, \citenamefont
  {Steffens}, \citenamefont {Abernathy}, \citenamefont {Bourges},\ and\
  \citenamefont {Greven}}]{Greven.Chan.2016}%
  \BibitemOpen
  \bibfield  {author} {\bibinfo {author} {\bibfnamefont {M.~K.}\ \bibnamefont
  {Chan}}, \bibinfo {author} {\bibfnamefont {C.~J.}\ \bibnamefont {Dorow}},
  \bibinfo {author} {\bibfnamefont {L.}~\bibnamefont {Mangin-Thro}}, \bibinfo
  {author} {\bibfnamefont {Y.}~\bibnamefont {Tang}}, \bibinfo {author}
  {\bibfnamefont {Y.}~\bibnamefont {Ge}}, \bibinfo {author} {\bibfnamefont
  {M.~J.}\ \bibnamefont {Veit}}, \bibinfo {author} {\bibfnamefont
  {G.}~\bibnamefont {Yu}}, \bibinfo {author} {\bibfnamefont {X.}~\bibnamefont
  {Zhao}}, \bibinfo {author} {\bibfnamefont {A.~D.}\ \bibnamefont
  {Christianson}}, \bibinfo {author} {\bibfnamefont {J.~T.}\ \bibnamefont
  {Park}}, \bibinfo {author} {\bibfnamefont {Y.}~\bibnamefont {Sidis}},
  \bibinfo {author} {\bibfnamefont {P.}~\bibnamefont {Steffens}}, \bibinfo
  {author} {\bibfnamefont {D.~L.}\ \bibnamefont {Abernathy}}, \bibinfo {author}
  {\bibfnamefont {P.}~\bibnamefont {Bourges}},\ and\ \bibinfo {author}
  {\bibfnamefont {M.}~\bibnamefont {Greven}},\ }\bibfield  {title} {\bibinfo
  {title} {{Commensurate antiferromagnetic excitations as a signature of the
  pseudogap in the tetragonal high-Tc cuprate
  $\mathrm{HgBa_2CuO_{4+\delta}}$}},\ }\href
  {https://doi.org/10.1038/ncomms10819} {\bibfield  {journal} {\bibinfo
  {journal} {Nature Communications}\ }\textbf {\bibinfo {volume} {7}},\
  \bibinfo {pages} {10819} (\bibinfo {year} {2016}{\natexlab{a}})},\ \Eprint
  {https://arxiv.org/abs/1402.4517} {1402.4517} \BibitemShut {NoStop}%
\bibitem [{\citenamefont {Chan}\ \emph
  {et~al.}(2016{\natexlab{b}})\citenamefont {Chan}, \citenamefont {Tang},
  \citenamefont {Dorow}, \citenamefont {Jeong}, \citenamefont {Mangin-Thro},
  \citenamefont {Veit}, \citenamefont {Ge}, \citenamefont {Abernathy},
  \citenamefont {Sidis}, \citenamefont {Bourges},\ and\ \citenamefont
  {Greven}}]{Greven.Chan.2016noa}%
  \BibitemOpen
  \bibfield  {author} {\bibinfo {author} {\bibfnamefont {M.~K.}\ \bibnamefont
  {Chan}}, \bibinfo {author} {\bibfnamefont {Y.}~\bibnamefont {Tang}}, \bibinfo
  {author} {\bibfnamefont {C.~J.}\ \bibnamefont {Dorow}}, \bibinfo {author}
  {\bibfnamefont {J.}~\bibnamefont {Jeong}}, \bibinfo {author} {\bibfnamefont
  {L.}~\bibnamefont {Mangin-Thro}}, \bibinfo {author} {\bibfnamefont {M.~J.}\
  \bibnamefont {Veit}}, \bibinfo {author} {\bibfnamefont {Y.}~\bibnamefont
  {Ge}}, \bibinfo {author} {\bibfnamefont {D.~L.}\ \bibnamefont {Abernathy}},
  \bibinfo {author} {\bibfnamefont {Y.}~\bibnamefont {Sidis}}, \bibinfo
  {author} {\bibfnamefont {P.}~\bibnamefont {Bourges}},\ and\ \bibinfo {author}
  {\bibfnamefont {M.}~\bibnamefont {Greven}},\ }\bibfield  {title} {\bibinfo
  {title} {{Hourglass Dispersion and Resonance of Magnetic Excitations in the
  Superconducting State of the Single-Layer Cuprate
  $\mathrm{HgBa_2CuO_{4+\delta}}$ Near Optimal Doping}},\ }\href
  {https://doi.org/10.1103/physrevlett.117.277002} {\bibfield  {journal}
  {\bibinfo  {journal} {Physical Review Letters}\ }\textbf {\bibinfo {volume}
  {117}},\ \bibinfo {pages} {277002} (\bibinfo {year} {2016}{\natexlab{b}})},\
  \Eprint {https://arxiv.org/abs/1610.01097} {1610.01097} \BibitemShut
  {NoStop}%
\bibitem [{\citenamefont {Pailhès}\ \emph {et~al.}(2004)\citenamefont
  {Pailhès}, \citenamefont {Sidis}, \citenamefont {Bourges}, \citenamefont
  {Hinkov}, \citenamefont {Ivanov}, \citenamefont {Ulrich}, \citenamefont
  {Regnault},\ and\ \citenamefont {Keimer}}]{Keimer.P.2004}%
  \BibitemOpen
  \bibfield  {author} {\bibinfo {author} {\bibfnamefont {S.}~\bibnamefont
  {Pailhès}}, \bibinfo {author} {\bibfnamefont {Y.}~\bibnamefont {Sidis}},
  \bibinfo {author} {\bibfnamefont {P.}~\bibnamefont {Bourges}}, \bibinfo
  {author} {\bibfnamefont {V.}~\bibnamefont {Hinkov}}, \bibinfo {author}
  {\bibfnamefont {A.}~\bibnamefont {Ivanov}}, \bibinfo {author} {\bibfnamefont
  {C.}~\bibnamefont {Ulrich}}, \bibinfo {author} {\bibfnamefont {L.~P.}\
  \bibnamefont {Regnault}},\ and\ \bibinfo {author} {\bibfnamefont
  {B.}~\bibnamefont {Keimer}},\ }\bibfield  {title} {\bibinfo {title}
  {{Resonant Magnetic Excitations at High Energy in Superconducting
  $\mathrm{YBa_2Cu_3O_{6.85}}$}},\ }\href
  {https://doi.org/10.1103/physrevlett.93.167001} {\bibfield  {journal}
  {\bibinfo  {journal} {Physical Review Letters}\ }\textbf {\bibinfo {volume}
  {93}},\ \bibinfo {pages} {167001} (\bibinfo {year} {2004})},\ \Eprint
  {https://arxiv.org/abs/cond-mat/0403609} {cond-mat/0403609} \BibitemShut
  {NoStop}%
\bibitem [{\citenamefont {Hayden}\ \emph {et~al.}(2004)\citenamefont {Hayden},
  \citenamefont {Mook}, \citenamefont {Dai}, \citenamefont {Perring},\ and\
  \citenamefont {Dogan}}]{Dogan.Hayden.2004}%
  \BibitemOpen
  \bibfield  {author} {\bibinfo {author} {\bibfnamefont {S.~M.}\ \bibnamefont
  {Hayden}}, \bibinfo {author} {\bibfnamefont {H.~A.}\ \bibnamefont {Mook}},
  \bibinfo {author} {\bibfnamefont {P.}~\bibnamefont {Dai}}, \bibinfo {author}
  {\bibfnamefont {T.~G.}\ \bibnamefont {Perring}},\ and\ \bibinfo {author}
  {\bibfnamefont {F.}~\bibnamefont {Dogan}},\ }\bibfield  {title} {\bibinfo
  {title} {{The structure of the high-energy spin excitations in a
  high-transition-temperature superconductor}},\ }\href
  {https://doi.org/10.1038/nature02576} {\bibfield  {journal} {\bibinfo
  {journal} {Nature}\ }\textbf {\bibinfo {volume} {429}},\ \bibinfo {pages}
  {531} (\bibinfo {year} {2004})}\BibitemShut {NoStop}%
\bibitem [{\citenamefont {Tranquada}\ \emph {et~al.}(2004)\citenamefont
  {Tranquada}, \citenamefont {Woo}, \citenamefont {Perring}, \citenamefont
  {Goka}, \citenamefont {Gu}, \citenamefont {Xu}, \citenamefont {Fujita},\ and\
  \citenamefont {Yamada}}]{Yamada.Tranquada.2004}%
  \BibitemOpen
  \bibfield  {author} {\bibinfo {author} {\bibfnamefont {J.~M.}\ \bibnamefont
  {Tranquada}}, \bibinfo {author} {\bibfnamefont {H.}~\bibnamefont {Woo}},
  \bibinfo {author} {\bibfnamefont {T.~G.}\ \bibnamefont {Perring}}, \bibinfo
  {author} {\bibfnamefont {H.}~\bibnamefont {Goka}}, \bibinfo {author}
  {\bibfnamefont {G.~D.}\ \bibnamefont {Gu}}, \bibinfo {author} {\bibfnamefont
  {G.}~\bibnamefont {Xu}}, \bibinfo {author} {\bibfnamefont {M.}~\bibnamefont
  {Fujita}},\ and\ \bibinfo {author} {\bibfnamefont {K.}~\bibnamefont
  {Yamada}},\ }\bibfield  {title} {\bibinfo {title} {{Quantum magnetic
  excitations from stripes in copper oxide superconductors}},\ }\href
  {https://doi.org/10.1038/nature02574} {\bibfield  {journal} {\bibinfo
  {journal} {Nature}\ }\textbf {\bibinfo {volume} {429}},\ \bibinfo {pages}
  {534} (\bibinfo {year} {2004})},\ \Eprint
  {https://arxiv.org/abs/cond-mat/0401621} {cond-mat/0401621} \BibitemShut
  {NoStop}%
\bibitem [{\citenamefont {Sato}\ \emph {et~al.}(2020)\citenamefont {Sato},
  \citenamefont {Ikeuchi}, \citenamefont {Kajimoto}, \citenamefont {Wakimoto},
  \citenamefont {Arai},\ and\ \citenamefont {Fujita}}]{Fujita.Sato.2020}%
  \BibitemOpen
  \bibfield  {author} {\bibinfo {author} {\bibfnamefont {K.}~\bibnamefont
  {Sato}}, \bibinfo {author} {\bibfnamefont {K.}~\bibnamefont {Ikeuchi}},
  \bibinfo {author} {\bibfnamefont {R.}~\bibnamefont {Kajimoto}}, \bibinfo
  {author} {\bibfnamefont {S.}~\bibnamefont {Wakimoto}}, \bibinfo {author}
  {\bibfnamefont {M.}~\bibnamefont {Arai}},\ and\ \bibinfo {author}
  {\bibfnamefont {M.}~\bibnamefont {Fujita}},\ }\bibfield  {title} {\bibinfo
  {title} {{Coexistence of Two Components in Magnetic Excitations of
  $\mathrm{La_{2-x}Sr_x CuO_4}$ (x = 0.10 and 0.16)}},\ }\href
  {https://doi.org/10.7566/jpsj.89.114703} {\bibfield  {journal} {\bibinfo
  {journal} {J. Phys. Soc. Jpn.}\ }\textbf {\bibinfo {volume} {89}},\ \bibinfo
  {pages} {114703} (\bibinfo {year} {2020})},\ \Eprint
  {https://arxiv.org/abs/https://doi.org/10.7566/JPSJ.89.114703}
  {https://doi.org/10.7566/JPSJ.89.114703} \BibitemShut {NoStop}%
\bibitem [{\citenamefont {Zhang}\ \emph
  {et~al.}(2023{\natexlab{b}})\citenamefont {Zhang}, \citenamefont {Chen},
  \citenamefont {Zhang},\ and\ \citenamefont {Weng}}]{Zhang.Weng.hg_2023}%
  \BibitemOpen
  \bibfield  {author} {\bibinfo {author} {\bibfnamefont {J.-X.}\ \bibnamefont
  {Zhang}}, \bibinfo {author} {\bibfnamefont {C.}~\bibnamefont {Chen}},
  \bibinfo {author} {\bibfnamefont {J.-H.}\ \bibnamefont {Zhang}},\ and\
  \bibinfo {author} {\bibfnamefont {Z.-Y.}\ \bibnamefont {Weng}},\ }\href@noop
  {} {} (\bibinfo {year} {2023}{\natexlab{b}}),\ \Eprint
  {https://arxiv.org/abs/2307.05671} {arXiv:2307.05671} \BibitemShut {NoStop}%
\bibitem [{\citenamefont {Chen}\ and\ \citenamefont
  {Weng}(2005)}]{Weng.Chen.2005}%
  \BibitemOpen
  \bibfield  {author} {\bibinfo {author} {\bibfnamefont {W.~Q.}\ \bibnamefont
  {Chen}}\ and\ \bibinfo {author} {\bibfnamefont {Z.~Y.}\ \bibnamefont
  {Weng}},\ }\bibfield  {title} {\bibinfo {title} {{Spin dynamics in a
  doped-Mott-insulator superconductor}},\ }\href
  {https://doi.org/10.1103/physrevb.71.134516} {\bibfield  {journal} {\bibinfo
  {journal} {Phys. Rev. B}\ }\textbf {\bibinfo {volume} {71}},\ \bibinfo
  {pages} {134516} (\bibinfo {year} {2005})}\BibitemShut {NoStop}%
\bibitem [{\citenamefont {Zhang}\ and\ \citenamefont
  {Weng}(2022)}]{Zhang.Weng_2022}%
  \BibitemOpen
  \bibfield  {author} {\bibinfo {author} {\bibfnamefont {J.-X.}\ \bibnamefont
  {Zhang}}\ and\ \bibinfo {author} {\bibfnamefont {Z.-Y.}\ \bibnamefont
  {Weng}},\ }\href@noop {} {} (\bibinfo {year} {2022}),\ \Eprint
  {https://arxiv.org/abs/2208.10519} {arXiv:2208.10519} \BibitemShut {NoStop}%
\bibitem [{\citenamefont {Hussey}\ \emph {et~al.}(2003)\citenamefont {Hussey},
  \citenamefont {Abdel-Jawad}, \citenamefont {Carrington}, \citenamefont
  {Mackenzie},\ and\ \citenamefont {Balicas}}]{Balicas.Hussey.2003}%
  \BibitemOpen
  \bibfield  {author} {\bibinfo {author} {\bibfnamefont {N.~E.}\ \bibnamefont
  {Hussey}}, \bibinfo {author} {\bibfnamefont {M.}~\bibnamefont {Abdel-Jawad}},
  \bibinfo {author} {\bibfnamefont {A.}~\bibnamefont {Carrington}}, \bibinfo
  {author} {\bibfnamefont {A.~P.}\ \bibnamefont {Mackenzie}},\ and\ \bibinfo
  {author} {\bibfnamefont {L.}~\bibnamefont {Balicas}},\ }\bibfield  {title}
  {\bibinfo {title} {{A coherent three-dimensional Fermi surface in a
  high-transition-temperature superconductor}},\ }\href
  {https://doi.org/10.1038/nature01981} {\bibfield  {journal} {\bibinfo
  {journal} {Nature}\ }\textbf {\bibinfo {volume} {425}},\ \bibinfo {pages}
  {814} (\bibinfo {year} {2003})}\BibitemShut {NoStop}%
\bibitem [{\citenamefont {Vignolle}\ \emph {et~al.}(2008)\citenamefont
  {Vignolle}, \citenamefont {Carrington}, \citenamefont {Cooper}, \citenamefont
  {French}, \citenamefont {Mackenzie}, \citenamefont {Jaudet}, \citenamefont
  {Vignolles}, \citenamefont {Proust},\ and\ \citenamefont
  {Hussey}}]{Hussey.Vignolle.2008}%
  \BibitemOpen
  \bibfield  {author} {\bibinfo {author} {\bibfnamefont {B.}~\bibnamefont
  {Vignolle}}, \bibinfo {author} {\bibfnamefont {A.}~\bibnamefont
  {Carrington}}, \bibinfo {author} {\bibfnamefont {R.~A.}\ \bibnamefont
  {Cooper}}, \bibinfo {author} {\bibfnamefont {M.~M.~J.}\ \bibnamefont
  {French}}, \bibinfo {author} {\bibfnamefont {A.~P.}\ \bibnamefont
  {Mackenzie}}, \bibinfo {author} {\bibfnamefont {C.}~\bibnamefont {Jaudet}},
  \bibinfo {author} {\bibfnamefont {D.}~\bibnamefont {Vignolles}}, \bibinfo
  {author} {\bibfnamefont {C.}~\bibnamefont {Proust}},\ and\ \bibinfo {author}
  {\bibfnamefont {N.~E.}\ \bibnamefont {Hussey}},\ }\bibfield  {title}
  {\bibinfo {title} {{Quantum oscillations in an overdoped high-$T_c$
  superconductor}},\ }\href {https://doi.org/10.1038/nature07323} {\bibfield
  {journal} {\bibinfo  {journal} {Nature}\ }\textbf {\bibinfo {volume} {455}},\
  \bibinfo {pages} {952} (\bibinfo {year} {2008})}\BibitemShut {NoStop}%
\bibitem [{\citenamefont {Rashba}\ \emph {et~al.}(1997)\citenamefont {Rashba},
  \citenamefont {Zhukov},\ and\ \citenamefont {Efros}}]{Efros.Rashba.1997}%
  \BibitemOpen
  \bibfield  {author} {\bibinfo {author} {\bibfnamefont {E.~I.}\ \bibnamefont
  {Rashba}}, \bibinfo {author} {\bibfnamefont {L.~E.}\ \bibnamefont {Zhukov}},\
  and\ \bibinfo {author} {\bibfnamefont {A.~L.}\ \bibnamefont {Efros}},\
  }\bibfield  {title} {\bibinfo {title} {{Orthogonal localized wave functions
  of an electron in a magnetic field}},\ }\href
  {https://doi.org/10.1103/physrevb.55.5306} {\bibfield  {journal} {\bibinfo
  {journal} {Physical Review B}\ }\textbf {\bibinfo {volume} {55}},\ \bibinfo
  {pages} {5306} (\bibinfo {year} {1997})},\ \Eprint
  {https://arxiv.org/abs/cond-mat/9603037} {cond-mat/9603037} \BibitemShut
  {NoStop}%
\end{thebibliography}%

\end{document}